\documentclass[10pt, journal,compsoc]{IEEEtran}

\usepackage{amsmath,amsfonts}
\usepackage{algorithmic}
\usepackage{algorithm}
\usepackage{array}
\usepackage[caption=false,font=normalsize,labelfont=sf,textfont=sf]{subfig}
\usepackage{textcomp}
\usepackage{stfloats}
\usepackage{url}
\usepackage{verbatim}
\usepackage{graphicx}
\usepackage{cite}
\usepackage{graphicx}
\usepackage[breaklinks=true]{hyperref}
\usepackage{caption}
\usepackage{xcolor}
\usepackage{marvosym}

\usepackage{subcaption}

\colorlet{hidden-red}{red}

\usepackage[textsize=footnotesize, linecolor=magenta, bordercolor=magenta,
    backgroundcolor=white, textwidth=40pt]{todonotes}

\usepackage{ragged2e}
\usepackage{booktabs}
\usepackage{hyperref}
\usepackage{tikz}
\usepackage[capitalize,noabbrev]{cleveref}
\usepackage{awesomebox}  

\usepackage[edges]{forest}
\usepackage{pifont}
\usepackage{tabularx}

\hypersetup{
  colorlinks=true,
  linkcolor=blue,
  filecolor=magenta,
  urlcolor=blue,
  citecolor=violet,
}
\crefname{figure}{Fig.}{Fig.}

\newlength{\extralength}
\newlength{\fulllength}
\newcommand{\phb}[1]{\vspace{.4em} \noindent\textbf{#1}\hspace{.5em}} 
\newcommand{\phm}[1]{\vspace{.4em} \noindent\textbf{#1}\hspace{.5em}} 

\usepackage{tikz}
\usepackage{etoolbox}
\newcommand{\circled}[2][]{\tikz[baseline=(char.base)]
    {\node[shape = circle, draw, inner sep = 1pt]
    (char) {\phantom{\ifblank{#1}{#2}{#1}}};%
    \node at (char.center) {\makebox[0pt][c]{#2}};}}
\robustify{\circled}

\begin{document}

\title{Efficient Training of Large Language Models on Distributed Infrastructures: A Survey}

\author{Jiangfei Duan\IEEEauthorrefmark{1}, Shuo Zhang\IEEEauthorrefmark{1}, Zerui Wang\IEEEauthorrefmark{1}, Lijuan Jiang, Wenwen Qu, Qinghao Hu, Guoteng Wang, \\ Qizhen Weng, Hang Yan, 
Xingcheng Zhang, Xipeng Qiu, Dahua Lin, Yonggang Wen, Xin Jin, \\ Tianwei Zhang and Peng Sun$^\text{\Letter}$
\IEEEcompsocitemizethanks{
\IEEEcompsocthanksitem Jiangfei Duan and Dahua Lin are with Shanghai AI Laboratory and the Chinese University of Hong Kong. 
\\ E-mail: {\it \{dj021, dhlin\}@ie.cuhk.edu.hk}
\IEEEcompsocthanksitem Shuo Zhang and Xipeng Qiu are  with Shanghai AI Laboratory and Fudan University. 
\\ E-mail: {\it \{zhangshuo, qiuxipeng\}@pjlab.org.cn}
\IEEEcompsocthanksitem Zerui Wang is with Shanghai AI Laboratory and Shanghai Jiao Tong University. 
\\ E-mail: {\it wangzerui@pjlab.org.cn}
\IEEEcompsocthanksitem Qinghao Hu, Yonggang Wen and Tianwei Zhang are with Nanyang Technological University. 
\\ E-mail: {\it \{qinghao.hu, ygwen, tianwei.zhang\}@ntu.edu.sg}
\IEEEcompsocthanksitem Xin Jin is with School of Computer Science, Peking University. 
\\ E-mail: {\it xinjinpku@pku.edu.cn}
\IEEEcompsocthanksitem Lijuan Jiang, Wenwen Qu, Guoteng Wang, Qizhen Weng, Hang Yan, Xingcheng Zhang and Peng Sun are with Shanghai AI Laboratory. 
\\ E-mail: {\it \{jianglijuan, quwenwen, wangguoteng, wengqizhen, yanhang, zhangxingcheng, sunpeng\}@pjlab.org.cn}
\IEEEcompsocthanksitem  \IEEEauthorrefmark{1} Equal Contribution.
}
}

%


\IEEEtitleabstractindextext{
\begin{abstract}
\justifying{
Large Language Models (LLMs) like GPT and LLaMA are revolutionizing the AI industry with their sophisticated capabilities. Training these models requires vast GPU clusters and significant computing time, posing major challenges in terms of scalability, efficiency, and reliability. This survey explores recent advancements in training systems for LLMs, including innovations in training infrastructure with AI accelerators, networking, storage, and scheduling. Additionally, the survey covers parallelism strategies, as well as optimizations for computation, communication, and memory in distributed LLM training. It also includes approaches of maintaining system reliability over extended training periods. By examining current innovations and future directions, this survey aims to provide valuable insights towards improving LLM training systems and tackling ongoing challenges. Furthermore, traditional digital circuit-based computing systems face significant constraints in meeting the computational demands of LLMs, highlighting the need for innovative solutions such as optical computing and optical networks.

}
\end{abstract}


}

\IEEEpubid{}

\maketitle

\section{Introduction}
\label{sec:intro}

Large Language Models (LLMs) are transforming the AI industry, demonstrating remarkable capabilities across a wide range of tasks and applications, including personal assistants~\cite{dong2023towards}, code copilot~\cite{chen2021evaluating}, chip design~\cite{liu2023chipnemo} and scientific discovery~\cite{jo2023promise}. The success of this revolution is built on the unprecedented scale of transformer-based LLMs like GPT~\cite{radford2019language}, LLaMA~\cite{touvron2023llama}, Gemini~\cite{GoogleGemini}, etc. Moreover, it is evidenced that the scaling of LLMs has not plateaued~\cite{kaplan2020scaling}. This trend has significantly shifted the design of underlying training systems and infrastructures, since LLM typically follows a relatively fixed architecture and its training exclusively occupies huge GPU clusters over extended periods. For example, the pretraining of LLaMA-3 takes approximately 54 days with 16K H100-80GB GPU on Meta's production cluster~\cite{llama3}.


LLM training highlights significant challenges for today's training system and infrastructure in terms of {``SER''}, i.e., \textit{Scalability, Efficiency}, and \textit{Reliability}. Scalability demands that both infrastructure and systems adapt seamlessly to massive clusters of tens of thousands of GPUs or AI accelerators, while preserving training correctness and model accuracy. This requires innovative solutions in hardware configuration, networking, and training framework. Efficiency focuses on maximizing resource utilization across the entire cluster, often measured by Model FLOPs Utilization (MFU). Achieving high MFU involves optimizing computation, minimizing communication overhead, and efficiently managing memory at an unprecedented scale. Reliability are critical given the prolonged duration of LLM training, usually lasting weeks to months. The system must maintain consistent performance and be resilient to various types of failures, including hardware malfunctions, network issues, and software errors. It should be capable of swiftly detecting and recovering from these failures without significant loss of progress or training quality. These interrelated challenges necessitate a holistic approach to system and infrastructure design, pushing the boundaries of large-scale distributed computing and opening new avenues for research and innovation in high-performance machine learning systems.


This survey paper aims to provide a comprehensive overview of the advancements in LLM training systems and infrastructure, addressing the aforementioned challenges. This survey spans from the distributed training infrastructure to training systems. We examine innovative approaches to infrastructure design, covering advancements in GPU clusters, high-performance networking, and distributed storage systems tailored for LLM workloads. We also explore the key aspects of distributed training systems, including parallelism strategies, computation, communication and memory optimizations that enhance the scalability and efficiency. We also delve into fault tolerance mechanisms for improving training reliability. By synthesizing recent advancements and identifying future research directions, this survey aims to provide researchers and practitioners with insights into the most promising avenues for improving LLM training systems. Our goal is to offer a valuable resource that not only addresses current challenges but also paves the way for future innovations in large-scale machine learning infrastructure.

\begin{figure*}[t]
    \centering
    \includegraphics[width=\textwidth]{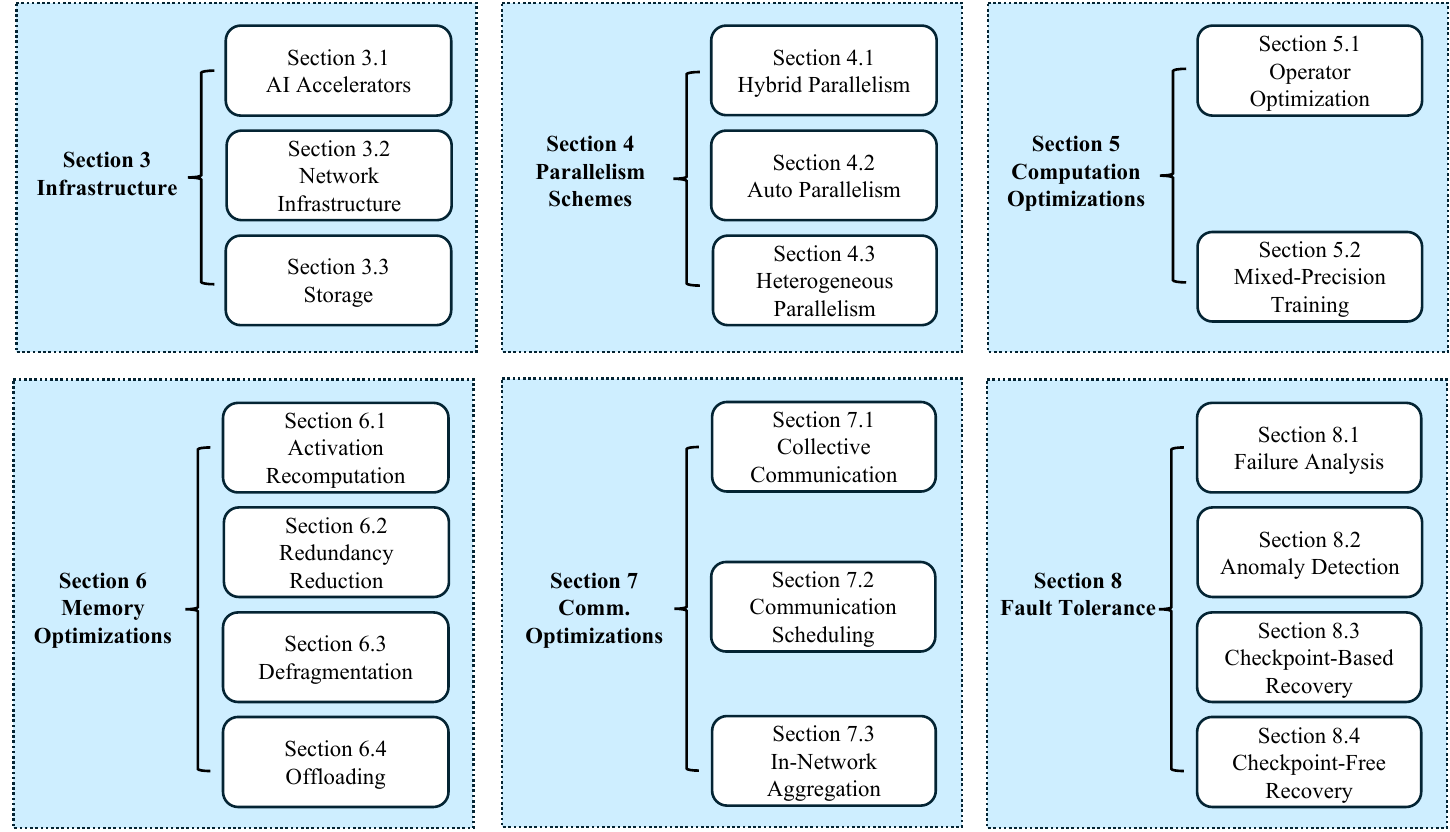}
    \caption{Overall structure of this survey.}
    \label{Fig:overall}
\end{figure*}

\phm{Organization.}
Fig.~\ref{Fig:overall} shows the organization of this survey. Section~\ref{sec:background} discusses the background information about LLM architecture, characteristics and challenges of LLM training. In Section~\ref{sec:infra}, we summarize the key aspects of the training infrastructure, including AI accelerators, network infrastructure and storage systems. In Section~\ref{sec:parallel}, we investigate the parallelism schemes for distributed LLM training. In Section~\ref{sec:comp}, we discuss computation optimizations to harness the unprecedented computational capabilities. In  Section~\ref{sec:mem}, we discuss techniques to optimize the memory footprint in LLM training. In Section~\ref{sec:comm}, we introduce communication optimizations to minimize the communication overhead. In Section~\ref{sec:resi}, we first present a failure analysis, the cover approaches to enable fast failure detection and recovery. Finally, we conclude the survey in Section~\ref{sec:conclusion}.

\section{Background}
\label{sec:background}

\subsection{Transformer-based LLMs}
\begin{figure}[t]
    \centering
    \includegraphics[width=\linewidth]{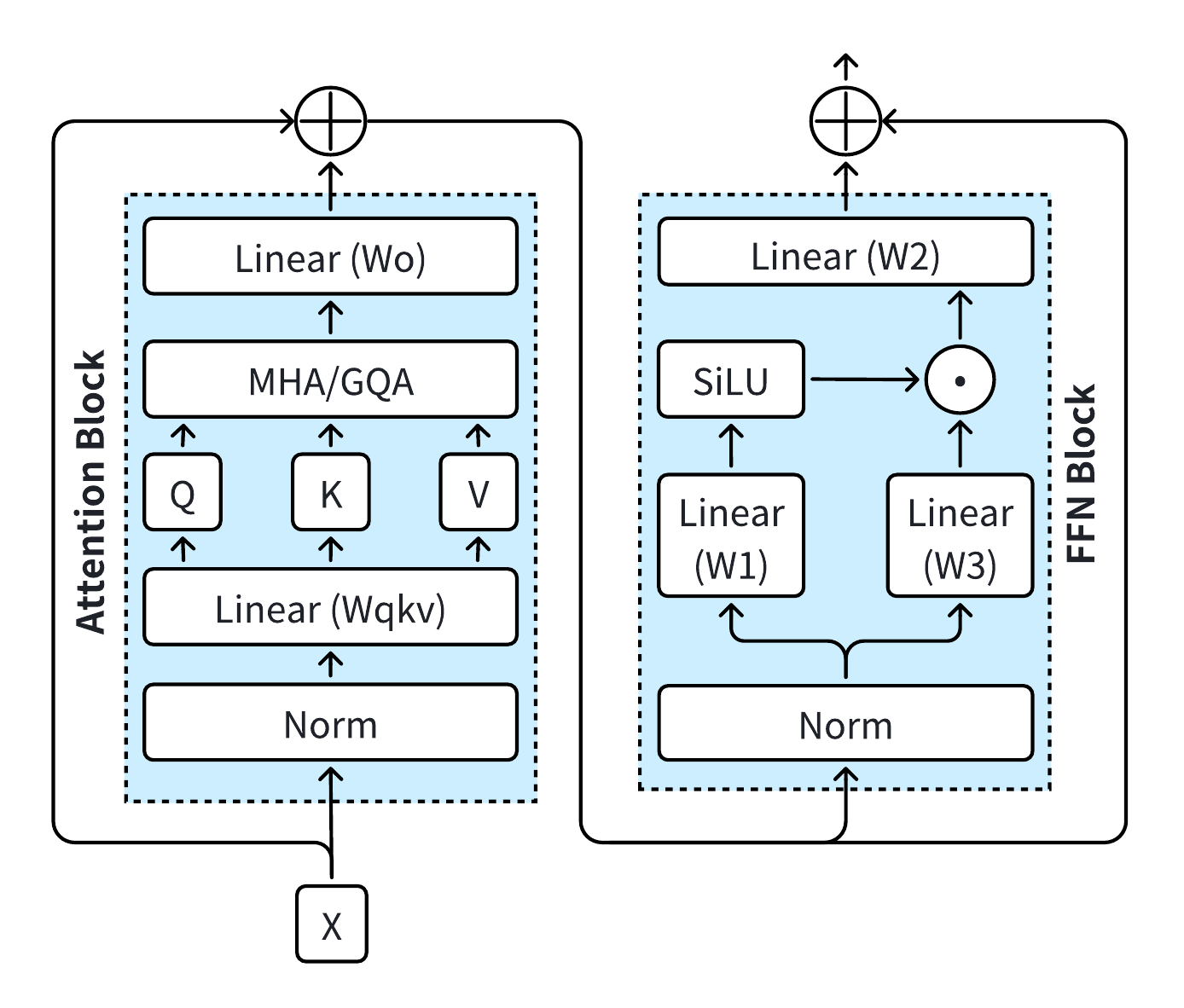}
  \caption{A typical Transformer layer contains an Attention block and a Feed-Forward Network (FFN) block.}
  \label{fig:transformer_arch}
\end{figure}

The current state-of-the-art Large Language Models (LLMs) are predominantly transformer-based. Their core architecture is built around the attention mechanism~\cite{vaswani2017attention}, which allows the model to dynamically weigh the importance of different words in a sentence. Fig.~\ref{fig:transformer_arch} depicts the typical architecture of a transformer layer~\cite{vaswani2017attention}, which can be stacked multiple times to construct an LLM. The input text is first tokenized into individual tokens, which are then converted into a token vector $X$ via an embedding layer. To preserve the sequential nature of the text, the token vector is embedded with positional information. The resulting token vector is then fed into the transformer layer, which consists of an Attention block and a Feed-Forward Neural Network (FFN) block.

Suppose the input token vector is $X=[x_1, x_2, \cdots, x_n]$. These tokens are first transformed into query $Q$, key $K$ and value $V$ tensors via linear transformation. The attention mechanism computes the attention output as follows:
\begin{equation}
\text{Attention}(Q, K, V) = \texttt{softmax}\left(\frac{QK^T}{\sqrt{d}}\right)V
\end{equation}
where $d$ is the dimension of the key tensor. This formula ensures that the LLM can focus on relevant parts of the input sequence by calculating a weighted sum of the values, where the weights are derived from the similarity between the queries and keys. After the attention layer, the output is passed through a FFN for further processing.

Nowadays, LLMs generally follow the original decoder-only transformer architecture but incorporate modifications to the attention mechanism and FFN to enhance efficiency and performance. The original attention mechanism, known as Multi-Head Attention (MHA)~\cite{vaswani2017attention}, suffers from quadratic computational complexity and high memory consumption due to the key-value cache. To address these issues, several variants such as Multi-Query Attention (MQA)~\cite{shazeer2019fast}, Group-Query Attention (GQA)~\cite{ainslie2023GQATraining} and Multi-Latent Attention (MLA)~\cite{deepseekai2024deepseekv2} have been proposed. A notable advancement of the FFN component is the Mixture-of-Experts (MoE)~\cite{jacobs1991adaptive, lepikhin2020gshard} architecture, which employs a sparsely activated FFN. In MoE, only a subset of the FFN layers (or experts) are activated for each input, significantly reducing the computational load while maintaining high model capacity.

\subsection{LLM Training Workloads Characteristics}

The characteristics of LLM training workloads diverge significantly from traditional deep learning workloads, primarily due to their complexity and scale. These unique characteristics affect training system design, performance, scalability, and resource utilization. Here, we highlight key differences and requirements for LLMs.

\phm{(1) Homogeneous Model Architecture.}~Unlike prior DL workloads that employed various model architectures (e.g., LSTM \cite{LSTM}, CNN \cite{CNN}) for different tasks, LLMs predominantly use the Transformer architecture \cite{vaswani2017attention}. Models like GPT \cite{radford2019language}, LLaMA \cite{touvron2023llama}, InternLM\cite{cai2024internlm2} and MOSS\cite{Sun2024MOSS} all share this common foundation. This architectural uniformity presents significant potential for optimizing system performance for a particular model architecture.

\phm{(2) Unprecedented Scale and Training Duration.}~LLM training operates at an unparalleled scale, typically updating models with hundreds of billions of parameters with terabyte-scale training datasets. Such magnitude necessitates distributed training across huge GPU clusters and presents challenges in maintaining high efficiency. Besides, the training of LLMs can span weeks or months, demanding robust fault tolerance mechanisms and efficient checkpointing strategies to safeguard against data loss and facilitate the resumption of interrupted training sessions.

\phm{(3) Specialized Software Optimization.}~
To accommodate the enormous model size of LLMs, specialized systems implement advanced techniques to optimize execution. For example, Megatron \cite{shoeybi2019megatron} and Alpa \cite{zheng2022AlpaAutomatinga} accelerate training through hybrid parallelism. DeepSpeed \cite{rasley2020deepspeed} reduces memory consumption by integrating state-sharding optimizers. 

\phm{(4) Shift in Training Paradigm.}~Traditional DL workloads follow a task-specific paradigm, training models on domain-specific data for particular tasks, such as translation. In contrast, LLMs adopt a self-supervised training approach on extensive datasets to create foundation models, which are then adapted for various downstream tasks. This paradigm shift represents a substantial change in the model development pipeline, including pretraining and alignment phases, and results in distinct workload characteristics compared to previous DL workloads. From the datacenter perspective, LLM development involves numerous small-scale workloads that are associated with pretraining, including alignment (i.e., fine-tuning) and periodical evaluation workloads \cite{hu2024CharacterizationLarge}.

\subsection{LLM Training Challenges}

The unique characteristics of LLM training workloads give rise to significant challenges in developing efficient training systems and infrastructure. These challenges primarily manifest in three critical areas: \textit{scalability, efficiency}, and \textit{reliability}. Each of these challenges stems directly from the massive scale of LLMs and the complexity of their training processes, requiring innovative solutions that push the boundaries of distributed computing and machine learning systems. Below, we detail these challenges and their implications for LLM training:


\phm{(1) Scalablity.} The success of LLMs is largely attributed to their scale, with performance often improving as LLMs grow larger~\cite{kaplan2020scaling}. However, the scaling of model size introduces substantial scalability challenges, since training LLMs requires increasingly large clusters of GPUs or specialized AI accelerators. First, building scalable infrastructure that can provide massive computations and memory capacity is essential. This involves designing and deploying large clusters of GPUs or specialized AI accelerators, high-performance networking to connect these devices, and distributed storage systems capable of handling enormous datasets and model checkpoints. The challenge lies in ensuring that these components work together efficiently at scale, managing heat dissipation, power consumption, and hardware failures in large-scale deployments. Second, designing scalable training systems that can effectively utilize massive accelerators in parallel is crucial. This includes designing parallelization strategies and communication algorithms that can achieve near-linear scalability with thousands of accelerators while maintaining consistent LLM accuracy.

\phm{(2) Efficiency.} The enormous computational requirements of LLM training translate to high training costs, making it imperative to maximize the efficiency of hardware and software systems. The efficiency can be measured with MFU (Model FLOPs Utilization), which quantifies how effectively the system uses available computing resources. However, achieving high efficiency at scale remains a significant challenge. For instance, LLaMA3 achieves only $38\%$ to $41\%$ MFU on 16K GPUs~\cite{llama3}, highlighting the difficulty in maintaining high utilization as systems scale. Maximizing efficiency demands the optimization in parallelism, computation, communication and memory. First, the parallelism of distributed LLM training demands careful design to minimize communication demands. Second, optimized computation operators and lower precision arithmetic are crucial to achieve high GPU FLOPS utilization. Third, communication overhead needs to minimized to reduce GPU idle time. Finally, efficient memory optimizations are required to hold LLMs in existing hardwares and reduce FLOPs waste of recomputation.

\phm{(3) Reliability.} Ensuring the reliability of LLM training  over extended periods is paramount. As training jobs can span weeks or months on large clusters of tens of thousands of GPUs, the probability of training failures increases, necessitating fast failure detection and recovery mechanism for resilient LLM training. First, LLM training jobs may crash due to various errors, making it hard to identify the exact fault reason across tens of thousands GPUs quickly. Second, the hang of LLM training jobs results in all GPUs becoming idle due to the synchronous nature of training, resulting in significant waste. Moreover, some intricate anomalies, like redundant link failures or stragglers, may not cause immediate crashes but can lead to training slowdowns. This instability can result in reduced training efficiency. To tackle these challenges, robust anomaly detection systems capable of detecting both catastrophic failures and performance degradations are essential. Additionally, implementing fault-tolerant training frameworks that can seamlessly handle node failures and network issues is crucial. 


\subsection{Related Survey}

This work focuses on the efficient training system and infrastructure of transformer-based LLMs, including the design of underlying distributed infrastructure, paradigm of parallelism, optimizations of computation and communication, efficient memory management and resilience of training system. We also investigate the efficient training systems of emerging workloads such as MoE, a promising efficient LLM variant, and fine-tuning, a necessary stage to align the capabilities of LLMs. However, this work does not cover the evolution of promising LLM architectures~\cite{zhou2023comprehensive, huang2023advancing} and the algorithms for training~\cite{zhao2023survey}, instruction tuning~\cite{zhang2023InstructionTuning} and alignment~\cite{wang2023aligning} towards powerful and safe LLMs. While previous works~\cite{wan2023EfficientLarge, liu2024understanding, xu2024SurveyResourceefficient} have discussed some aspects of LLM training systems, their primary focus was not on the design of efficient training systems and infrastructure. Wan et al.~\cite{wan2023EfficientLarge} aims to provide a holistic view of efficient LLM advances in model- and data-centric methods. Liu et al.~\cite{liu2024understanding} covers the training and inference deployment techniques of LLMs. Xu et al.~\cite{xu2024SurveyResourceefficient} targets on discussing the resource-efficient strategies for LLM development in both algorithmic and systemic aspects. This work also discuss the approaches for quantized LLM training and efficient LLM fine-tuning, but we focus on the systemic approaches. The algorithmic approaches to compress and fine-tune LLMs are discussed by Zhu et al.~\cite{zhu2023survey} and Han et al.~\cite{han2024parameter}. The discussion scope of this work does not include advanced optimization algorithms~\cite{he2021large} and distributed DNN training systems~\cite{mayer2020scalable}. While Liang et al.~\cite{liang2023survey} have extensively reviewed auto parallelism approaches, their focus is on general DNNs rather than LLMs specifically.

\begin{figure}[t]
    \centering
    \includegraphics[width=\linewidth]{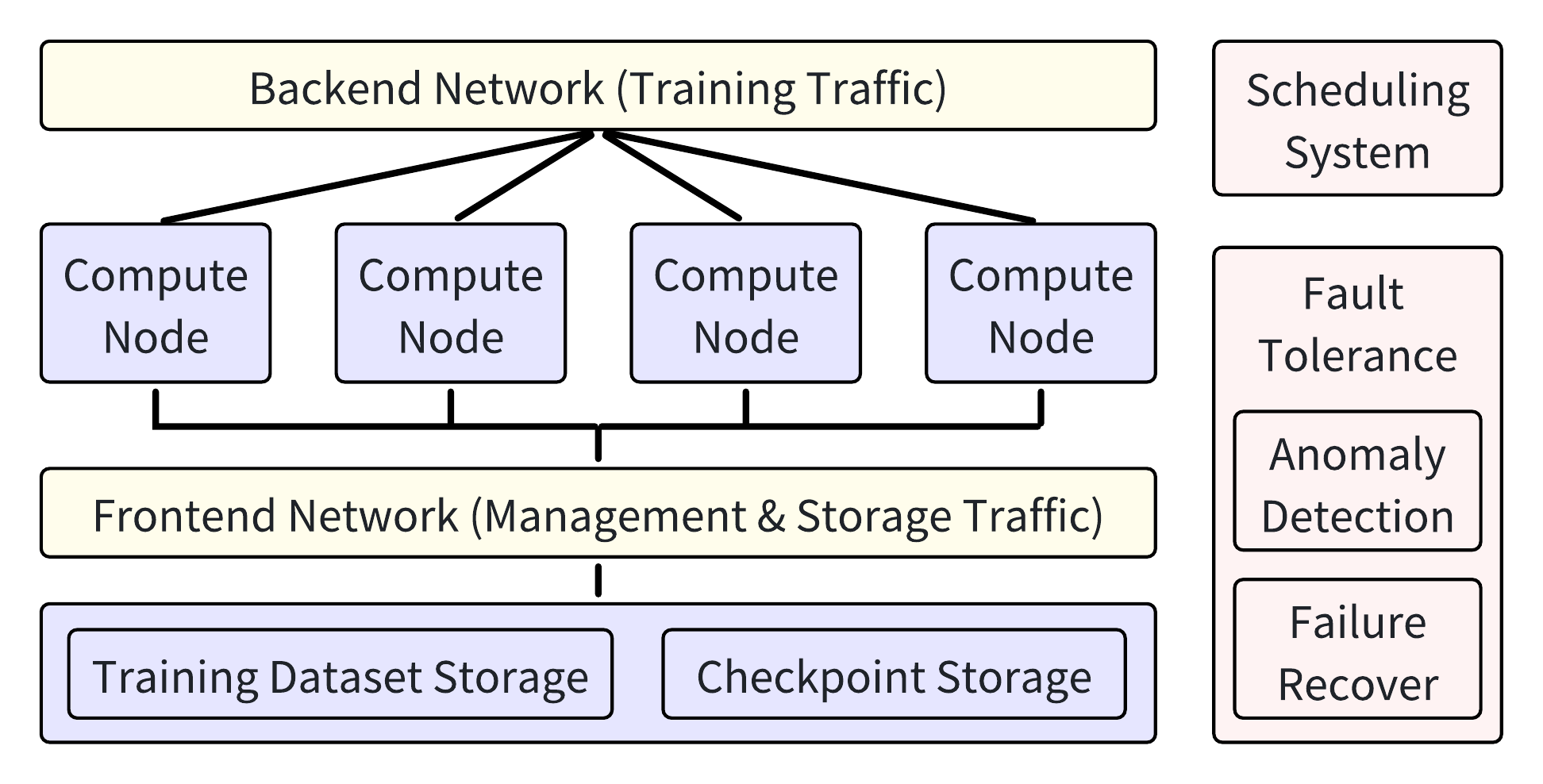}
    \caption{Infrastructure overview for distributed LLM training. }
    \label{fig:infra_arch}
\end{figure}

\section{Infrastructure for LLM Training}
\label{sec:infra}

\tikzstyle{my-box}=[
    rectangle,
    draw=black,
    rounded corners,
    text opacity=1,
    minimum height=1.5em,
    minimum width=5em,
    inner sep=2pt,
    align=center,
    fill opacity=.5,
    line width=0.5pt,
]
\tikzstyle{leaf}=[my-box, minimum height=1.5em,
    fill=hidden-red!10, text=black, align=left,font=\normalsize,
    inner xsep=2pt,
    inner ysep=4pt,
    line width=0.8pt,
]

\begin{figure*}[htpb]
    \centering
    \resizebox{\textwidth}{!}{
        \begin{forest}
        forked edges,
        for tree={
            grow=east,
            reversed=true,
            anchor=base west,
            parent anchor=east,
            child anchor=west,
            base=center,
            font=\large,
            rectangle,
            draw=black,
            rounded corners,
            align=left,
            text centered,
            minimum width=4em,
            edge+={black, line width=1pt},
            s sep=3pt,
            inner xsep=2pt,
            inner ysep=3pt,
            line width=0.8pt,
            ver/.style={rotate=90, child anchor=north, parent anchor=south, anchor=center},
        },
        where level=1{text width=12em,font=\normalsize,}{},
        where level=2{text width=12em,font=\normalsize,}{},
        [
        {Infrastructure for LLM Training}, ver
                [
                {AI Accelerators}, fill=blue!10
                    [
                    {NVIDIA GPUs}, fill=yellow!10
                        [
                        Ampere~\cite{ampere-architecture}{, }
                        Hopper~\cite{hopper-architecture}{, }
                        Blackwell~\cite{blackwell-architecture}
                        ,
                        leaf, 
                        text width= 32em
                        ]
                    ]
                    [
                    {Other AI Accelerators}, fill=yellow!10
                        [
                        AMD GPU ~\cite{swaminathan2023amd}{, }
                        GAUDI~\cite{habana2023gaudi}{, }
                        TPU~\cite{jouppi2023TPUv4}{, }
                        Graphcore IPU~\cite{emani2023comprehensive}{, }\\
                        Cerebras CS-2~\cite{fricker2022CS2} 
                        ,
                        leaf, 
                        text width= 32em
                        ]
                    ]
                ]
                [
                {Network Infrastructure}, fill=blue!10
                    [
                    {Chip-to-Chip}, fill=yellow!10
                        [
                        \textbf{Cube-Mesh Topology:} NVLink-1.0~\cite{NVIDIA2017DGX1}\\
                        \textbf{Fully-Connected Topology:} NVSwitch~\cite{nvswitch}{, }
                        Infinity Fabric~\cite{naffziger2020AMDChipletArchitecture}\\
                        \textbf{2D/3D-Torus Topology:} TPUv2~\cite{tpuv2}{, }
                        TPUv3~\cite{tpuv3}{, }
                        TPUv4~\cite{jouppi2023TPUv4}
                        ,
                        leaf, 
                        text width= 32em
                        ]
                    ]
                    [
                    {Node-to-Node}, fill=yellow!10
                        [
                        GPUDirect-RDMA~\cite{nvidia_gpudirect}{, }
                        InfiniBand EDR/HDR/NDR ~\cite{infiniband}{, }\\
                        RoCE-v1~\cite{Infiniband2010A16}{, }
                        RoCE-v2~\cite{Infiniband2010A17}{, }
                        iWARP~\cite{rdmaconsortium}
                        ,
                        leaf, 
                        text width= 32em
                        ]
                    ]
                    [
                    {Network Topology}, fill=yellow!10
                        [
                        \textbf{HPC Topology:} Clos~\cite{clos1953study}{, }
                        BCube~\cite{guo2009bcube}{, }
                        DCell~\cite{guo2008dcell}{, }
                        Jellyfish~\cite{singla2012jellyfish}{, } \\
                        Torus~\cite{duato2003interconnection}{, }
                        Dragonfly~\cite{kim2008technology}{, }
                        Dragonfly+~\cite{shpiner2017dragonfly+}\\
                        \textbf{Training-Optimized Topology:} Rail-Optimized~\cite{rail-optimized-topology}{, } HPN~\cite{qian2024alibaba}{, } \\
                        Rail-Only~\cite{wang2023railonly}{, } 
                        BiGraph~\cite{qian2024alibaba}{, } 
                        HammingMesh  \cite{hoefler2022hammingmesh} \\
                        {Reconfiguable Topology:} SiP-ML~\cite{Khani2021sipml}{, }
                        TopoOpt~\cite{weiyang2023topoopt}{, }
                        TPUv4~\cite{jouppi2023TPUv4}
                        ,
                        leaf, 
                        text width= 32em
                        ]
                    ]
                    [
                    {Load Balancing \& CC}, fill=yellow!10
                        [
                        \textbf{Load Balancing: } ECMP~\cite{ecmp}{, }
                        Multi-Flow with Enhanced-ECMP~\cite{llama3}{,} \\
                        Packet spraying~\cite{packetspraying}{, } 
                        Ethereal~\cite{addanki2024challenging}{, }
                        HPN~\cite{qian2024alibaba}{, }
                        MegaScale~\cite{jiang2024MegaScaleScaling} \\
                        \textbf{Congestion Control (CC): }PFC~\cite{PFC}{, }
                        TIMELY~\cite{mittal2015timely}{, } 
                        Swift~\cite{kumar2020swift}{, } \\
                        DCQCN~\cite{dcqcn,zhu2016ecn}{, } 
                        HPCC~\cite{li2019hpcc}{, } 
                        EQDS~\cite{olteanu2022edge}{, }
                        RoCC~\cite{taheri2020rocc}{, } \\
                        MLTCP~\cite{rajasekaran2024mltcp}{, } 
                        CASSINI~\cite{rajasekaran2023cassini}{, }
                        MLT~\cite{wang2024mlt}
                        ,
                        leaf, 
                        text width= 32em
                        ]
                    ]
                ]
                [
                {Storage Systems}, fill=blue!10
                    [
                    {Checkpoint Storage}, fill=yellow!10
                        [
                        Tectonic~\cite{pan2021facebook}{, }
                        HDFS~\cite{hdfs}{, }
                        Ceph Object Storage~\cite{weil2006ceph}
                        ,
                        leaf, 
                        text width= 32em
                        ]
                    ]
                    [
                    {Training Data Storage}, fill=yellow!10
                        [
                        Lustre \cite{schwan2003lustre}{, }
                        GPFS \cite{schmuck2002gpfs}{, }
                        BeeGFS \cite{chowdhury2019characterization}{, }
                        Alluxio \cite{li2014tachyon}{, }
                        JuiceFS \cite{JuiceFS}{, }\\
                        Quiver \cite{kumar2020quiver}{, }
                        Fluid \cite{gu2022fluid}
                        ,
                        leaf, 
                        text width= 32em
                        ]
                    ]
                ]
                [
                {Scheduling Systems}, fill=blue!10
                    [
                    {Workload Scheduling}, fill=yellow!10
                        [
                        Tiresias~\cite{gu2019tiresias}{, }
                        THEMIS~\cite{mahajan2020themis}{, }
                        ElasticFlow~\cite{gu2023ElasticFlowElastic}{, }
                        Gavel~\cite{Gavel}{, } \\
                        Gandiva$_\text{fair}$~\cite{Gandivafair}{, }
                        FGD~\cite{FGD}{, }
                        Lucid~\cite{hu2023LucidNonintrusive}{, }
                        Pollux~\cite{Pollux}{, }
                        Sia~\cite{Sia}{, } \\
                        Crius~\cite{Crius}{, }
                        Hydro~\cite{Hydro}{, }
                        Acme~\cite{hu2024CharacterizationLarge}
                        ,
                        leaf, 
                        text width= 32em
                        ]
                    ]
                    [
                    {Resource Scheduling}, fill=yellow!10
                        [
                        Cassini~\cite{rajasekaran2023cassini}{, }
                        HIRE~\cite{HIRE}{, }
                        SiloD~\cite{zhao2023SiloDCodesign}{, }
                        Synergy~\cite{mohan2022LookingGPUs}{, }
                        EnvPipe~\cite{EnvPipe}{, }\\
                        Zeus~\cite{you2023ZeusUnderstanding}{, }
                        Perseus~\cite{chung2023perseus}
                        ,
                        leaf, 
                        text width= 32em
                        ]
                    ]
                ]
        ]
        \end{forest}
    }
    \caption{Studies on infrastructure optimizations for distributed LLM training.}
    \label{taxonomy:infra}
\end{figure*}
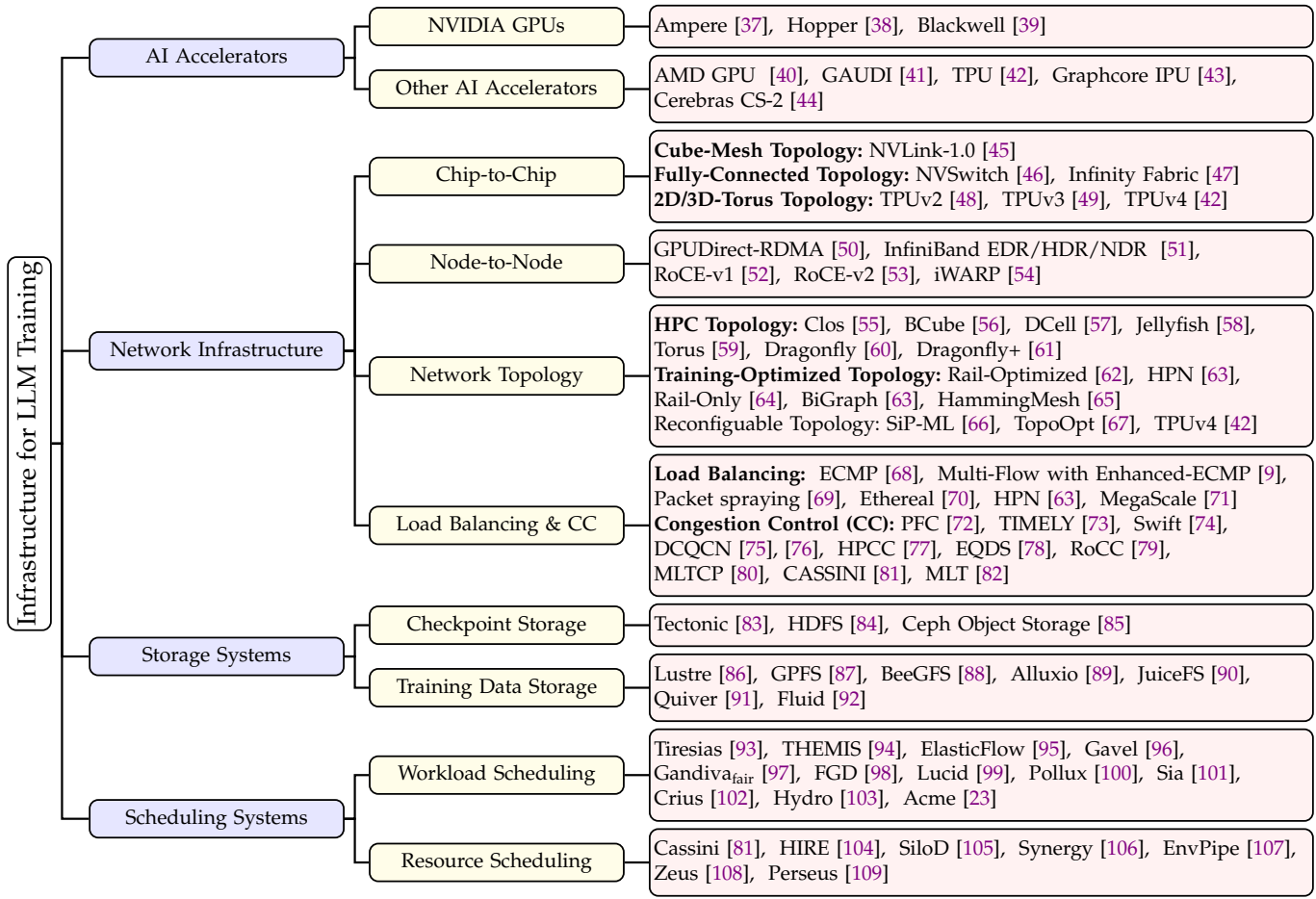

In this section, we explore the infrastructure design for training LLMs, encompassing accelerators, networks, storage, and scheduling systems (Fig.~\ref{fig:infra_arch}).

\subsection{AI Accelerators}
\label{subsec:accelerators}


The rapid progress of LLMs is significantly benefiting from the evolution of GPUs and AI accelerators, which are crucial for improving the model training  performance. 

\subsubsection{NVIDIA Graphics Processing Unit (GPU)}
\label{subsubsec:gpu}

NVIDIA GPUs have become an essential component for distributed LLM training  due to their superior ability to handle parallel computations. These processors are built with a multitude of compact, high-efficiency cores that can execute numerous tasks simultaneously. The design of GPUs is ideally matched for the matrix and vector operations in LLM training. They offer support for various numerical precision formats such as FP32, TF32, FP16, BF16, FP8, INT8, and even FP4. This allows researchers to well balance the training speed and accuracy, making LLM training more efficient~\cite{miao2023towards}. NVIDIA's GPU programming language (i.e. CUDA) makes it easier to manage how tasks are split and processed in parallel on GPUs. This helps researchers harness the full power of GPUs for training advanced LLMs.

A typical GPU comprises an array of Streaming Multiprocessors (SMs), with each SM housing several cores that share an instruction unit but are capable of executing distinct threads in parallel. The shared memory within each SM allows for effective data exchange and synchronization among threads, which is essential for optimizing the memory access patterns required for LLM computations.
Furthermore, GPUs are furnished with high-bandwidth memory (HBM), which accelerates data transfer and mitigates memory access bottlenecks in computationally intensive tasks. The latest GPU architectures, such as NVIDIA’s Ampere \cite{ampere-architecture}, Hopper \cite{hopper-architecture} and Blackwell \cite{blackwell-architecture}, are continuously pushing the boundaries of LLM computation. They offer enhanced memory bandwidth and capacity, increased floating-point operations per second (FLOPS), and specialized mixed-precision computing units like Tensor Cores. Notably, NVIDIA’s Hopper architecture introduces a significant advancement with the Transformer Engine \cite{choquette2022nvidia}, a feature that leverages mixed FP8 and FP16 precisions to expedite the training of Transformer-based LLMs. 

\subsubsection{Other AI Accelerators}
\label{subsubsec:OtherAIAccelerators}

Distributed LLM training on AMD GPUs has become a reality, particularly on Frontier \cite{schneider2022exascale}, the world's first exascale supercomputer. Each Frontier node is equipped with 8 MI250X \cite{swaminathan2023amd} AMD GPUs, each with 64 GB of HBM and a theoretical FP16 peak performance of 191.5 TFLOPS. This configuration presents an unparalleled opportunity for training trillion-parameter models efficiently. The key to unlocking this potential lies in adapting existing CUDA-based tools and frameworks to the ROCm platform \cite{dash2024optimizing,yin2023evaluation}. Notably, ROCm-enabled versions of FlashAttention \cite{dao2022flashattention} and FlashAttention2 \cite{dao2023FlashAttention2Faster}  have been developed, allowing for the efficient execution of attention. 

Various  AI accelerators with powerful computing and software optimizations have been developed to train LLMs. 
GAUDI \cite{habana2023gaudi} offers a heterogeneous compute architecture comprising two Matrix Multiplication Engines and a cluster of  fully programmable tensor processor cores, enabling efficient handling of LLM training operations. This processor can support the training of a GPT-3 model with 175 billion parameters using 384 GAUDI2 cards  \cite{zhang2023benchmarking}. 
Google TPUv4 \cite{jouppi2023TPUv4} supercomputer has 4096 chips, supporting LLM training at an average of approximately $60\%$ of peak FLOPS.
Graphcore Bow Pod64 \cite{emani2023comprehensive}, a one-rack setup with 64 Bow-class IPUs, achieves 22 petaFLOPS. It supports GPT-3 model training with 256 IPUs. 
Cerebras CS-2\cite{fricker2022CS2} is a wafer-scale deep
learning accelerator comprising 850,000 processing cores, each providing 48KB of dedicated SRAM memory. It is used to train Cerebras-GPT, a family of open compute-optimal language models \cite{dey2023cerebras}.
%

\subsection{Network Infrastructure}
\label{subsec:networkinfra}

Communication overhead is a major obstacle to scaling LLM training \cite{zhang2020isnetwork,dai2024high}. For example, reducing model gradients during training can lead to over 90\% of the training time being spent on communication~\cite{zhao2022Multiresourceinterleaving}. To address this, the research community has focused on improving communication infrastructure for LLM training.


\begin{figure*}[t]
    \centering
    \includegraphics[width=\linewidth]{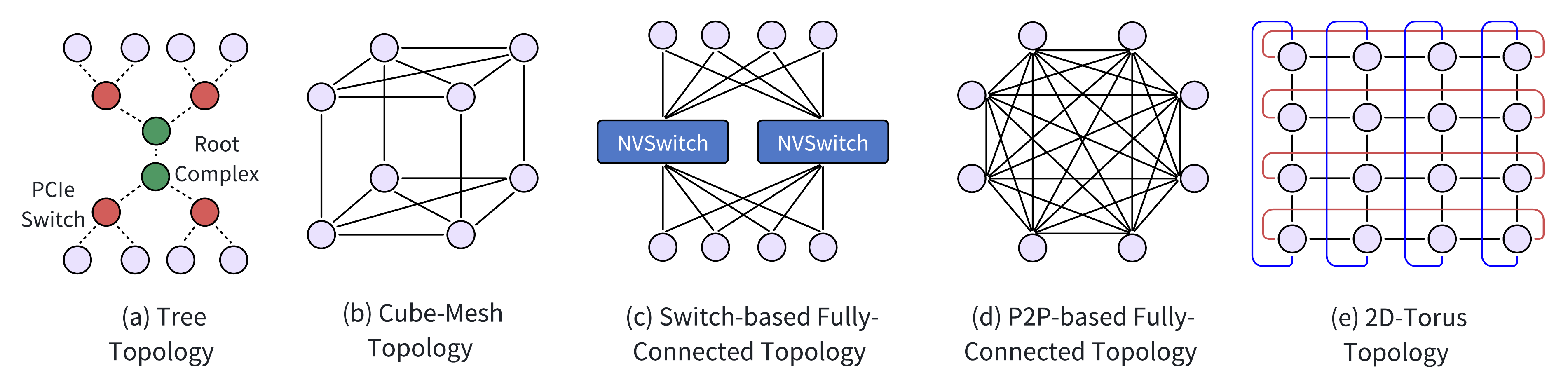}
    \caption{Five chip-to-chip topologies: tree topology, cube-mesh topology, switch-based fully-connected topology, P2P-based fully-connected topology, and 2D-torus topology.}
    \label{c2c_topo}
\end{figure*}

\subsubsection{Chip-to-Chip Communications}
\label{subsubsec:chip-to-chip}

Chip-to-chip communication is pivotal for data transfer between AI accelerators within a node, significantly impacting the efficiency of LLM training. Traditionally, this communication has relied on PCI Express (PCIe) \cite{pcieRC}, which employs a tree topology—a hierarchical structure where multiple devices connect to a single  root complex. Over the years, PCIe has improved its bandwidth: PCIe 3.0 offers approximately 1 GB/s per lane, totaling about 16 GB/s for a configuration with 16 lanes; PCIe 4.0 doubles the bandwidth to 2 GB/s per lane, while PCIe 5.0 further increases it to 4 GB/s per lane. Despite these enhancements, PCIe's inherent limitations in bandwidth, latency, and scalability render it suboptimal for LLM training \cite{li2019evaluating}.
To address these limitations, specialized chip-to-chip interconnects like NVLink \cite{NVIDIA2017DGX1} are increasingly preferred for LLM training. Compared to PCIe, these advanced interconnects provide significantly higher bandwidth and lower latency by utilizing various topologies: \textit{cube-mesh}, \textit{fully-connected}, and \textit{3D-torus}. Additionally, shared memory models, specialized communication protocols, and synchronization mechanisms also play crucial roles.

\phm{Cube-Mesh Topology.} NVLink-1.0 \cite{NVIDIA2017DGX1} offers a direct and high-speed connection between GPUs, with each link providing 160 GB/s of bidirectional bandwidth. This architecture enables the formation of planar mesh structures for four GPUs and a cube-mesh topology for eight GPUs, which can be configured into a DGX-1 server. This cube-mesh configuration, although not an all-to-all connection, significantly enhances data communication efficiency and training performance on GPUs.

\phm{Fully-Connected Topology.} Many interconnects utilize either switch-based or P2P-based fully connected topologies to improve chip-to-chip communication performance. NVIDIA uses NVSwitch \cite{nvswitch} to achieve switch-based all-to-all interconnections among GPUs. In the DGX-2 \cite{NVIDIA2018DGX2} system, six NVSwitches fully connect each of the sixteen GPUs to all others, providing a bidirectional bandwidth of 300 GB/s between any two GPUs. This bandwidth increased to 600 GB/s with NVSwitch 2.0  and further to 900 GB/s with NVSwitch 3.0. Intel, AMD, and Huawei Ascend utilize P2P-based fully-connected topology for their accelerators, where each chip directly connects to every other chip within the same node using Ethernet or Infinity Fabric \cite{naffziger2020AMDChipletArchitecture}. Compared to the switch-based topology, the bandwidth between two GPUs in a P2P-based topology is limited by the bandwidth of the directly connected link.

\phm{2D/3D-Torus Topology.} 
Google's TPU systems utilizes Torus network topology \cite{duato2003interconnection} for chip-to-chip communication. It establishes connectivity by linking each TPU chip to its four adjacent neighbors in a grid, with wraparound edges, thus forming a toroidal structure. This architectural design ensures low latency and high bandwidth due to the availability of multiple direct paths between chips.  Specifically, the TPUv2 \cite{tpuv2} supercomputer employs a 16x16 2D torus configuration, encompassing 256 chips, interconnected via high-speed Inter-Chip Interconnect (ICI) links. The TPUv3 \cite{tpuv3} supercomputer utilizes a 32x32 2D torus,  comprising 1024 chips.  Advancing from the 2D torus design, the TPUv4 \cite{jouppi2023TPUv4} supercomputer organizes compute resources into multi-machine cubes with a 3D torus topology. Each TPU machine contains four  chips arranged in a 2x2x1 mesh, interconnected through the ICI links. Sixteen of these TPU machines are combined to form a data center rack, where  ICI links within the rack interconnect to create a 4x4x4 mesh, resulting in a 3D torus structure. This advanced configuration significantly enhances communication efficiency and scalability, particularly beneficial for LLM training.

\subsubsection{Node-to-Node Communications}
\label{subsubsec:node-to-node}

Remote Direct Memory Access (RDMA)\cite{rdmaconsortium}  enables high-speed and low-latency data transfer between nodes. RDMA allows direct memory access from the memory of one computer to another without involving either node's operating system.  GPUDirect-RDMA\cite{nvidia_gpudirect} enhances this process by enabling direct communication between GPUs across different nodes, bypassing the CPU entirely.  This technology is particularly beneficial for LLM training, as it accelerates the synchronization of model parameters and gradients. The two most prevalent RDMA technologies are InfiniBand \cite{infiniband} and RDMA over Converged Ethernet (RoCE)\cite{Infiniband2010A16}.  

InfiniBand is a high-speed, low-latency networking technology widely utilized in HPC (High Performance Computing) environments, such as the Eagle supercomputer~\cite{elster2022hopper}. This technology necessitates a dedicated network infrastructure, reflecting its design focus on delivering superior performance. Over the years, InfiniBand has significantly evolved in terms of bandwidth capabilities, advancing from EDR (Enhanced Data Rate) at 100 Gbps to HDR (High Dynamic Range) at 200 Gbps, and more recently to NDR (Next Data Rate) at 400 Gbps per link \cite{nvidia2020infiniband}.
RoCE leverages the existing Ethernet infrastructure to deliver RDMA capabilities. This approach offers a more cost-effective and easier-to-deploy solution, particularly in data centers that already utilize Ethernet. RoCE is available in two versions: RoCE-v1 \cite{Infiniband2010A16}, which operates as an Ethernet link layer protocol, and RoCE-v2 \cite{Infiniband2010A17}, which operates over UDP.
Industry leaders such as ByteDance and Meta have employed these technologies to scale LLM training.
Another RDMA protocol, the Internet Wide Area RDMA Protocol (iWARP) \cite{rdmaconsortium}, enables RDMA over TCP/IP networks. However, due to its comparatively limited performance, iWARP is not commonly used for  distributed LLM training\cite{roce_vs_iwarp_2017}.

\subsubsection{Network Topology}
\label{subsubsec:topology}

\begin{figure*}[t]
    \centering
    \includegraphics[width=\linewidth]{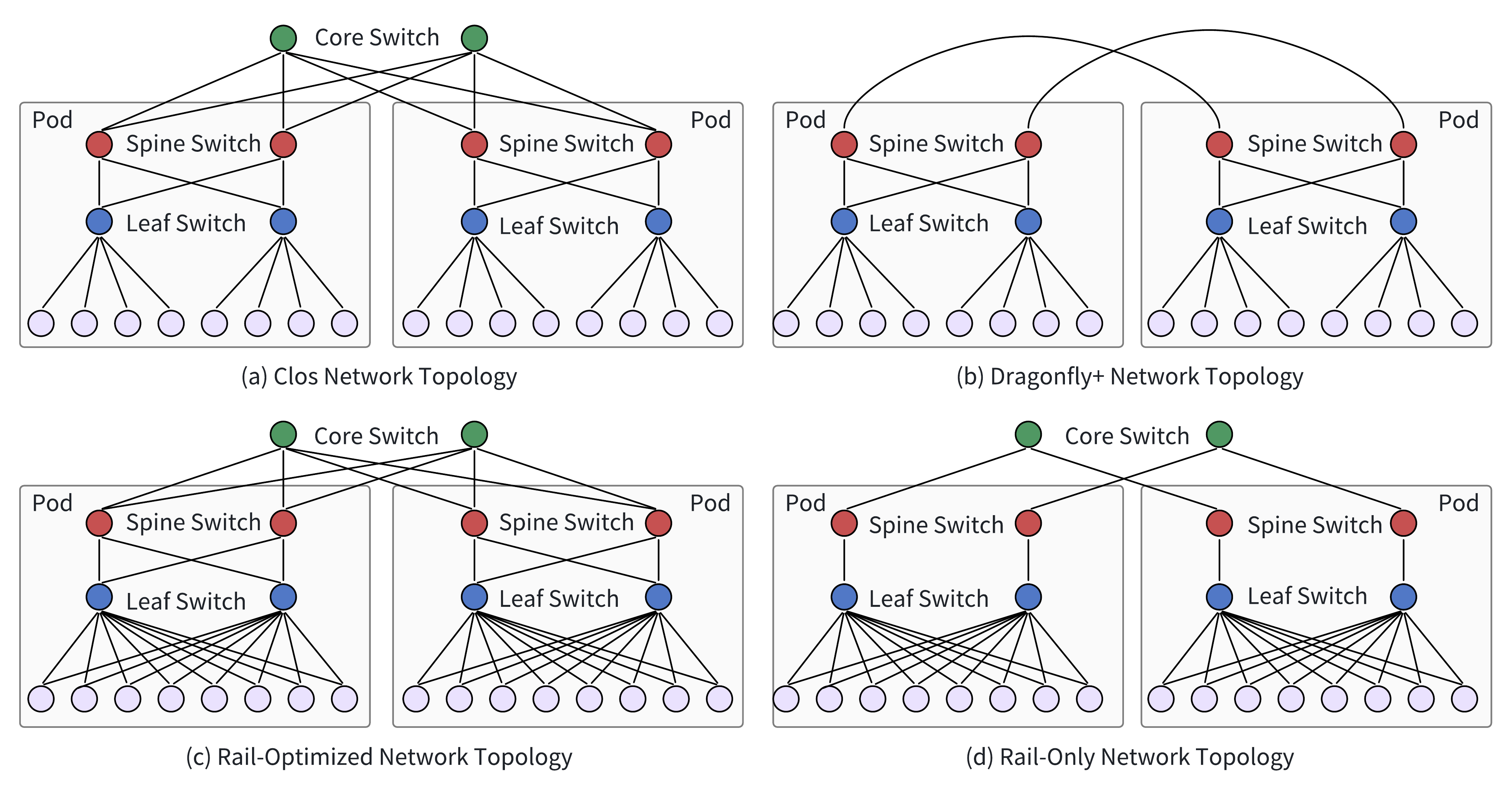}

    \caption{Four typical network topologies in large-scale GPU clusters: Clos topology, Dragonfly+ topology, rail-optimization topology, and rail-only topology.}
    \label{topo_all}
\end{figure*}

In LLM training clusters, the network architecture is structured into frontend and backend components (Fig.~\ref{fig:infra_arch}). The frontend network handles a variety of traffic, such as job management, model inference, and storage activities, while the backend network is dedicated to the high-volume traffic generated during the training process. Our primary focus in optimizing LLM training lies in improving the performance and efficiency of this backend network, so as to scale AI accelerators to tens of thousands.

\phb{HPC Network Topology.}
Conventional topologies for HPC environments can also be used in AI clusters for distributed training, such as Clos \cite{clos1953study}, BCube\cite{guo2009bcube}, DCell\cite{guo2008dcell}, Jellyfish\cite{singla2012jellyfish}, Torus \cite{duato2003interconnection}, Dragonfly \cite{kim2008technology} and Dragonfly+ \cite{shpiner2017dragonfly+}. The Clos network architecture, commonly known as a Fat-Tree topology,  is widely used in LLM training clusters. In a Clos-based cluster, each server, equipped with one or more NICs, is organized into racks connected to leaf switches. These leaf switches link to spine switches, providing inter-rack connectivity and forming a pod. The pods are further interconnected with core switches, facilitating any-to-any communication across servers within the cluster.
For example, Meta’s last-generation GPU cluster architecture, supporting up to 24,000 GPUs, consists of eight pods with full-fat bandwidth between them and use a 7:1 oversubscription ratio on the core layer \cite{llama3}. Meta uses 24,000 GPUs to this cluster for training Llama 3 405B.

\phb{Training-Optimized Topology.} 
Many network topologies are co-designed with distributed training algorithms.
The rail-optimized topology~\cite{rail-optimized-topology} enhances the connection from GPUs to leaf switches. Within each rail, GPUs that share the same index across different servers are interconnected via the same leaf switch. This configuration improves collective communication performance by reducing network interference between data flows. The SuperPod architecture utilizes a rail-optimized network and is capable of connecting over 16,000 GPUs \cite{DGX-SuperPOD-Architectur}. ByteDance employs a three-layer rail-optimized network to connect more than 10,000 GPUs in its MegaScale system design \cite{jiang2024MegaScaleScaling}. However, rail-optimized network designs could be less efficient because they necessitate connecting GPUs to distant switches, which requires costly and power-hungry optical transceivers. These optical components increase power consumption and heat, leading to higher rates of network failures, which is significant for distributed LLM training. Alibaba further optimizes the rail-optimized topology with a 2-tier, dual-plane architecture named HPN \cite{qian2024alibaba}. This architecture employs the latest 51.2Tbps single-chip switch and supports 1,000 GPUs in a tier1 network and up to 15,000 GPUs within one pod.

The network traffic  analysis of GPT/OPT-175B model training reveals that 99\% of GPU pairs do not carry any traffic, and less than 0.25\% of GPU pairs handle pipeline/tensor parallel and data parallel traffic \cite{wang2023railonly}. Based on these findings, the rail-only topology \cite{wang2023railonly} eliminates the connection between different rails on rail-optimized network. Each rail is connected by a dedicated but separate Clos network. Communication across GPUs on different rails is managed by forwarding the data through internal chip-to-chip interconnects. This method could effectively reduce costs while maintaining performance. HammingMesh \cite{hoefler2022hammingmesh} organizes GPUs into groups with a 2D-torus topology and connects these 2D-torus groups via sparsely connected switches. This design aims to save costs without compromising training performance. Given GPUs that are only connected by PCIe,  BiGraph \cite{dong2020eflops} proposes a new network architecture that exports intra-node GPU communication to outside the node, bypassing PCIe bandwidth bottlenecks. It features a two-layer network interconnected via a Clos architecture, with unique shortest paths for communication that support application-controlled traffic routing.

\phb{Reconfiguable Topology.}
Reconfigurable network  can be dynamically adjusted to optimize communication patterns  for improving the training performance. They typically utilize optical switching and customized configurations to improve bandwidth utilization, flexibility, and scalability of the network infrastructure. 
Driven by Silicon Photonic (SiP) interfaces, SiP-ML~\cite{Khani2021sipml} advances two principal architectures: SiP-OCS and SiP-Ring. SiP-OCS incorporates a fully connected configuration that maximizes the bandwidth by employing commercially accessible optical circuit switches, linking GPUs to all switches via Tbps SiP interfaces. Conversely, the SiP-Ring utilizes a switchless ring configuration, reducing the reconfiguration latency by integrating micro-ring resonators within the SiP interfaces. Wang et al. proposed TopoOpt~\cite{weiyang2023topoopt} for co-optimizing the network topology and parallelization strategies in distributed  training. 
This approach not only optimizes the computational and communication demands but also addresses the physical layer of network topology. 
TPUv4 ~\cite{jouppi2023TPUv4} features Optical Circuit Switches (OCS), allowing for dynamic reconfiguration of the 3D-Torus based interconnect topology, optimizing the data flow for the varied and intensive communication patterns that characterize LLM training. For instance, with 512 chips, TPUv4 offers flexibility in 3D-Torus topologies such as 4x4x32 or 8x8x8.

\subsubsection{Load Balancing \& Congestion Control}
\label{subsubsec:lbandcc}

\phb{Load Balancing.} The network traffic of  LLM training is characterized by a small number of elephant flows. Specifically, LLM training demonstrates periodic bursts of network traffic due to gradient synchronization \cite{qian2024alibaba}. Each burst demands significant network bandwidth. Moreover, each compute node involved in LLM training generates very few connections \cite{qian2024alibaba}. 
The conventional load-balancing technique, ECMP (Equal-Cost Multi-Path routing) \cite{ecmp}, uses hash algorithms to evenly distribute traffic across equivalent paths, such as those from leaf switches to spine switches in a Clos topology. However, this hash-based scheme is inefficient for handling LLM training traffic, which consists of a small number of elephant flows. When multiple elephant flows are routed to the same link, it can result in congestion and high latency.

Various strategies have been developed to address the load balancing challenges in large-scale GPU clusters. During the Llama 3 405B training, the collective library establishes 16 network flows between two GPUs, rather than a single flow, thereby reducing traffic per flow and enhancing load balancing opportunities \cite{llama3}. Additionally, the Enhanced-ECMP (E-ECMP) protocol effectively distributes these 16 flows across different network paths by hashing on additional fields in the RoCE header of packets.
Packet spraying \cite{packetspraying} distributes packets from a flow across all available parallel links, which can lead to out-of-order packets. NICs need to process out-of-order RDMA packets. Based on the traffic pattern of LLM training, Ethereal \cite{addanki2024challenging} demonstrates that greedily assigning paths to each flow can uniformly distribute the load across all network paths and resolve the ECMP hash conflict problem. In a large-scale GPUs cluster, HPN \cite{qian2024alibaba} achieves efficient load balancing by identifying precise disjoint equal paths and balancing the load within the collective communication library. MegaScale \cite{jiang2024MegaScaleScaling} shows that a rail-optimized topology can also mitigate ECMP hashing conflicts.

\phb{Congestion Control.} Lossless transmission is crucial in RDMA clusters. Priority-based Flow Control (PFC) \cite{PFC} is a flow control mechanism that prevents packet loss. When congestion occurs in a PFC-enabled queue on a downstream device, the device instructs the upstream device to halt traffic in the queue, thereby ensuring zero packet loss. Since PFC is a coarse-grained mechanism, it can lead to head-of-line blocking \cite{guo2016rdma}, which significantly diminishes network throughput. To address these challenges, various general-purpose congestion control schemes have been developed. These techniques include TIMELY \cite{mittal2015timely}, Data Center Quantized Congestion Notification (DCQCN) \cite{dcqcn,zhu2016ecn}, Swift \cite{kumar2020swift}, High Precision Congestion Control (HPCC) \cite{li2019hpcc}, Edge-Queued Datagram Service (EQDS) \cite{olteanu2022edge}, and Robust Congestion Control (RoCC) \cite{taheri2020rocc}. These schemes monitor network congestion, adjust data rate to alleviate congestion, and recover the rate to minimize throughput reduction. 



When there are concurrent training jobs, many congestion control schemes leverage the bursty and periodic traffic patterns to interleave the network traffic effectively.  MLTCP \cite{rajasekaran2024mltcp} interleaves the communication phases of jobs that compete for bandwidth based on a key insight: training flows should adjust their congestion window size based on the number of bytes sent in each training iteration.
CASSINI \cite{rajasekaran2023cassini} optimizes job placement on network links by considering the communication patterns of different jobs.
MLT~\cite{wang2024mlt} leverages the characteristics of LLM training, where the gradients of earlier layers are less important than those of later layers, and larger gradients are more significant than smaller ones. Consequently, in the event of communication congestion, MLT prioritizes queuing or discarding packets based on the importance of the gradients within them at the switch level to mitigate communication congestion issues.

\subsection{Storage}

The storage system plays a critical role in distributed LLM training and need to meet several key requirements. Firstly, it should align with the computing power of GPUs to maximize their utilization and avoid resource wastage caused by storage bottlenecks. Secondly, it should support the storage of large-scale structured and unstructured training datasets and be scalable in distributed processing environments. Additionally, the storage and retrieval of model checkpoints present challenges in LLM training, requiring the system to meet the writing and reading bandwidth dictated by the model size and training duration. Lastly, the storage system should satisfy traditional enterprise-level requirements, such as data protection, high availability, and security.

\subsubsection{Storage Systems for Checkpoint} 


The model checkpoint size is enormous in LLM training. As the number of parameters increases, so does the volume of data that needs to be written, demanding greater write bandwidth from the storage system. For example, the checkpoint size is 980GB for an LLM with 70B parameters.
Numerous storage systems have been deployed in large-scale GPU data centers to manage model checkpoints. Meta's distributed filesystem, Tectonic~\cite{pan2021facebook}, enables thousands of GPUs to save and load model checkpoints simultaneously, providing efficient and scalable storage solutions for extensive training operations \cite{metainfra}. At ByteDance, HDFS~\cite{hdfs} is employed for centralized model checkpoint maintenance, ensuring consistency and reliability at scale \cite{jiang2024MegaScaleScaling}.
To mitigate bandwidth bottlenecks during checkpoint recovery, a common approach designates a single worker to read the checkpoint partition from HDFS and then broadcast it to other workers sharing the same data. 
Distributed object stores, such as Ceph Object Storage~\cite{weil2006ceph}, offer easier scalability. This advantage stems from their lack of a hierarchical directory tree or namespace, simplifying consistency maintenance. Due to these benefits, object stores have become widely adopted for model checkpoint storage.

\subsubsection{Storage Systems for Training Data}

The raw dataset for LLM training is substantial. LLaMA~3 was trained on over 15 trillion tokens, which is more than seven times larger than LLaMA~2's dataset \cite{touvron2023llama}. Each token requires about 2 bytes, equaling roughly 30 TB of data. Preparing datasets for training involves extensive preprocessing steps, including data crawling and cleaning, requiring significant experimentation. Typically, the data processed during these steps  exceeds 100 times the final size of the training dataset \cite{qiu2024wanjuan}.  For instance, the WanJuan-CC dataset \cite{qiu2024wanjuan} selectively extracts approximately 68 billion   documents, generating around 1 trillion high-quality tokens, equivalent to a data size of 2 TB, after discarding $99\%$ of the raw data.  Therefore, the total data volume for LLM training is expected to exceed tens of PB.

Parallel file systems such as Lustre \cite{schwan2003lustre}, GPFS \cite{schmuck2002gpfs}, and BeeGFS \cite{chowdhury2019characterization} are frequently deployed on leading high-performance computing systems to ensure efficient I/O, persistent storage, and scalable performance. These systems are also widely used in training clusters for data loading, providing the necessary infrastructure to handle large-scale training data efficiently. Additionally, it is crucial for file systems to enable engineers to perform interactive debugging on jobs utilizing thousands of GPUs, as code changes need to be immediately accessible to all nodes \cite{metainfra}. 

During the training of most LLMs, each token is typically encountered only once. However, employing data caching remains crucial to mitigate I/O bottlenecks during data loading. This strategy involves prefetching training data from slower backend storage to faster cache storage.  Alluxio \cite{li2014tachyon} and JuiceFS \cite{JuiceFS} enhance LLM  training by efficiently caching training data from underlying storage systems such as HDFS or object storage.  Quiver \cite{kumar2020quiver} supports the transparent reuse of cached data across multiple jobs and users operating on the same dataset. Fluid \cite{gu2022fluid} leverages Alluxio for data caching, incorporating a mechanism that enables on-the-fly autoscaling of the cache based on I/O conditions.

\subsection{Scheduling}
LLM training workloads generally run on large-scale multi-tenant infrastructures (e.g., GPU clusters, public clouds) where users share cluster resources. Effective scheduling mechanisms are crucial for managing these workloads, ensuring efficient resource utilization and task execution \cite{SurveyDLSched}. Unlike \emph{task-level} scheduling (e.g., pipeline scheduling \cite{harlap2018pipedream, narayanan2019pipedream, li2021chimera}), which focuses on fine-grained optimization of single-job execution (\S\ref{subsubsec_pipeline_parallelism}), \emph{cluster-level} scheduling aims to optimize resource allocation and task scheduling across the entire cluster. We categorize existing \emph{cluster-level} scheduling systems into two types, workload scheduling and resource scheduling, according to their primary optimization aspects.

\subsubsection{Workload Scheduling}

Schedulers tailored for DL training workloads have been actively explored in recent years \cite{xiao2020AntManDynamic,xiao2018GandivaIntrospective, Philly, Helios, weng2022MLaaSwild,gu2019tiresias,mahajan2020themis,gu2023ElasticFlowElastic}. To enhance resource utilization, three advanced features are commonly implemented: (1) \emph{heterogeneous-aware} schedulers (e.g., Gavel \cite{Gavel}, Gandiva$_\text{fair}$ \cite{Gandivafair}) focus on optimizing job allocation across different GPU generations; (2) \emph{job-packing} schedulers (e.g., FGD \cite{FGD}, Lucid \cite{hu2023LucidNonintrusive}) enable fine-grained GPU sharing to fully facilitate hardware capability; (3) \emph{adaptive-scaling} schedulers (e.g., Pollux \cite{Pollux}, Sia \cite{Sia}) dynamically adjust the number of GPUs as well as the training hyperparameters to accelerate training progress. However, these schedulers are designed for general DL workloads and may not be directly applicable to LLMs due to the unique characteristics of LLM workloads \cite{hu2024CharacterizationLarge}.

To better manage LLM workloads, several recent studies have proposed systems tailored to LLMs. Crius \cite{Crius} jointly considers hybrid parallelism (\S\ref{subsec_hybrid_parallelism}) and hardware-affinity within heterogeneous clusters. It investigates the workflow efficiency of integrating adaptive parallelism configuration at the cluster scheduling level, offering significant opportunities to improve the efficiency of training multiple LLMs concurrently. To achieve highly efficient hyperparameter tuning for LLMs, Hydro \cite{Hydro} scales down the model to a smaller surrogate model for hyperparameter search, and then fuses multiple models into a single entity to enhance the hardware utilization. Additionally, Hydro extends the resources for tuning workloads by interleaving them with pipeline-enabled LLM pretraining tasks, effectively utilizing the pipeline bubbles. Acme \cite{hu2024CharacterizationLarge} further characterizes the workload mixture of the LLM development workflow and proposes a system to efficiently schedule associated jobs related to LLM training, including decoupled evaluation scheduling for timely model quality feedback as well as LLM-involved failure diagnosis and automatic recovery.

\subsubsection{Resource Scheduling}

In addition to workload scheduling, associated resource scheduling (e.g., CPU, memory, and networking) is another critical aspect of cluster-level management. For networking, Cassini \cite{rajasekaran2023cassini} enables the interleaving of bandwidth demands during the up and down phases of different jobs by using an affinity graph to determine time-shift values for adjusting communication phases. HIRE \cite{HIRE} introduces an innovative in-network computing scheduling system for datacenter switches, significantly reducing the network detours and tail placement latency. For storage, SiloD \cite{zhao2023SiloDCodesign} treats data cache and remote I/O as first-class resources for joint allocation, showing significant throughput improvements. For CPU and memory, Synergy \cite{mohan2022LookingGPUs} enhances training efficiency by optimizing CPU core allocations instead of relying on GPU-proportional allocation. Additionally, some works focus on energy conservation. EnvPipe \cite{EnvPipe} leverages the bubble time in the pipeline parallelism, stretching the execution time of pipeline units by lowering the SM frequency to save energy. Zeus \cite{you2023ZeusUnderstanding} automatically configures the batch size and GPU power limit for improving energy efficiency during training. Perseus\cite{chung2023perseus} introduces an efficient graph cut-based iterative algorithm to obtain the iteration time-energy Pareto front for large model training job.

\section{Parallelism Schemes for LLM Training}
\label{sec:parallel}


\tikzstyle{my-box}=[
    rectangle,
    draw=black,
    rounded corners,
    text opacity=1,
    minimum height=1.5em,
    minimum width=5em,
    inner sep=2pt,
    align=center,
    fill opacity=.5,
    line width=0.8pt,
]
\tikzstyle{leaf}=[my-box, minimum height=1.5em,
    fill=hidden-red!10, text=black, align=left,font=\normalsize,
    inner xsep=2pt,
    inner ysep=4pt,
    line width=0.8pt,
]

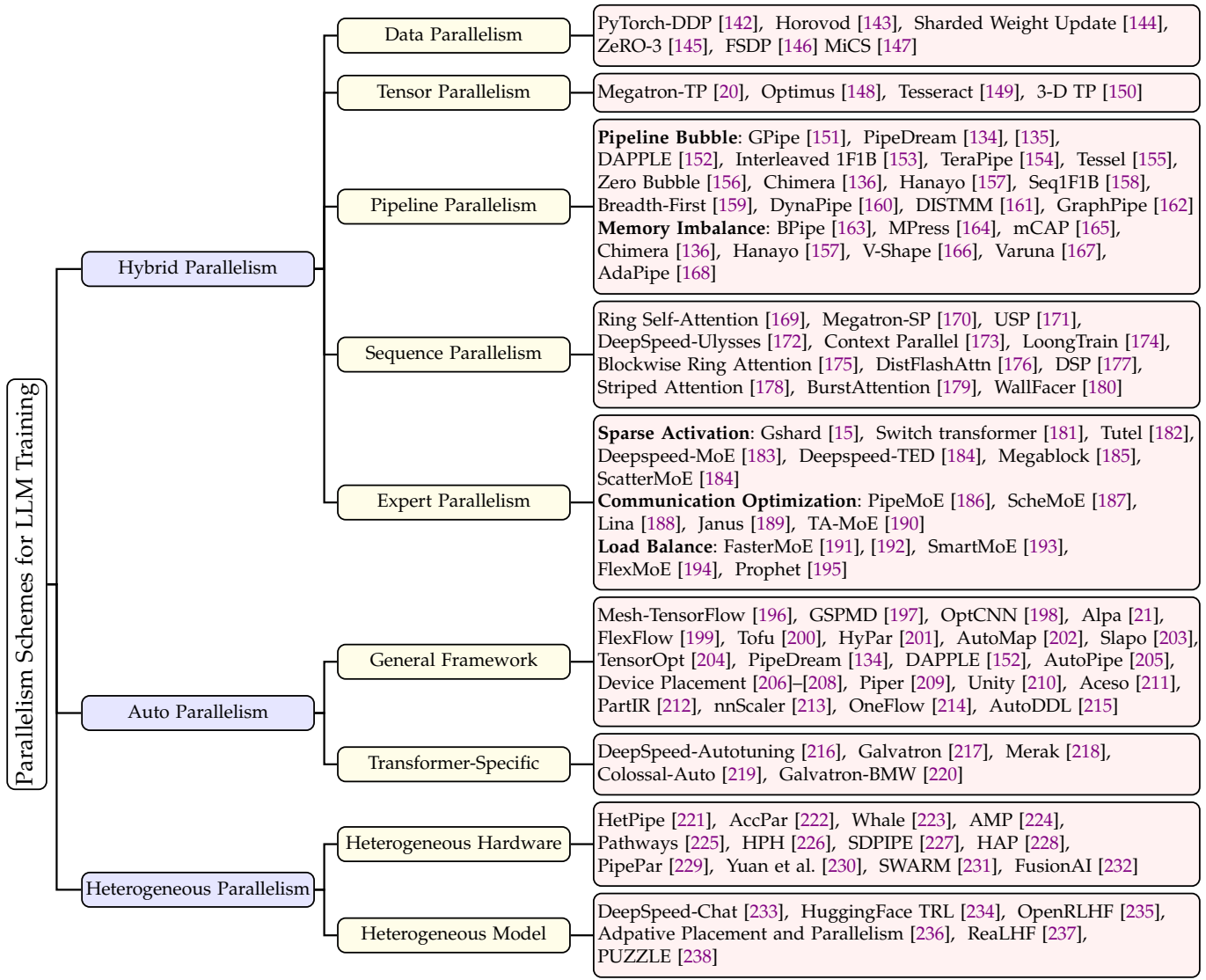
\begin{figure*}[htpb]
    \centering
    \resizebox{\textwidth}{!}{
        \begin{forest}
        forked edges,
        for tree={
            grow=east,
            reversed=true,
            anchor=base west,
            parent anchor=east,
            child anchor=west,
            base=center,
            font=\large,
            rectangle,
            draw=black,
            rounded corners,
            align=left,
            text centered,
            minimum width=4em,
            edge+={black, line width=1pt},
            s sep=3pt,
            inner xsep=2pt,
            inner ysep=3pt,
            line width=0.8pt,
            ver/.style={rotate=90, child anchor=north, parent anchor=south, anchor=center},
        },
        where level=1{text width=12em,font=\normalsize,}{},
        where level=2{text width=12em,font=\normalsize,}{},
        [
        {Parallelism Schemes for LLM Training}, ver
                [
                {Hybrid Parallelism}, fill=blue!10
                    [
                    {Data Parallelism}, fill=yellow!10
                        [
                        PyTorch-DDP~\cite{li2020PyTorchDistributed}{, }
                        Horovod~\cite{alex2018horovod}{, }
                        Sharded Weight Update~\cite{xu2020automatic}{, } \\
                        ZeRO-3~\cite{rajbhandari2020zero}{, }
                        FSDP~\cite{zhao2023PyTorchFSDP}
                        MiCS~\cite{zhang2022mics},
                        leaf, 
                        text width= 32em
                        ]
                    ]
                    [
                    {Tensor Parallelism}, fill=yellow!10
                        [
                        Megatron-TP~\cite{shoeybi2019megatron}{, }
                        Optimus~\cite{xu2023efficient}{, }
                        Tesseract~\cite{wang2022tesseract}{, }
                        3-D TP~\cite{bian2021maximizing},
                        leaf, 
                        text width= 32em
                        ]
                    ]
                    [
                    {Pipeline Parallelism}, fill=yellow!10
                        [
                        \textbf{Pipeline Bubble}:
                        GPipe~\cite{huang2019gpipe}{, }
                        PipeDream~\cite{harlap2018pipedream, narayanan2019pipedream}{, } \\
                        DAPPLE~\cite{fan2021dapple}{, } 
                        Interleaved 1F1B~\cite{narayanan2021efficient}{, } 
                        TeraPipe~\cite{li2021terapipe}{, }
                        Tessel~\cite{lin2024tessel}{, } \\
                        Zero Bubble~\cite{qi2023zero}{, } 
                        Chimera~\cite{li2021chimera}{, }
                        Hanayo~\cite{liu2023hanayo}{, } 
                        Seq1F1B~\cite{ao2024seq1f1b}{, } \\
                        Breadth-First~\cite{lamy2023breadth}{, } 
                        DynaPipe~\cite{jiang2024dynapipe}{, } 
                        DISTMM~\cite{huang2024distmm}{, }
                        GraphPipe~\cite{jeon2024graphpipe} \\
                        \textbf{Memory Imbalance}:
                        BPipe~\cite{kim2023bpipe}{, }
                        MPress~\cite{zhou2023mpress}{, }
                        mCAP~\cite{dreuning2022mcap}{, } \\
                        Chimera~\cite{li2021chimera}{, } 
                        Hanayo~\cite{liu2023hanayo}{, } 
                        V-Shape~\cite{qi2024pipeline}{, }
                        Varuna~\cite{athlur2022VarunaScalable}{, }\\
                        AdaPipe~\cite{sun2024adapipe},
                        leaf, 
                        text width= 32em
                        ]
                    ]
                    [
                    {Sequence Parallelism}, fill=yellow!10
                        [
                        Ring Self-Attention\cite{li2022SequenceParallelism}{, }
                        Megatron-SP~\cite{korthikanti2022ReducingActivation}{, } 
                        USP~\cite{fang2024unified}{, }\\
                        DeepSpeed-Ulysses~\cite{jacobs2023DeepSpeedUlysses}{, } 
                        Context Parallel~\cite{MegatronCP}{, }
                        LoongTrain~\cite{gu2024loongtrain}{, } \\
                        Blockwise Ring Attention~\cite{liu2023ring}{, }
                        DistFlashAttn~\cite{li2023LightSeqSequence}{, }
                        DSP~\cite{zhao2024dsp}{, } \\
                        Striped Attention~\cite{brandon2023striped}{, } 
                        BurstAttention~\cite{ao2024burstattention}{, }
                        WallFacer~\cite{liu2024wallfacer}, 
                        leaf, 
                        text width= 32em
                        ]
                    ]
                    [
                    {Expert Parallelism}, fill=yellow!10
                        [
                        \textbf{Sparse Activation}:
                            Gshard\cite{lepikhin2020gshard}{, }
                            Switch transformer\cite{fedus2022switch}{, }
                            Tutel\cite{hwang2023tutel}{, }\\
                            Deepspeed-MoE\cite{rajbhandari2022deepspeed}{, }
                            Deepspeed-TED\cite{tan2024scattered}{, }
                            Megablock\cite{gale2023megablocks}{, }\\
                            ScatterMoE\cite{tan2024scattered}\\
                        \textbf{Communication Optimization}:
                            PipeMoE\cite{shi2023pipemoe}{, } 
                            ScheMoE\cite{shi2024schemoe}{, } \\
                            Lina\cite{li2023accelerating}{, }
                            Janus\cite{liu2023janus}{, }
                            TA-MoE\cite{chen2022ta}\\
                        \textbf{Load Balance}:  
                            FasterMoE\cite{he2021fastmoe, he2022fastermoe}{, }
                            SmartMoE\cite{zhai2023smartmoe}{, } \\
                            FlexMoE\cite{nie2023flexmoe}{, }
                            Prophet\cite{wang2023prophet},
                        leaf, 
                        text width= 32em
                        ]
                    ]
                ]
                [
                {Auto Parallelism}, fill=blue!10
                    [
                    {General Framework}, fill=yellow!10
                        [
                        Mesh-TensorFlow~\cite{shazeer2018mesh}{, }
                        GSPMD~\cite{xu2021gspmd}{, }
                        OptCNN~\cite{jia2018exploring}{, } 
                        Alpa~\cite{zheng2022AlpaAutomatinga}{, } \\
                        FlexFlow~\cite{jia2019beyond}{, } 
                        Tofu~\cite{wang2019supporting}{, } 
                        HyPar~\cite{song2019hypar}{, } 
                        AutoMap~\cite{schaarschmidt2021automap}{, }
                        Slapo~\cite{chen2024slapo}{, } \\
                        TensorOpt~\cite{cai2021tensoropt}{, }
                        PipeDream~\cite{harlap2018pipedream}{, } 
                        DAPPLE~\cite{fan2021dapple}{, } 
                        AutoPipe~\cite{liu2022autopipe}{, } \\
                        Device Placement~\cite{mirhoseini2017device, gao2018spotlight, tarnawski2020efficient}{, }
                        Piper~\cite{tarnawski2021piper}{, } 
                        Unity~\cite{unger2022unity}{, }
                        Aceso~\cite{liu2024aceso}{, } \\
                        PartIR~\cite{alabed2024partir}{, } 
                        nnScaler~\cite{lin2024nnscaler}{, }
                        OneFlow~\cite{yuan2021oneflow}{, }
                        AutoDDL~\cite{chen2024autoddl},
                        leaf, 
                        text width= 32em
                        ]
                    ]
                    [
                    {Transformer-Specific}, fill=yellow!10
                        [
                        DeepSpeed-Autotuning~\cite{deepspeed2autotuning}{, }
                        Galvatron~\cite{miao2022galvatron}{, }
                        Merak~\cite{lai2023MerakEfficient}{, } \\
                        Colossal-Auto~\cite{liu2023ColossalAutoUnified}{, }
                        Galvatron-BMW~\cite{wang2024improving},
                        leaf, 
                        text width= 32em
                        ]
                    ]
                ]
                [
                {Heterogeneous Parallelism}, fill=blue!10
                    [
                    {Heterogeneous Hardware}, fill=yellow!10
                        [
                        HetPipe~\cite{park2020hetpipe}{, }
                        AccPar~\cite{song2020accpar}{, }
                        Whale~\cite{jia2022whale}{, }
                        AMP~\cite{li2022amp}{, } \\
                        Pathways~\cite{barham2022PathwaysAsynchronous}{, } 
                        HPH~\cite{duan2022hph}{, }
                        SDPIPE~\cite{miao2023sdpipe}{, } 
                        HAP~\cite{zhang2024hap}{, } \\
                        PipePar~\cite{zhang2023pipepar}{, }
                        Yuan et al.~\cite{yuan2022decentralized}{, } 
                        SWARM~\cite{ryabinin2023SWARMParallelism}{, } 
                        FusionAI~\cite{tang2023fusionai},
                        leaf, 
                        text width= 32em
                        ]
                    ]
                    [
                    {Heterogeneous Model}, fill=yellow!10
                        [
                        DeepSpeed-Chat~\cite{yao2023DeepSpeedChatEasy}{, }
                        HuggingFace TRL~\cite{trl}{, }
                        OpenRLHF~\cite{hu2024openrlhf}{, } \\
                        Adpative Placement and Parallelism~\cite{xiao2023AdaptivePlacement}{, }
                        ReaLHF\cite{mei2024realhf}{, } \\
                        PUZZLE\cite{lei2024puzzle}
                        ,
                        leaf, 
                        text width= 32em
                        ]
                    ]
                ]
                 %
                %
        ]
        \end{forest}
    }
    \caption{Studies on parallelism schemes for distributed LLM training.}
    \label{taxonomy:parallelism}
\end{figure*}

The continual growing scales of LLMs demand substantial computational resources and memory capacity. Distributed training, leveraging large-scale HPC clusters, has emerged as a crucial approach to efficiently train these models. In this section, we investigate various parallelism schemes proposed to enhance the utilization of HPC clusters for LLM training. We categorize these approaches into three main groups: Hybrid Parallelism, Auto Parallelism, and Heterogeneous Parallelism. Hybrid Parallelism combines multiple handcrafted parallelization strategies, such as data parallelism, tensor parallelism, pipeline parallelism, sequence parallelism, and expert parallelism. Auto Parallelism automatically determines the optimal parallelization strategy based on the model and hardware characteristics. Heterogeneous Parallelism exploits the heterogeneity in hardware or model for efficient training. This includes techniques that leverage different types of accelerators or leverage the heterogeneity within a single model (e.g., RLHF training) to improve the overall training efficiency on HPC clusters.

Most of today’s state-of-the-art parallelization strategies adopt a Single Program Multiple Data (SPMD) programming model, akin to the MPI paradigm~\cite{clarke1994mpi}, where the same program runs across multiple processors, each working on different pieces of data~\cite{barham2022PathwaysAsynchronous}. For example, data, model and sequence parallelism utilizes the SPMD programming model. This approach ensures uniformity and consistency in operations, making it well-suited for large-scale, distributed training environments. Some strategies explore to break the restriction of SPMD and further improve the resource utilization with the Multiple Program Multiple Data (MPMD) model, where different programs (or different parts of a program) run on different processors, handling different parts of the data or model~\cite{barham2022PathwaysAsynchronous}. For example, pipeline parallelism runs different parts of an LLM on different devices. In addition, auto parallelism and heterogeneous parallelism can leverage both SPMD and MPMD models to increase the resource utilization. Therefore, we discuss these approaches according to the dimensions along which parallelism occurs and whether the computing devices employed are homogeneous or heterogeneous, rather than focusing on the underlying programming models.

\subsection{Hybrid Parallelism}
\label{subsec_hybrid_parallelism}

Hybrid parallelism typically combines multiple handcrafted parallelization strategies to partition different parallelizable dimensions of an LLM. These strategies include data parallelism, tensor parallelism, pipeline parallelism and sequence parallelism, as illustrated in \Cref{fig:hybrid_parallelism}. The combination of data parallelism, tensor parallelism, and pipeline parallelism is also referred to as 3D parallelism.

\subsubsection{Data Parallelism}
\label{subsubsec_data_parallelism}

\begin{figure*}[t]
    \centering
    \includegraphics[width=0.95\linewidth]{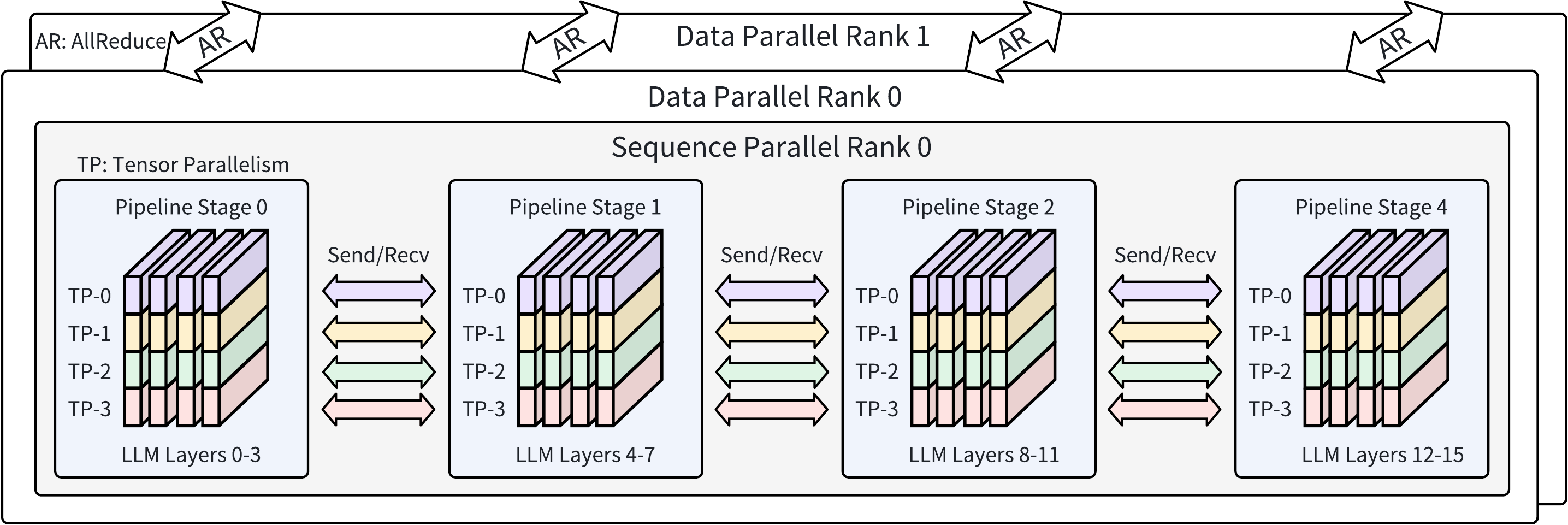}
  \caption{An example of 3D-parallelism with data parallelism, tensor parallelism, and pipeline parallelism.}
  \label{fig:hybrid_parallelism}
\end{figure*}

Data parallelism is the most commonly used parallelization strategy for distributed training due to its high scalability and ease of implementation. It follows the Single Program Multiple Data (SPMD) programming model. Data parallelism partitions the input training data along the batch dimension, where each GPU processes its assigned segment of data, as shown in \Cref{fig:hybrid_parallelism}(a). Throughout the training process, the data first undergoes forward computation with the full model weights layer by layer, and then performs backward computation in the reverse order. Each layer produces gradients that will be aggregated across all GPUs using collective communication operations for optimizer updates. 

Data parallelism incorporates various sharding strategies, which significantly influence the memory footprint and communication overhead. Supposing the global world size is $W$ (i.e. the number of devices), a sharding factor $F$ is introduced to control the sharding strategy used~\cite{zhao2023PyTorchFSDP}, defined as the number of devices across which parameters are partitioned ($1 \le F \le W$). We have the following scenarios. 

\noindent\textbf{Full replication} ($F=1$): this sharding strategy is simplified as vanilla data parallelism. Pytorch-DDP~\cite{paszke2019pytorch} and Horovod~\cite{alex2018horovod} fully replicate the model across all devices and use All-Reduce for gradient aggregation. They also divide the gradients into small buckets to overlap the gradient communication with backward computation. 

\noindent\textbf{Full sharding} ($F=W$). This sharding strategy comes with the lowest memory consumption but the most communication overhead ($1.5\times$ over vanilla data parallelism). Full sharding strategy fully shards the model, where each device holds only $\frac{1}{W}$ of the model parameters. The full weights and gradients are communicated and recovered on-demand before the computation, and immediately discarded afterwards. ZeRO-3~\cite{rajbhandari2020zero} employs per-parameter sharding to shard the full model and utilizes All-Gather and Reduce-Scatter for unsharding and sharding communication, respectively. Sharded Weight Update~\cite{xu2020automatic} also employs per-parameter sharding but focuses more on sharding the redundant parameter update computation across all the devices. FSDP (Fully Sharded Data Parallelism)\cite{zhao2023PyTorchFSDP} achieves the same functionality by sharding model parameters in the grain of module units and provides more user-friendly API.

\noindent\textbf{Hybrid sharding} ($1<F<W$). In this strategy~\cite{zhao2023PyTorchFSDP}, all the devices are divided into a $N \times M$ device mesh. The model parameters are sharded along the $N$ dimension of the mesh, and replicated along the $M$ dimension. MiCS~\cite{zhang2022mics} invokes All-Gather collective to gather the sharded parameters and All-Reduce to aggregate the gradients. FSDP~\cite{zhao2023PyTorchFSDP} replaces the All-Reduce with Reduce-Scatter to reduce the memory and communication overhead. Compared to full replication and full sharding, hybrid sharding is more flexible to provide a trade-off between memory consumption and communication overhead via adjsuting $F$ based on the model architecture and hardware constraints. 



\subsubsection{Tensor parallelism}
\label{subsubsec_tensor_parallelism}

Tensor parallelism (\Cref{fig:hybrid_parallelism}(b)), also known as intra-layer model parallelism, is a technique proposed to enable the training of LLMs across multiple GPUs. It partitions the parameter tensors of each layer along multiple dimensions, effectively distributing the model parameters across the available GPUs. Tensor parallelism communicates intermediate activation tensors, the size of which is much smaller than that of parameters and gradients communicated in data parallelism, except for long context LLM training scenarios. However, in tensor parallelism, it is challenging to overlap the communication with computation, necessitating the use of high-bandwidth connections. Consequently, tensor parallelism is more commonly employed in a single GPU node. 

Tensor parallelism can be divided into 1-D~\cite{shoeybi2019megatron}, 2-D~\cite{xu2023efficient}, 2.5-D~\cite{wang2022tesseract} and 3-D~\cite{bian2021maximizing} parallelism according to the dimensionality of the partition. There are two parameter matrices in both the MLP and self-attention module of transformer-based LLMs. Megatron-LM~\cite{shoeybi2019megatron} first adopts 1-D tensor parallelism to partition the first parameter matrix along its column, and the second parameter matrix along its row. It replicates the input and output tensors of each partitioned module and introduces two All-Reduce collective communication across all GPUs to fit an LLM into multiple GPUs. 
Inspired by Scalable Universal Matrix Multiplication Algorithm (SUMMA)~\cite{van1997summa} and Cannon's algorithm~\cite{cannon1969cellular} for 2-D parallel matrix multiplication, Optimus~\cite{xu2023efficient} further partitions the input and parameter tensors in 2 dimensions to improve the communication and memory efficiency of 1-D tensor parallelism. Tesseract~\cite{wang2022tesseract} extends the 2.5-D matrix multiplication method~\cite{solomonik2011communication}, which is proposed to improve the efficiency of Cannon's algorithm, to LLM training and proposes the 2.5-D tensor parallelism to overcome the abundance of unnecessary communication resulting from the increasing model size. 3-D tensor parallelism~\cite{bian2021maximizing} adopts and improves the 3-D parallel matrix multiplication algorithm~\cite{agarwal1995three} for linear operations and achieves a perfect load balance across multiple devices for LLM training. 

\subsubsection{Pipeline Parallelism}
\label{subsubsec_pipeline_parallelism}


Pipeline parallelism (\Cref{fig:hybrid_parallelism}(c))~\cite{huang2019gpipe}, also known as inter-layer model parallelism, is proposed to accommodate large models across multiple GPUs, particularly across different nodes. Pipeline parallelism partitions the layers of a model into multiple stages, where each stage consists of a consecutive set of layers in the model and is mapped to a set of GPUs. Unlike tensor parallelism, which typically demands high-bandwidth connections like NVLink for communication, pipeline parallelism only necessitates the exchange of intermediate tensors at designated cutting points, resulting in less frequent communication requirements. Therefore, pipeline parallelism is suitable for scaling up the LLM training across multiple GPU nodes that are connected with small bandwidth. For example, Strati et al.~\cite{strati2024ml} adopt pipeline parallelism to fully utilize geo-distributed resources to overcome the shortage of GPUs. Due to the data dependency of different stages, pipeline parallelism typically splits the input data into multiple micro-batches for pipelining to enable the efficient training of giant models. However, it comes with two significant problems. First, the pipeline bubble problem reduces the utilization of GPUs due to the time spent on waiting for the output of the previous stage. Second, there exists memory consumption imbalance across different stages since the former stages need to hold more active micro-batches than the latter for better pipelining and higher utilization. We detail each problem below.

\phb{Pipeline Bubble.} 
Efficient micro-batch scheduling algorithms are proposed to reduce pipeline bubbles. GPipe~\cite{huang2019gpipe} introduces a fill-drain schedule that injects all micro-batches at once for forward pass execution, followed by backward passes. Gpipe incurs significant pipeline bubbles due to the warm up and cool down of both forward and backward passes. PipeDream~\cite{harlap2018pipedream, narayanan2019pipedream} introduces a 1F1B (1 Forward 1 Backward) schedule, which executes the backward pass of a micro-batch as soon as the corresponding forward pass has completed, to reduce the pipeline bubbles in the asynchronous scenario. DAPPLE~\cite{fan2021dapple} employs an early backward schedule to first inject a fixed number of micro-batches at the beginning of each stage and then interleave forward and backward passes with round robin. Interleaved 1F1B~\cite{narayanan2021efficient} adapts the 1F1B schedule but assigns multiple stages to each GPU (i.e. looping pipeline placement). The pipeline bubble is reduced at the cost of higher communication and peak memory consumption. Chimera~\cite{li2021chimera} introduces a bidirectional pipeline to reduce bubbles with weight duplication. Hanayo~\cite{liu2023hanayo} further proposes a wave-like pipeline that assigns multiple symmetrical stages to one GPU to improve the pipeline utilization. Zero bubble~\cite{qi2023zero} splits the backward computation into two parts: activation and parameter gradient computation. It schedules the forward and activation gradient computation with 1F1B and then fills the bubbles with parameter gradient computation, which reduce bubbles with higher peak memory consumption. Breadth-First~\cite{lamy2023breadth} runs all micro-batches at once in looping pipeline placement to reduce the communication overhead when combined with sharded data parallelism.
TeraPipe~\cite{li2021terapipe} splits the micro-batch along the sequence dimension and exploits more fine-grained token parallelism to reduce pipeline bubbles. However, The memory overhead of TeraPipe is large since it is based on the GPipe schedule. Seq1F1B~\cite{ao2024seq1f1b} splits the sequence into chunks and utilizes the 1F1B schedule to reduce the peak memory consumption while achieving low pipeline bubble rates. DynaPipe~\cite{jiang2024dynapipe} uses a dynamic micro-batching approach to the multi-task training of LLMs with variable-length input. It introduces a memory-aware adaptive scheduling algorithm and ahead-of-time communication planning to further reduce the pipeline bubble rates. Tessel~\cite{lin2024tessel} is a two-phase approach, including repetitive pattern construction and schedule completion, to automatically search for the efficient pipeline schedule for a specified partition strategy. DISTMM~\cite{huang2024distmm} launches doubled micro-batches to bypass the dependency barrier caused by the large batch requirement of multi-modal training, thus reducing idle cycles. GraphPipe~\cite{jeon2024graphpipe} preserves the DNN graph topology and partitions it into stages that can be concurrently executed to improve pipeline utilization and reduce memory consumption.

\phm{Memory Imbalance.} 
Pipeline parallelism typically injects more micro-batches into the beginning stages to improve the pipeline utilization, resulting in higher activation memory consumption in these stages. To resolve this problem, BPipe~\cite{kim2023bpipe} and MPress~\cite{zhou2023mpress} employ D2D (device to device) transfer to swap intermediate activation tensors from high-load GPU to light-load GPU during runtime. MPress also combines the activation recomputation to reduce the memory footprint. Chimera~\cite{li2021chimera} introduces a bidirectional pipeline that combines two pipelines in different directions together to achieve more balanced memory consumption. Each GPU holds two symmetric stages, leading to weight duplication. Hanayo~\cite{liu2023hanayo} turns the bidirectional pipeline into two data parallel pipelines to remove the weight duplication and achieves a balanced memory consumption by assigning multiple stages to one GPU symmetrically. V-Shape~\cite{qi2024pipeline} partitions the model into stages twice the number of devices, where the two halves of stages are placed in reverse order. By varying the offsets between stages, V-Shape makes a trade-off between peak memory consumption and bubble utilization. mCAP~\cite{dreuning2022mcap} utilizes an incremental-profiling approach to partition models evenly across GPUs with respect to peak memory usage. 

Peak memory consumption limits the number of active micro-batches in pipeline parallelism, thus its efficiency. Activation recomputation can be employed to reduce the peak memory consumption effectively. Varuna~\cite{athlur2022VarunaScalable} combines pipeline parallelism and activation recomputation to achieve this goal. It designs a static rule-based schedule enumerated for a given pipeline with an opportunistic policy to hide jitter and reduce bubbles. The static schedule is generated based on constraints including activation recomputation timing, activation memory management, and backward computation prioritization. To resolve the memory imbalance with low recomputation overhead, AdaPipe~\cite{sun2024adapipe} adopts adaptive recomputation to support different recomputation strategies for different stages, and adaptive partitioning based on the 1F1B schedule to balance the computation of each stage.


\subsubsection{Sequence parallelism}
\label{subsubsec:SequenceParallelism}


The context window of today's LLMs grows rapidly, and the most powerful LLM can support millions of tokens~\cite{GoogleGemini}. Such ultra long sequence leads to significant memory and computation requirements for LLM training: linearly increasing memory footprint of activations and quadratic complexity of attention mechanism. Recomputing activations in the backward can reduce the peak memory consumption but also introduce significant overheads (30\% with full recompute). Large tensor parallelism degree incurs significant communication overhead. Sequence parallelism (\Cref{fig:hybrid_parallelism}(d))~\cite{li2022SequenceParallelism, korthikanti2022ReducingActivation} is proposed to accommodate the long sequence training and distribute the computation in multiple GPUs efficiently within the memory capacity. It divides the input data into multiple chunks along the sequence dimension and each chunk is fed to one GPU for computation. Since sequence parallelism replicates the model parameters, it is typically combined with tensor and pipeline parallelism to scale up the LLM training. When used together with tensor parallelism, sequence parallelism distributes the memory and computation of attention on multiple GPUs, but incurs redundant memory consumption and computation in the non-tensor parallel regions of a transformer layer. Megatron-SP~\cite{korthikanti2022ReducingActivation} splits these computations along the sequence dimension to reduce redundant the activation computation and memory consumption without increasing the communication.

Although sequence parallelism partitions the memory, computation and communication across multiple GPUs, the quadratic causal attention still presents remarkable challenges in training efficiency, including the key-value tensor communication overhead, IO-awareness attention computation overhead and load imbalance among GPUs due to the causal attention mask. 
Most sequence parallelism approaches for attention are ring-based~\cite{li2022SequenceParallelism, MegatronCP, liu2023ring, li2023LightSeqSequence, brandon2023striped, ao2024burstattention}. Ring Self-Attention~\cite{li2022SequenceParallelism} leverages sequence parallelism and calculates the self-attention with ring-style communication to scale up the context window of LLM training. It first transmits the key tensors among GPUs to calculate the attention scores in a circular fashion, and then calculates the self-attention output based on the attention scores and value tensors transmitted in a similar fashion. DistFlashAttn~\cite{li2023LightSeqSequence} transmits the key-value tensor chunks concurrently to utilize IO-awareness FlashAttention~\cite{dao2022flashattention} kernel and balance the computation of different GPUs by filling the idle cycles of earlier tokens with later tokens.  Megatron Context Parallel~\cite{MegatronCP} also leverages the FlashAttention kernel and removes the unnecessary computation resulted from low-triangle causal masking. It further balances the computation among GPUs by exchanging half of the chunk with the symmetric GPU. DistFlashAttn and Context Parallel also overlap the key-value tensor communication and attention computation with separate CUDA streams. Striped Attention~\cite{brandon2023striped} resolves the imbalance by assigning each GPU a subset of tokens distributed uniformly throughout the sequence, instead of contiguous chunks. BurstAttention~\cite{ao2024burstattention} computes the attention with FlashAttention on each GPU and utilizes double buffers to overlap the communication and computation. Blockwise Ring Attention~\cite{liu2023ring} extends Ring Self-Attention~\cite{li2022SequenceParallelism} to blockwise attention, which computes the attention in small blocks to reduce the memory footprint. Inspired from N-body simulation, WallFacer~\cite{liu2024wallfacer} first divides GPUs into subgroups and replicates query and key-value tensors in each subgroup via asynchronous AllGather. The attention computation leverages multiple ring-style P2P communication to enhance efficiency. A final asynchronous ReduceScatter is needed to distribute the attention output.

DeepSpeed-Ulysses~\cite{jacobs2023DeepSpeedUlysses} differs from previous ring-based approaches by splitting the head dimension instead of sequence dimension and leverages All-to-All to shift the partition dimension from sequence to head. DeepSpeed-Ulysses can be seamlessly combined with existing attention implementation, e.g., FlashAttention, and the workload among GPUs is naturally balanced. However, the parallelism degree of DeepSpeed-Ulysses is restricted by the number of heads, especially for LLMs using MQA~\cite{shazeer2019fast} and GQA~\cite{ainslie2023GQATraining}. LoongTrain~\cite{gu2024loongtrain} and USP~\cite{fang2024unified} are concurrent work that integrate the advantages of DeepSpeed-Ulysses and Ring Attention. They organize the GPUs into 2-dimension mesh, forming hybrid ulysses- and ring-style process groups. During training, they first perform All-to-All among ulysses groups to switch partition from sequence to head dimension, and then perform attention computation with Ring-Attention among ring groups. LoongTrain further proposes Double-Ring-Attention to fully utilize available bandwidth for inter-node communication and overlap communication with computation. DSP~\cite{zhao2024dsp} dynamically switches the parallelism dimension according to the computation stage in multi-dimensional transformers, like DiT~\cite{peebles2023scalable}.

\subsubsection{Expert Parallelism}
\label{subsubsec:ExpertParallelism}

\begin{figure}[t]
    \centering
    \includegraphics[width=0.95\linewidth]{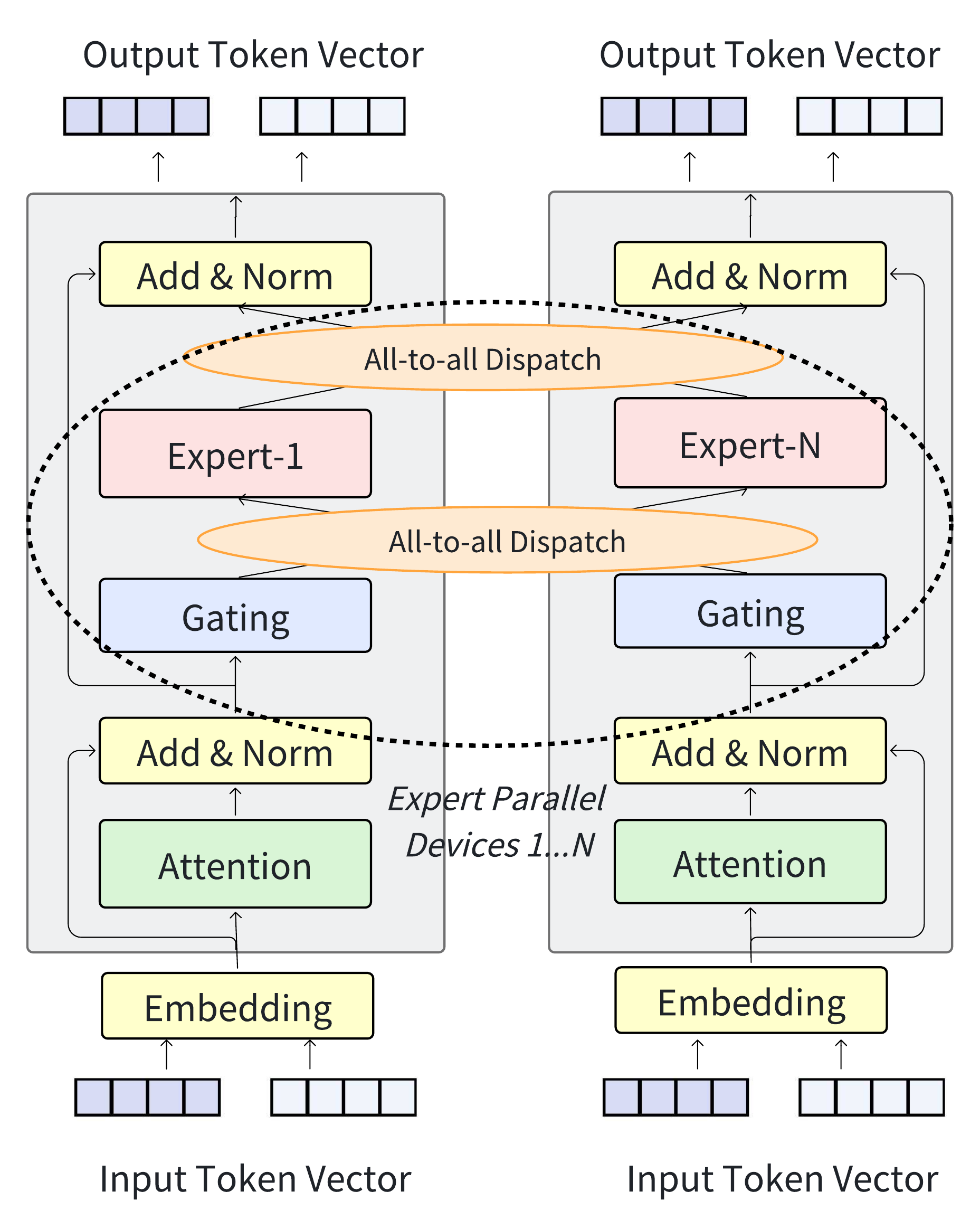}
  \caption{Expert parallelism. The dotted line highlights the MoE components within the transformer model, where each device maintains one expert for expert parallelism and collaborate based on All-to-All communication.}
  \label{fig:expert_parallelism}
\end{figure}

The Mixture-of-Experts (MoE) is currently the most popular sparse model in LLMs. While significantly increasing the number of parameters in LLMs, MoE does not greatly increase the computational cost with conditional computations\cite{bengio2013estimating}. The basic framework of MoE, as shown in Fig.~\ref{fig:expert_parallelism}, consists of multiple expert networks that handle different subsets of training data and a gate network that applies routing algorithm to assign the input tokens to different experts networks. MoE enables the training of large models with parameters beyond the trillion scale and is claimed to be used in popular LLM models such as Mixtral 8x7B~\cite{jiang2024MixtralExperts} and DeepSeek2~\cite{deepseekai2024deepseekv2}. 

\phb{Sparse Activation.}
With the increasing model size, all experts cannot be accommodated and trained on a single device.
Therefore, GShard \cite{lepikhin2020gshard} extends the idea of MoE to Transformers in distributed settings, where experts are distributed across different workers and collaborates with All-to-All communication, as shown in Fig.~\ref{fig:expert_parallelism}.
Subsequent research for expert parallelism generally follows the same paradigm.
For example, the Switch Transformer \cite{fedus2022switch} incorporates the design of distributed MoE training on the T5 model. But unlike the top-2 routing algorithm used in GShard, the Switch Transformer routes each token to only the top-1 expert to maximize computational efficiency.
Additionally, DeepSpeed-MoE \cite{rajbhandari2022deepspeed} proposes a new distributed MoE architecture that applies shared experts in each worker and places more experts in deeper layers to balance communication costs with training accuracy.

Expert parallelism can be effectively integrated with conventional 3D parallelism. For example, GShard, Switch Transformer, and DeepSpeed-MoE all treat expert parallelism as an orthogonal dimension of hybrid parallelism. For efficient hybrid training, DeepSpeed-TED \cite{singh2023hybrid} presents a hybrid parallel algorithm that combines data, tensor, and expert parallelism to enable the training of MoE models. The authors partition the MoE parameters into “tiles” of a predefined size to avoid high optimizer memory spikes and propose communication optimizations like Duplicate Token Dropping (DTD) and activation checkpointing to eliminate duplicate data in All-to-All communication. However, it is challenging to choose the optimal hybrid parallelism plan due to the dynamic nature of MoE, and switching between different parallelism strategies during runtime also incurs substantial overhead. Therefore, some research like Tutel \cite{hwang2023tutel} designs an adaptive parallelism switching algorithm that applies the same distribution model layout for all possibly optimal strategies, and can dynamically switch the parallelism strategy at every iteration without any extra overhead.

Since General Matrix Multiplications (GeMMs) require the size of all experts' inputs to be consistent, existing MoE training frameworks often perform token dropping and padding to match the same expert capacity, which wastes computation. Megablocks \cite{gale2023megablocks} optimizes grouped GeMMs by implementing Block Sparse Matrix Multiplication and supports different batch sizes for expert computations in a single kernel to avoid unnecessary token dropping in MoE training. Another framework that supports grouped GeMMs is ScatterMoE \cite{tan2024scattered}, which implements the ParallelLinear kernel that fuses grouped GeMMs and scattered read and write operations to reduce the memory footprint for top-k ($k \geq 2$) gating.


\phb{Communication Optimization.}
All-to-all communication in expert parallelism can seriously affect the training efficiency of MoE, especially in poor network environments. Existing distributed training systems try to optimize the performance of MoE by overlapping communication tasks with computing tasks so that some communication costs can be hidden. For example, Tutel \cite{hwang2023tutel} divides the input tensors into groups along the expert capacity dimension and overlaps computation and communication among different groups to hide All-to-All overhead. FasterMoE \cite{he2021fastmoe, he2022fastermoe} uses a similar strategy to Tutel but splits the tensor along the expert dimension. Additionally, Tutel \cite{hwang2023tutel} also optimizes the All-to-All kernel implementation by aggregating small messages into a single large chunk inside the nodes before exchanging data among different nodes. This optimization is also used in FasterMoE and ScheMoe \cite{shi2024schemoe}.
Based on the overlap strategy in Tutel, PipeMoE \cite{shi2023pipemoe} models the execution time of both communication and computation tasks based on the workloads and designs an adaptive algorithm to find the optimal number of partitions to minimize training time. ScheMoE \cite{shi2024schemoe} considers data compression approaches before All-to-All communication and modularizes time-consuming operations, including data compression, collective communication, and expert computation. ScheMoE then proposes an adaptive optimal scheduling algorithm to pipeline communication and computing operations to improve training efficiency.

Expert parallelism usually interacts with other parallel strategies in MoE training. It is possible to reduce communication overhead by fine-grained task scheduling. For example, Lina \cite{li2023accelerating} analyzes the All-to-All overhead of MoE during distributed training and inference systematically and finds that All-to-All latency is prolonged when it overlaps with AllReduce operations. Lina proposes prioritizing All-to-All over AllReduce to improve its bandwidth and reduce its blocking period in distributed training. Additionally, Lina incorporates tensor partitioning and pipelining to perform micro-op scheduling similar to Tutel. Lina also dynamically schedules resources based on expert popularity during inference to minimize overhead.
Janus \cite{liu2023janus} designs a data-centric paradigm that keeps data in place and moves experts between GPUs based on a Parameter Server. The data-centric paradigm uses fine-grained asynchronous communication and allows experts to move between GPUs using non-blocking communication primitives such as pull. Janus implements a topology-aware strategy to effectively pull experts between nodes and supports expert prefetching to pull all external experts to local CPU memory.

There are some research optimizes MoE training from model-system co-design perspective. For example, TA-MoE \cite{chen2022ta} proposes a topology-aware routing strategy for large-scale MoE training. TA-MoE abstracts the dispatch problem into an optimization objective to obtain the target dispatch pattern under different topologies and designs a topology-aware auxiliary loss based on the dispatch pattern. This approach adaptively routes the data to fit the underlying topology without sacrificing model accuracy.


\phb{Load Balance.}
Due to the sparse and conditional computing nature of MoE, a popular expert may receive more tokens than others in expert parallelism (usually caused by a poor routing algorithm), leading to serious load imbalance and affecting the training efficiency of MoE. FasterMoE \cite{he2022fastermoe} proposes the shadowing experts approach, which dynamically broadcasts the parameters of popular experts to all other GPUs based on the workload of previous iterations. By spreading the workload of popular experts across different devices, the shadowing experts approach reduces the impact of skewed expert popularity.
SmartMoE \cite{zhai2023smartmoe} adopts a two-stage approach to search for the optimal parallel plan for load balance. First, SmartMoE designs a data-sensitive performance model that divides parallel plans into pools, where the cost of switching parallel modes within a pool is relatively low. 
Then, SmartMoE can switch to the appropriate parallelism (referred to as expert placement in SmartMoE) to keep load balance during the online stage.
FlexMoE \cite{nie2023flexmoe} found that the distribution of expert-to-device mapping does not shift significantly over a short period, so it introduces fine-grained replicated expert parallelism that duplicates heavy experts across multiple devices. FlexMoE monitors data workload and uses three placement adjustment primitives (i.e., expand, shrink, migrate) to generate optimal placement solutions if the balance ratio is exceeded.
Prophet \cite{wang2023prophet} presents a systematic, fine-grained, and efficient load balancing training method for large-scale MoE models. Taking the MoE model, device pool, and token distribution as inputs, Prophet’s planner iteratively searches and evaluates expert placements and finally outputs a well-balanced expert placement. Additionally, Prophet hides the overhead of these resource allocation operations using a layer-wise scheduling strategy.


\subsection{Auto Parallelism} 
\label{subsec:AutoParallelism}

Given an arbitrary DNN model and a GPU cluster, there exists a vast array of options for parallelism, encompassing the partitioning of individual layers and their partitioning degrees. It is a time-consuming and knowledge-intensive process to design handcrafted hybrid parallelism approaches that can maximize the training efficiency, requiring expert understanding of the model architecture, hardware characteristics, and intricate trade-offs involved in parallelization strategies. Moreover, the efficient implementation of optimal parallelization strategies often necessitates substantial human efforts. To address these challenges, auto parallelism emerges as a promising solution, which seeks to automatically determine the most effective parallelization strategy for a given DNN model on a specific GPU cluster. By leveraging sophisticated algorithms and heuristics, auto parallelism systems can analyze the model architecture, hardware specifications, and performance characteristics to identify the optimal combination of parallelism techniques, such as data, tensor, and pipeline parallelism. This approach streamlines the process of optimizing the distributed training across various models and infrastructures, enhancing the overall efficiency and reducing the manual effort. Furthermore, auto parallelism can adapt to changing hardware configurations and model architectures, automatically adjusting the parallelization strategy to maintain the optimal performance. In the following, we categorize existing auto parallelism systems into general and transformer-specific frameworks, according to the targeted model architecture.


\subsubsection{General Framework}
\label{subsubsec:GeneralFramework}

General auto parallelism frameworks focus on automatically parallelizing general DNNs on a specific computation cluster. These frameworks typically follow a three-step process: (1) defining the search space of parallelization strategies; (2) developing performance models to measure the training efficiency of different strategies; (3) designing algorithms to efficiently identify the optimal parallelization strategy. Below we investigate different approaches according to the search space they cover.

Some works have explored the search space of hybrid data and pipeline parallelism strategies for DNN training optimization. These approaches focus on partitioning DNNs automatically and designing pipeline schedules to improve the pipeline utilization. PipeDream~\cite{harlap2018pipedream} measures the efficiency of pipeline partitions with the execution time of the slowest stage and develops a dynamic programming algorithm to partition the DNN evenly by minimizing the slowest stage. DAPPLE~\cite{fan2021dapple} builds an analytical model to estimate the execution time of one partition strategy and uses dynamic programming to determine the optimal pipeline partition. AutoPipe~\cite{liu2022autopipe} constructs a simulator to simulate the pipeline execution and proposes a heuristic algorithm to obtain the balanced partition. AutoPipe also automatically splits the micro-batch to reduce the latency of the warm-up stage. Some device placement approaches~\cite{mirhoseini2017device, gao2018spotlight, tarnawski2020efficient} use reinforcement learning to predict the optimal operator placement for pipeline parallelism.

Researchers also explore the automated data and model parallelism by partitioning operators along different dimensions. OptCNN~\cite{jia2018exploring} partitions operators along all divisible dimensions in their output tensor and utilizes an analytical performance model to pick the optimal parallelization strategy, including the parallelizable dimensions and degree of parallelism, which defines how to parallelize an individual layer across different devices. FlexFlow~\cite{jia2019beyond} further extends the search space to Sample-Operator-Attribute-Parameter (SOAP), which includes almost all the divisible dimensions in input and output tensors, and introduces a novel execution simulator for accurate performance modeling. FlexFlow efficiently finds an optimal parallelization strategy with MCMC sampling. Tofu~\cite{wang2019supporting} and HyPar~\cite{song2019hypar} develop dynamic programming algorithms that minimize the total communication cost rather than the end-to-end performance, to identify the optimal partition for each operator in the hybrid data and model parallelism space. TensorOpt~\cite{cai2021tensoropt} optimizes the parallelization strategy under a given memory budget with a frontier tracking algorithm. AutoMap~\cite{schaarschmidt2021automap} employs Monte Carlo Tree Search (MCTS) to select a sequence of partitioning rules defined by PartIR~\cite{alabed2024partir} for a set of selected important operators via a learned scorer. The whole parallelization strategy is propagated from the strategy via the selected operators.

Recent works also design approaches for automated data, model and pipeline parallelism. Piper~\cite{tarnawski2021piper} designs a two-level dynamic programming approach to find the optimal hybrid data, tensor and pipeline parallelism combined with activation recomputation. It first divides the model into small partitions for the pipeline and then splits operators within each partition. Alpa~\cite{zheng2022AlpaAutomatinga} formulates a comprehensive space by viewing parallelisms as two hierarchical levels: inter-operator and intra-operator parallelism. Then it automatically derives an efficient parallel execution plan at each parallelism level. Unity~\cite{unger2022unity} jointly optimizes the parallelization and algebraic transformations by representing them as substitutions on a unified parallel computation graph. Aceso~\cite{liu2024aceso} proposes an iterative bottleneck alleviation approach to significantly reduce the search time. It identifies a performance bottleneck at every step and adjusts the strategy to mitigate the bottleneck until convergence. nnScaler~\cite{lin2024nnscaler} introduces three primitives to enable the composition of the search space with arbitrary partitioning and spatial-temporal scheduling of the partitioned model. Domain experts can apply constraints to the primitives to build effective and small search spaces, which can be automatically explored with low overheads. AutoDDL~\cite{chen2024autoddl} customizes a coordinate descent algorithm by  iteratively updating the SBP~\cite{yuan2021oneflow} distributions for each layer and quickly discover an optimal strategy with near-optimal communication cost.

General auto parallelism frameworks demand efficient system support for various parallelization strategies, in addition to fast optimization algorithms for optimal parallelization strategy discovery. This is because parallelism often involves complex computation and communication operators, especially for model parallelism that partitions operators. Prior works have developed efficient systems that enable a wide range of parallelization strategies, either by building upon modern DL frameworks~\cite{zheng2022AlpaAutomatinga, lin2024nnscaler} or implementation from scratch~\cite{jia2019beyond}. Mesh-TensorFlow~\cite{shazeer2018mesh} observes the intrinsic complexity of implementing a parallelization strategy, and first proposes to abstract the device cluster into a multi-dimensional mesh, and abstract parallelism into partitioning the iteration space (i.e. tensor dimensions). By mapping the tensor and mesh dimensions, a hybrid data and model parallelism strategy can be easily implemented with high performance. For example, data and model parallelisms split the batch and hidden dimensions, respectively. GSPMD~\cite{xu2021gspmd} further provides a unified way to achieve various general parallelism schemes with simple tensor sharding annotations based on JAX~\cite{jax2018github} and XLA~\cite{GoogleXLA}. OneFlow~\cite{yuan2021oneflow} proposes SBP (Split, Broadcast, Partial-value) abstraction for partition and allows users to specify the placement and SBP signature for tensors to implement different parallelization strategies. PartIR~\cite{alabed2024partir} decouples the model from its partitioning and designs a compiler stack for users to compose SPMD sharding strategies incrementally via a schedule. Similar to TVM~\cite{chen2018tvm}, Slapo~\cite{chen2024slapo} defines a comprehensive set of schedule primitives for parallelization and subgraph optimization like operator fusion and activation checkpointing. These schedules are decoupled from execution and preserves the original model structure for progressive optimization.

\subsubsection{Transformer-Specific Framework}
\label{subsubsec:TransformerSpecificFramework}

As LLMs are based on the transformer architecture, recent works tailor automated systems for transformers. DeepSpeed-Autotuning~\cite{deepspeed2autotuning} automatically tunes the system knobs to figure out good performance-relevant configurations in the user-defined tuning space, including the degree of parallelism. Galvatron~\cite{miao2022galvatron} designs a dynamic programming algorithm to generate the most efficient hybrid data, tensor and pipeline parallelism strategy. Merak~\cite{lai2023MerakEfficient} introduces an automatic model partitioner for non-intrusive automatic parallelism and a high-performance 3D parallel runtime engine to enhance the utilization of available resources. Colossal-AI~\cite{liu2023ColossalAutoUnified, li2023colossal} provides a unified interface for modular usage of hybrid data, tensor, sequence and pipeline parallelism. Galvatron-BMW~\cite{wang2024improving} extends the space of Galvatron to include sharded data parallelism and activation recomputation, and searches for the optimal strategy considering both the memory consumption and computation while maximizing the hardware utilization.



\subsection{Heterogeneous Parallelism} 
\label{subsec:HeterogeneousParallelism}

The escalating computational demands of LLM training have spurred advancements in heterogeneous hardware, which harnesses diverse computing resources and globally distributed devices. This heterogeneity is also reflected in model architectures, particularly with Reinforcement Learning from Human Feedback (RLHF). Utilizing heterogeneous hardware and diverse model architectures has become essential for the efficient training of LLMs.

\subsubsection{Heterogeneous Hardware} 
\label{subsubsec:HeterogeneousHardware}

The massive computational requirements of LLM training have driven the evolution of accelerators, leading to clusters with mixed device types and uneven interconnect bandwidths. Additionally, modern data and computing clusters are often distributed globally due to factors such as power shortages. These phenomena have motivated the adoption of heterogeneous parallelisms, which leverage diverse computing resources and geographically distributed devices to accelerate LLM training.

Some works leverage heterogeneous computing resources, such as CPUs, GPUs, and specialized accelerators, to enhance the performance of LLMs. The distinct computation, memory capacity and interconnect bandwidth of these devices introduce challenges for efficient LLM pretraining. HetPipe~\cite{park2020hetpipe} partitions the heterogeneous cluster into multiple virtual works. Each virtual work processes mini-batches with the pipeline parallelism, and different virtual works employ asynchronous data parallelisms to improve the throughput. AccPar~\cite{song2020accpar} proposes flexible tensor partitioning to balance the computation of different accelerators and uses dynamic programming to automatically decide tensor partitioning among heterogeneous devices for DNNs. Whale~\cite{jia2022whale} proposes a unified abstraction to ease the efforts for parallel training of giant models on heterogeneous clusters. It seamlessly adapts to heterogeneous GPUs through automatic graph optimizations and balances the workloads with hardware information. AMP~\cite{li2022amp} utilizes a heterogeneous-aware performance model to find the optimal hybrid data, tensor and pipeline parallelism strategy. HPH~\cite{duan2022hph} arranges different GPUs into stages according to the compute-communication ratio in descending order and formulates the model partitioning as an integer programming problem to minimize the iteration time. Pathways~\cite{barham2022PathwaysAsynchronous} employs a sharded dataflow model and asynchronous gang-scheduling to efficiently execute ML models on heterogeneous cluster. SDPIPE~\cite{miao2023sdpipe} introduces a semi-decentralized scheme that decentralizes the communication model synchronization and centralizes the process of group scheduling for pipeline parallelism to utilize heterogeneous devices. HAP~\cite{zhang2024hap} uses an A$^*$-based search algorithm to generate the optimal tensor sharding strategy, sharding ratio across heterogeneous devices and the communication methods for distributed training. PipePar~\cite{zhang2023pipepar} proposes a dynamic programming algorithm to partition the model into stages for pipeline considering both the heterogeneity of GPUs and network bandwidths.

Some other works explore geo-distributed devices, featuring the low network bandwidth, to enhance the training efficiency.
Yuan et al.~\cite{yuan2022decentralized} partition the LLMs into computational tasklets and propose a novel scheduling algorithm to efficiently utilize a group of heterogeneous devices connected by a slow heterogeneous network for hybrid data and pipeline parallelism. SWARM parallelism~\cite{ryabinin2023SWARMParallelism} partitions the model into equal-sized stages and prioritizes routing inputs to stable peers with lower latency for workload balance. It also adaptively moves devices across stages to  maximize the training throughput. FusionAI~\cite{tang2023fusionai} splits the training computation graph (DAG) into subgraphs (sub-DAG) and generates a load balanced task schedule to utilize heterogeneous consumer GPUs connected with low bandwidth for pipeline training. Communication compression approaches, like CocktailSGD~\cite{wang2023cocktailsgd}, can also be leveraged to train LLMs efficiently in low-bandwidth clusters. 







\subsubsection{Heterogeneous Model} 
\label{subsubsec:HeterogeneousModel}



\begin{figure}[t]
    \centering
    \includegraphics[width=0.95\linewidth]{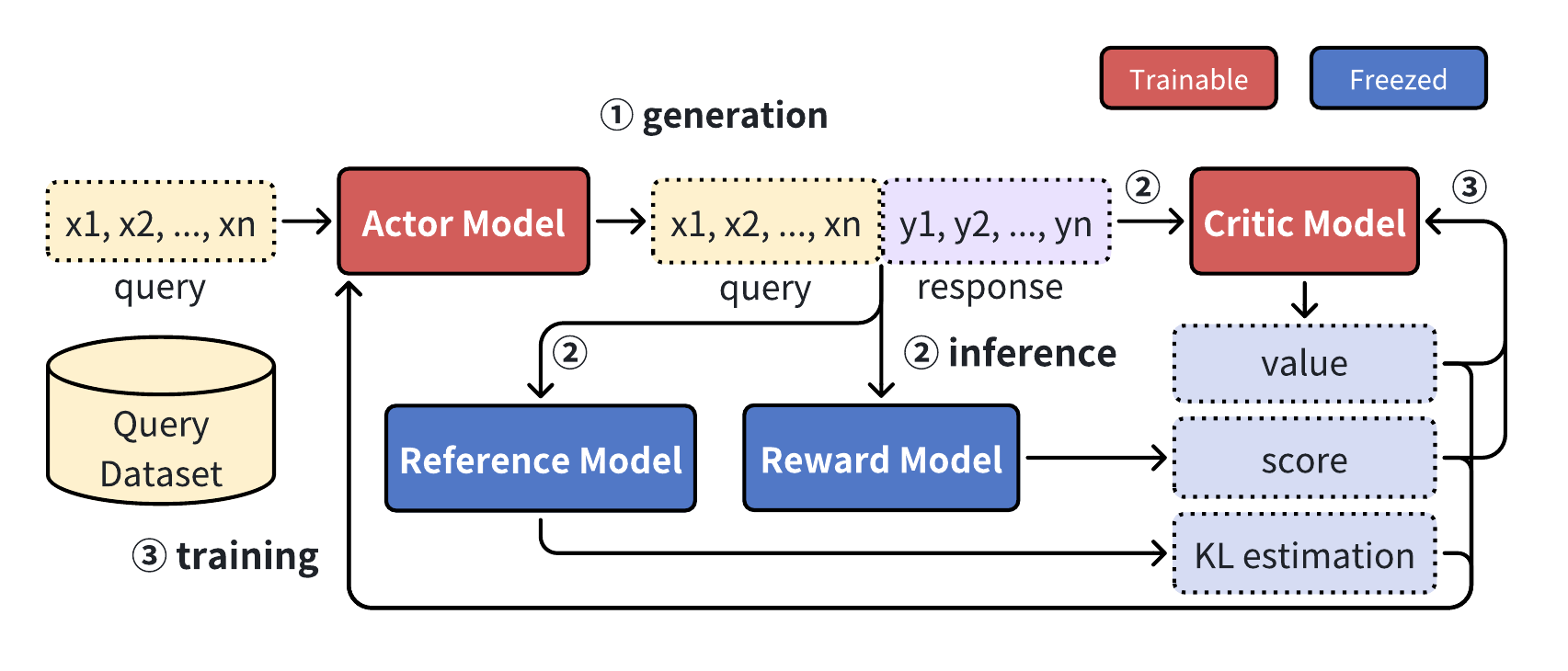}
  \caption{An example of RLHF. \textbf{Inference process}: \circled{1} The actor model generates a response from a given query. \circled{2} The critic model, reward model, and reference model use the query and response pairs to generate the value, score, and KL divergence required for training through inference. \textbf{Training process}: \circled{3} The actor model and critic model use the data collected in the inference process to update their weights through gradient descent.} 
  \label{fig:rlhf_alg}
\end{figure}

During the LLM training process, heterogeneity is not only reflected in the hardware, but also in the model. Training may involve the interaction of several different models. A specific example is Reinforcement Learning from Human Feedback (RLHF). RLHF is a training method that aims to align AI systems more closely with human preferences~\cite{ouyang2022Traininglanguage}, leveraging human's advantages in judging appropriate behavior rather than demonstrating. This method has received widespread attention, especially for fine-tuning large language models. However, due to the particularity of the Proximal Policy Optimization (PPO)\cite{schulman2017ppo} algorithm, the model heterogeneity is introduced into the RLHF training, making the training process of RLHF very different from pre-training and supervised fine-tuning.

In principle, RLHF consists of three different stages: the stage 1 is supervised fine-tuning, the stage 2 is the training of the reward model, and the stage 3 is PPO training. Model heterogeneity is presented in stage 3, as shown in Fig.~\ref{fig:rlhf_alg}. The PPO training stage consists of two different processes, namely the inference process that generates data, and the training process that updates the weights of the actor model and critic model. PPO training is performed via the collaboration of these two processes. Moreover, the training stage introduces higher memory cost, as we need to serve several copies of auto-regressive generation models and reward models at the same time, and more time costs, because we must wait for the experience generation to be completed before updating the weights.

Many frameworks have been proposed for RLHF training. For instance, DeepSpeed-Chat~\cite{yao2023DeepSpeedChatEasy} uses Hybrid Engine to seamlessly switch model partitioning between training and inference, such as using tensor parallelism to improve throughput during inference and using ZeRO~\cite{rajbhandari2020zero} or LoRA\cite{hu2022LoRALowRank} to improve memory utilization during training, providing outstanding system efficiency for RLHF training. HuggingFace TRL~\cite{trl} can make full use of various parameter-efficient fine-tuning (PEFT) methods, such as LoRA or QLoRA\cite{dettmers2023qlora}, to save memory cost, and use a dedicated kernel designed by unsloth\cite{unsloth} to increase the training speed of RLHF. ColossalAI-Chat~\cite{li2023colossal} is another end-to-end RLHF training framework that also supports LoRA and supports the use of ZeRO \cite{rajbhandari2020zero} to reduce memory redundancy.

However, the above work adopts a flattening strategy for model placement, that is, placing the four models in RLHF on the same device, and then using methods such as ZeRO or LoRA to minimize memory cost. But using only ZeRO will lead to memory bottlenecks when training larger models, while using efficient parameter fine-tuning strategies such as LoRA will damage model performance. To solve this problem, OpenRLHF~\cite{hu2024openrlhf} uses Ray~\cite{moritz2018ray} and vLLM~\cite{kwon2023EfficientMemory} to distribute the reward models to different devices, avoiding placing all four models in PPO on the same device. Similarly, Adpative Placement and Parallelism (APP) framework~\cite{xiao2023AdaptivePlacement} proposed two other model placement strategies, namely Interleaving Strategy and Separation Strategy. It captures the fact that the generation part and the training part can run independently during PPO training, and some serialization can be eliminated by placing them on different devices, which introduces additional communication but can overlap well with computing. 

Meanwhile, there are some works that apply the parallel strategies in the first two stages to the stage 3 of RLHF in a fine-grained scheduling manner. For example, ReaLHF\cite{mei2024realhf} switches the most suitable parallel mode for different sub-stages in stage 3 by redistributing parameters, which greatly increases the optimization space. PUZZLE\cite{lei2024puzzle} reschedules the order of task execution according to the affinity of different stages, so that stages with better affinity can effectively cover execution and improve training efficiency.



\section{Computation Optimizations}
\label{sec:comp}

Today's AI accelerators offer unprecedented computational capabilities in terms of FLOPs. However, effectively utilizing these FLOPs to their full potential requires sophisticated optimization techniques. This section introduces systems and techniques of computation optimizations to effectively utilize GPU FLOPs. We first elaborate operator optimizations including the core attention operator optimizations and automatic optimizations via compilers. Remarkable performance for operator and computing graphs is gained based on exploiting massive parallelism and efficient multi-level memory access concerning the underlying hardware features. Second, mixed-precision training is detailed where computations are accelerated benefiting from reduced precision. 16-Bit floating point mixed training has been the de facto method in most training systems. Low-bit fixed points as low as 1-bit have been studied and employed for high training efficiency. 

\tikzstyle{my-box}=[
    rectangle,
    draw=black,
    rounded corners,
    text opacity=1,
    minimum height=1.5em,
    minimum width=5em,
    inner sep=2pt,
    align=center,
    fill opacity=.5,
    line width=0.8pt,
]
\tikzstyle{leaf}=[my-box, minimum height=1.5em,
    fill=hidden-red!10, text=black, align=left,font=\normalsize,
    inner xsep=2pt,
    inner ysep=4pt,
    line width=0.8pt,
]

\begin{figure*}[htpb]
    \centering
    \resizebox{\textwidth}{!}{
        \begin{forest}
        forked edges,
        for tree={
            grow=east,
            reversed=true,
            anchor=base west,
            parent anchor=east,
            child anchor=west,
            base=center,
            font=\large,
            rectangle,
            draw=black,
            rounded corners,
            align=left,
            text centered,
            minimum width=4em,
            edge+={black, line width=1pt},
            s sep=3pt,
            inner xsep=2pt,
            inner ysep=3pt,
            line width=0.8pt,
            ver/.style={rotate=90, child anchor=north, parent anchor=south, anchor=center},
        },
        where level=1{text width=15em,font=\normalsize,}{},
        where level=2{text width=15em,font=\normalsize,}{},
        [
        {Computation Optimizations for LLM Training}, ver
                [
                {Operator Optimizations}, fill=blue!10
                    [
                    {Manual Optimizations}, fill=yellow!10
                        [
                        FlashAttention~\cite{dao2022flashattention}{, }
                        FlashAttention-2~\cite{dao2023FlashAttention2Faster}{, } \\
                        FlashAttention-3~\cite{shah2024flashattention} {, } 
                        BPT~\cite{liu2024blockwise}{, } 
                        SWattention~\cite{wu2024swattention}{, } \\
                        Bikshand et al.~\cite{bikshandi2023case} {,}
                        ByteTransformer~\cite{zhai2023bytetransformer}
                        ,
                        leaf, 
                        text width= 26em
                        ]
                    ]
                    [
                    {Automatic Optimizations}, fill=yellow!10
                        [
                        \textbf{Kernel-level:}
                        Halide~\cite{ragan2013halide}{, }
                        TVM~\cite{chen2018tvm}{, }
                        Roller~\cite{zhu2022roller}{, } \\
                        Triton~\cite{tillet2019triton}{, } 
                        ALCOP~\cite{huang2023alcop} \\
                        \textbf{Graph-level:}
                        Chimera~\cite{zheng2023chimera}{, } 
                        Welder~\cite{shi2023welder}{, } 
                        Slapo~\cite{chen2024slapo}{, } \\
                        TorchDynamo \& TorchInductor~\cite{ansel2024pytorch}{, } 
                        JIT-Q~\cite{ibrahim2024jit} 
                        ,
                        leaf, 
                        text width= 26em
                        ]
                    ]
                ]
                [
                {Mixed-precision Training}, fill=blue!10
                    [
                    {16-Bit Floating Point}, fill=yellow!10
                        [
                        FP16 Mixed-Precision Training~\cite{micikevicius2017mixed}{, }
                        Campo~\cite{he2022campo}{, } \\
                        BF16 Mixed-Precision Training ~\cite{kalamkar2019study}{, }
                        THC~\cite{li2024thc}
                        ,
                        leaf, 
                        text width= 26em
                        ]
                    ]
                    [
                    {Sub-8-Bit Floating Point}, fill=yellow!10
                        [
                        Wang et al.~\cite{wang2018training}{, }
                        Sun et al.~\cite{sun2019hybrid}{, }
                        FP8-LM~\cite{peng2023FP8LMTraining} {, } \\
                        Rouhani et al.~\cite{rouhani2023microscaling}
                        ,
                        leaf, 
                        text width= 26em
                        ]
                    ]
                    [
                    {Low-Bit Fixed Point}, fill=yellow!10
                        [
                        \textbf{INT8:}
                        Jetfire~\cite{xi2024jetfire} \\
                        \textbf{INT4:}
                        Xi et al.~\cite{xi2023training} \\
                        \textbf{1-Bit:}
                        BitNet~\cite{wang2023bitnet}{, }
                        BitNet b1.58~\cite{ma2024era}
                        ,
                        leaf, 
                        text width= 26em
                        ]
                    ]
                ]
        ]
        \end{forest}
    }
    
    \caption{Studies on computation optimizations for distributed LLM training.}
    \label{taxonomy:computation}
\end{figure*}
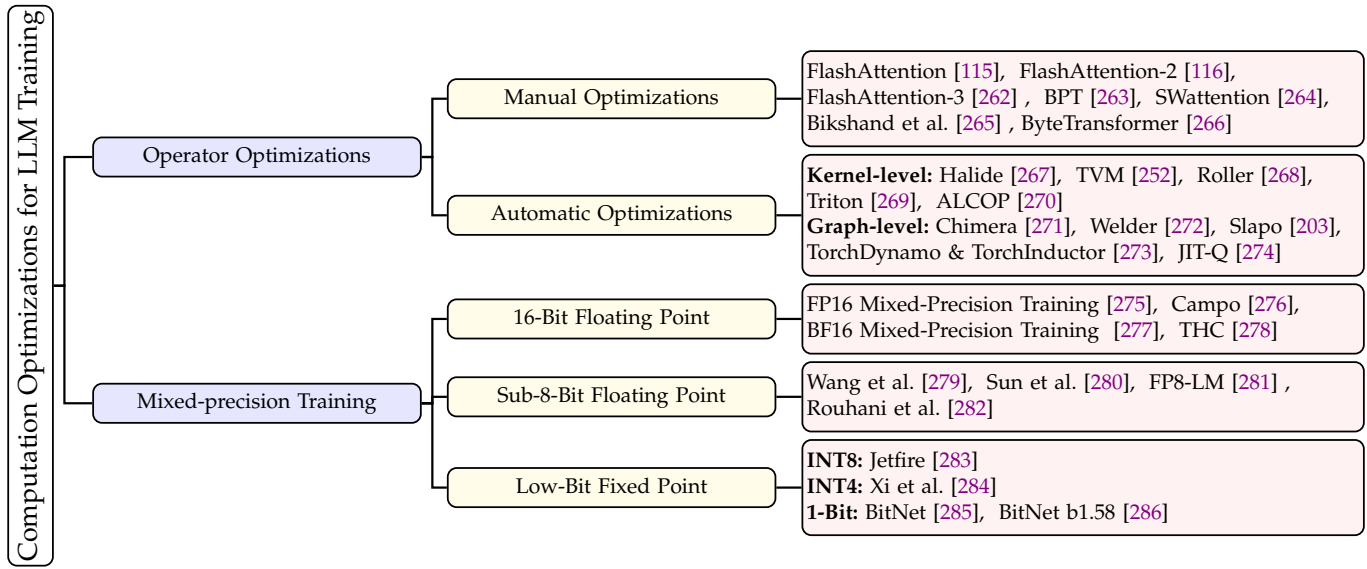




\subsection{Operator Optimizations}
\label{subsec:OperatorOptimizations}
Operator optimizations can be categorized into manual and automatic optimizations. Manual optimizations mainly focus on the attention operator, while automatic optimizations are applied more broadly.

\subsubsection{Manually Optimized Attention Operator}
\label{subsubsec:ManuallyOptimizedAttentionOperator}




Attention, as the core of transformer architectures, plays a crucial role in the training efficiency of LLMs. 
Given a query $q$, and lists of keys $k_1, k_2, ..., k_n$ and values $v_1, v_2, ..., v_n$, where $q, k_i, v_i \in \mathbb{R}^d$, the attention is computed as follows,
$$s_i=dot(q, k_i), \enspace s^{'}_{i}=softmax(s_i)=\frac{e^{s_i}}{\sum_{j}{e^{s_j}}}, \enspace o_i=\sum_{i}v_{i}s^{'}_{i}.$$
The self-attention exhibits quadratic time and memory complexity relative to sequence length. The substantial memory consumption and frequent access to high-bandwidth memory (HBM) imposed by self-attention constrain both the performance and the context length of transformer models. Extensive work is presented to optimize self-attention. We focus on exact attention optimizations while lossy optimizations, like linear attention, are out of our scope.

Memory-efficient attention is primarily proposed to mitigate the large memory cost.
Rabe et al.~\cite{rabe2021self} prove that self-attention needs $O(logn)$ memory complexity instead of $O(n^2)$. 
By employing lazy softmax, the division by $\sum_{j}{e^{s_j}}$ in softmax can be delayed to the very end of the attention operation. Thus the summation could be processed incrementally which requires just a scalar (i.e. $O(1)$) to maintain the intermediate result but not change the output. The self-attention requires extra $O(logn)$ memory complexity to keep the additional index into the list of queries to compute the results to all queries sequentially.

The FlashAttention series further demonstrate fast and memory-efficient exact attention with IO-awareness, high parallelism, and balanced workloads on GPU. 
In FlashAttention~\cite{dao2022flashattention}, an IO-aware tiling algorithm is proposed
 to reduce the number of memory reads/writes between slow HBM and fast on-chip SRAM based on the online softmax. More specifically, the softmax could be calculated one block at a time by tracking the normalization statistics including the maximum score and the sums of exponentiated scores. The tiling algorithm thus fuses all the computation operation chain in self-attention including matrix multiply, softmax, matrix multiply, etc, in one cuda kernel for reduced HBM access. 
FlashAttention-2~\cite{dao2023FlashAttention2Faster} 
further improves the low-occupancy and unnecessary shared memory reads/writes in FlashAttention with additional parallelism in sequence length dimension and improved warp-level scheduling for data sharing inside a thread block. 
Besides, the popular training systems~\cite{gu2024loongtrain} generally employ FlashAttention-2 for high performance. 
FlashAttention-3~\cite{shah2024flashattention} speeds up attention on H100 GPU by excavating the newly presented hardware capabilities as the former FlashAttention implementations are based on A100 GPU. An interleaved block-wise GEMM and softmax algorithm is redesigned based on FlashAttention-2 to hide the non-GEMM operations in softmax with the asynchronous WGMMA instructions for GEMM. Besides, 
by leveraging the asynchrony of the Tensor Cores and Tensor Memory Accelerator (TMA), overall computation is overlapped with data movement via a warp-specialized software pipelining scheme. 
Blockwise Parallel Transformer (BPT)~\cite{liu2024blockwise} further reduces the substantial memory requirements 
by extending the tiling algorithm in FlashAttention to fuse the feedforward network.

The attention operation is also optimized on various architectures by leveraging hardware-specific features. For instance, SWattention~\cite{wu2024swattention} designs a two-level blocking attention algorithm to exploit the underlying hardware of the new Sunway architecture, building upon FlashAttention.
Similarly, Bikshand et al.~\cite{bikshandi2023case} implement FlashAttention-2 on the H100 GPU using the Cutlass library. They utilize the TMA and WarpGroup Matrix-Multiply-Accumulate (WGMMA) instructions to optimize data copying and GEMM operations, respectively. Additionally, tensor layout transformations and software pipelining of data copying and computations between the two GEMMs are carefully designed based on the Cutlass library.

Attention mechanisms are also optimized for variable-length sequences, which are common in distributed LLM training. These variable-length sequences can incur significant memory and computation costs if padded to the maximum length. 
FlashAttention-2 efficiently handles variable-length inputs by parallelizing the sequence length dimension inseparably. ByteTransformer~\cite{zhai2023bytetransformer} focuses on padding-free transformers for variable-length inputs, maintaining a position array during computation. This array records the mapping relationship of valid tokens between the original tensor and the intermediate packed tensor.
The fused Multihead Attention algorithm for long sequences employs optimized grouped GEMM for unpadded tensors. This optimization reduces the memory and computation overhead associated with padding, thereby enhancing performance.

\subsubsection{Automatic Optimizations via Compilers}
\label{subsubsec:AutomaticOptimizationsviaCompilers}

DNN compilers play an important role in optimizing key computations in LLM training. Highly efficient kernels of operators are generated automatically which mitigates the burden of library-based kernel optimizations on diverse hardware vendors to a great extent. Operator fusion is performed by analyzing the computation graphs automatically in the training process. 


\phb{Efficient Operator Kernel Generation.} 
Halide~\cite{ragan2013halide} and TVM~\cite{chen2018tvm} generate high-performance operator implementations automatically, 
relying on multiple effective schedule primitives that exploit parallelism and data locality 
on various backends. Furthermore, Roller~\cite{zhu2022roller} optimizes the cost of searching for optimal alternatives in the large search space of kernel implementations. 
It primarily generates a tile kernel consisting of Load, Store, and Compute interfaces, following which the complete operator kernel is constructed
by a scale-up-then-scale-out approach. 
Triton~\cite{tillet2019triton} provides a C-based language and compiler that facilitates expressing 
and optimizing tile tensor programs for competitive performance. In particular, 
effective optimizations such as hierarchical tiling and shared memory allocation are supported via machine-dependent compiling passes.
ALCOP~\cite{huang2023alcop} performs automatic load-compute pipelining to overlap the high-latency memory access with computations for operators on GPUs. Multi-stage pipelining is utilized by pipeline buffer detection as well as sophisticated index analysis and substitution in complicated loop structures. 

\phb{Graph-level Optimizations for Operator Fusion.}
With the disparity of speed of computing cores and memory bandwidth enlarging, modern DNNs are restricted by memory access. Data reuse among inter-operators is excavated via operator fusion using compilers.
Plenty of compiler works~\cite{jia2019taso,zheng2020ansor,niu2021dnnfusion,zheng2022astitch} performs operator fusion by setting expert rules. Particularly, Chimera~\cite{zheng2023chimera} works on optimizing compute-intensive operator chains. The operator chain is firstly decomposed into a series of computation blocks and the optimal block execution order is then selected to maximize data reuse according to an analytical model. In addition, replaceable microkernels are designed to leverage hardware-specific intra-block optimizations. Welder~\cite{shi2023welder} lowers the computing graph into a tile-level data-flow graph whose nodes are operator tiles and edges are marked with the memory level of the tensor data reused by the connected nodes. Operator fusion combinations that maximize data reuse across different levels of memory hierarchies are searched at the tile level. 

Pytorch2~\cite{ansel2024pytorch} presents two extensions, i.e. a Python-level JIT compiler TorchDynamo and the corresponding compiler backend TorchInductor, to enable more robust graph compilation on various back-ends for remarkable performance improvement without sacrificing the flexibility of Python. 
Slapo~\cite{chen2024slapo} proposes a schedule language to decouple model execution from definition. Declaring a set of schedule
primitives, users could convert the model for high-performance kernels.
JIT-Q~\cite{ibrahim2024jit} proposes just-in-time quantization for weights which enables storing only a high-precision copy of weights during training and creates low-precision weight copies based on the in-memory ALU augmentations of the commercial PIM (processing-in-memory) solutions.

\subsection{Mixed-precision Training}
\label{subsec:Mixed-precisionTraining}


Low-precision training is an effective methodology to reduce the computation, storage, and communication costs in training large-scale models. Nowadays LLM training generally leverages FP16 and BF16 data types.
In particular, BF16 
can represent the same range of values as that of FP32. BF16 training is utilized in models such as BLOOM~\cite{le2023bloom} since the loss slowly diverges when the loss scalar becomes too low in FP16~\cite{cherti2023reproducible}. However,
fast bfloat16 support is only available on TPUs, or GPUs
developed with or after the NVIDIA Ampere series. Furthermore,
mixed-precision training and techniques such as loss scaling are exploited to ensure numerical stability due to the limited dynamic range represented by reduced precision. 
8-Bit or even lower-bit training is also becoming the focus of quantitative research.

\subsubsection{16-Bit Floating Point}
\label{subsubsec:16-BitFloatingPoint}

Popular training systems often employ FP16/BF16 mixed-precision strategies to reduce precision during training, as highlighted by works like Megatron-LM~\cite{shoeybi2019megatron} and Colossal-AI~\cite{li2023colossal}. The FP16 mixed-precision training scheme~\cite{micikevicius2017mixed} utilizes the IEEE half-precision format to store weights, activations, and gradients for forward and backward arithmetic operations. To maintain model accuracy at reduced precision, a single-precision copy of weights is kept for accumulation at each optimizer step. Loss scaling is also applied to preserve the values of small-magnitude gradients.
Campo~\cite{he2022campo} optimizes the casting cost incurred by conversions between FP32 and FP16 through automatic graph rewriting. This is crucial since the casting cost can sometimes negate the performance benefits of low precision. Campo also employs offline-trained linear regression models to predict casting costs and execution times for FP32 and FP16 operations. 
BF16~\cite{kalamkar2019study} is also widely used in mixed-precision training across various fields~\cite{yang2019high,fischer2018automatic}. It has the same representational range as FP32 and does not require hyperparameter tuning for convergence.
In addition, THC~\cite{li2024thc} addresses computational overhead in parameter server architectures by eliminating the need for decompression and compression. THC enables direct aggregation of compressed gradient values through the Uniform Homomorphic Compression property, thus enhancing efficiency.

\subsubsection{Sub-8-Bit Floating Point}
\label{subsubsec:8-BitFloatingPoint}

With the newly released chips characterized by lower precision data types such as FP8, mixed-precision training is designed to train with lower precision. 
Newly designed data formats combined with the techniques to ensure numerical stability are
primarily leveraged to enable FP8 training for deep learning neural networks. 
Wang et al.~\cite{wang2018training} use a new FP8 floating point format for numerical representation of data as well as computations. Chunk-based computations and stochastic rounding are utilized in the floating point accumulation and weight update process, respectively, to preserve model accuracy.
Sun et al.~\cite{sun2019hybrid} propose hybrid 8-bit floating point training across the whole spectrum of deep learning models without accuracy degradation. The novel hybrid FP8 formats utilize different exponent bits and mantissa bits for forward and backward propagation, respectively, since forward and backward passes have different optimal balances between range and precision. Besides, the techniques such as loss scaling are used to avoid accuracy degradation. 
With the maturation of more accelerators with FP8 data types, an FP8 automatic mixed-precision framework (FP8-LM)~\cite{peng2023FP8LMTraining} for training LLMs based on NVIDIA H100 GPU~\cite{choquette2023nvidia} is proposed, where 8-bit gradients, optimizer states, and distributed parallel training are gradually incorporated and FP8 low-bit parallelism including tensor, pipeline, and sequence parallelism is specified. Besides, precision decoupling and automatic scaling are designed to solve the data underflow or overflow issues due to the narrower dynamic range and reduced precision. FlashAttention-3 also employs block GEMM quantization and incoherent processing that exploits hardware support for FP8 low-precision on H100 GPU. Furthermore, Rouhani et al.~\cite{rouhani2023microscaling} train LLMs at sub-8-bit weights, activations, and gradients with minimal accuracy loss by utilizing micro scaled data formats that associate scaling factors with fine-grained sub-blocks of a tensor.


\subsubsection{Low-Bit Fixed Point}
\label{subsubsec:Low-BitFixedPoint}

Low-bit fixed point training is also studied for LLM training. 
Jetfire~\cite{xi2024jetfire} maintains an INT8 data flow where inputs and outputs are loaded and stored in INT8 data formats to accelerate both compute-bound linear operators and memory-bound non-linear operators.
In addition, tiling algorithms are utilized to excavate shared memory data access with a per-block quantization method, where higher precision computations are performed,  i.e. INT32 for WMMA tensor core operations for linear operators and FP32 for non-linear operations, to maintain the accuracy of pretrained transformers.
Xi et al.~\cite{xi2023training} propose a novel INT4 training algorithm for transformer models. 
In the forward propagation, the activation matrix is firstly transformed into a block diagonal Hadamard matrix to alleviate the accuracy degradation caused by outliers in the activation, and the transformed matrix is then quantized. In the backward propagation, bit splitting and leverage score sampling are exploited to choose informative gradients for quantization based on the structural sparsity of activation gradients.

Recently, low-precision training for LLMs has advanced to using 1-bit precision. BitNet~\cite{wang2023bitnet} employs a novel low-bit precision matrix multiplication within transformer blocks, utilizing weights that are 1-bit and activations that are 8-bit. The model weights are centralized around zero to maximize capacity within the limited numerical range, then binarized to +1 or -1 using the signnum function. To ensure training stability and accuracy, the gradients, optimizer states, and a high-precision latent weight copy are maintained for parameter updates.
Building on BitNet, BitNet b1.58~\cite{ma2024era} further enhances modeling capability by lowering model weights to ternary values \{-1, 0, 1\}. The weight matrix is scaled by its average absolute value, and each value is rounded to the nearest integer among -1, 0, and +1.

\section{Memory Optimizations}
\label{sec:mem}

\tikzstyle{my-box}=[
    rectangle,
    draw=black,
    rounded corners,
    text opacity=1,
    minimum height=1.5em,
    minimum width=5em,
    inner sep=2pt,
    align=center,
    fill opacity=.5,
    line width=0.8pt,
]
\tikzstyle{leaf}=[my-box, minimum height=1.5em,
    fill=hidden-red!10, text=black, align=left,font=\normalsize,
    inner xsep=2pt,
    inner ysep=4pt,
    line width=0.8pt,
]

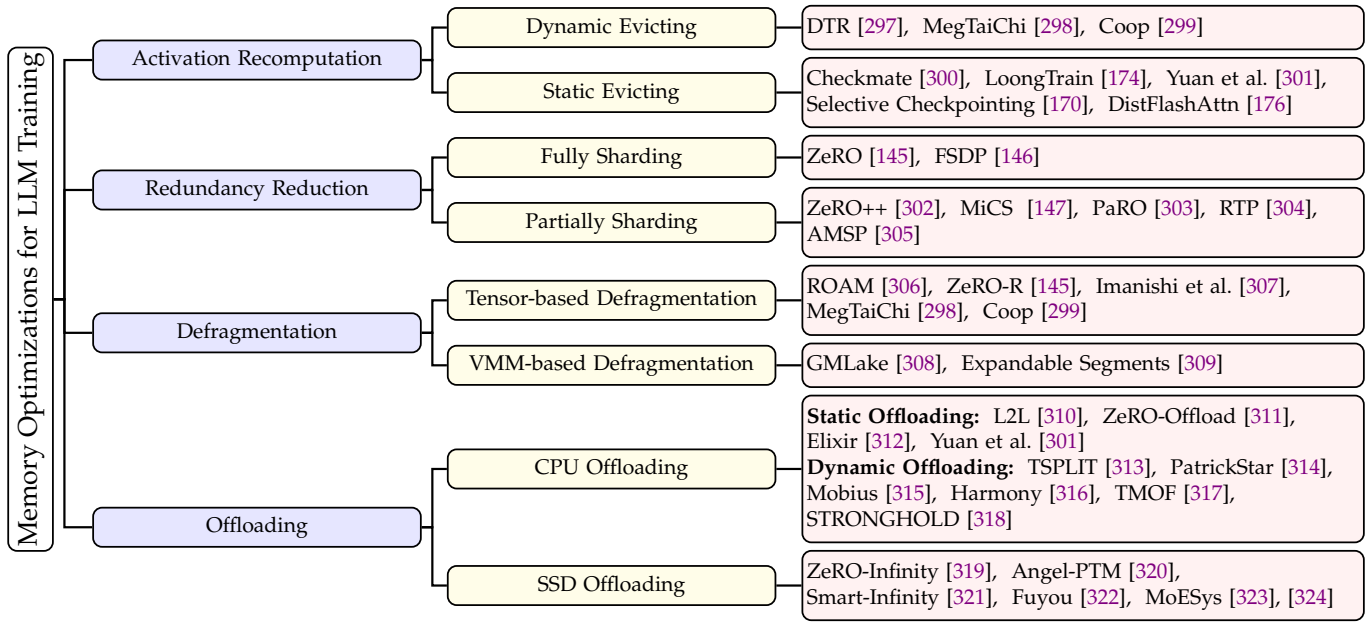
\begin{figure*}[htpb]
    \centering
    \resizebox{\textwidth}{!}{
        \begin{forest}
        forked edges,
        for tree={
            grow=east,
            reversed=true,
            anchor=base west,
            parent anchor=east,
            child anchor=west,
            base=center,
            font=\large,
            rectangle,
            draw=black,
            rounded corners,
            align=left,
            text centered,
            minimum width=4em,
            edge+={black, line width=1pt},
            s sep=3pt,
            inner xsep=2pt,
            inner ysep=3pt,
            line width=0.8pt,
            ver/.style={rotate=90, child anchor=north, parent anchor=south, anchor=center},
        },
        where level=1{text width=15em,font=\normalsize,}{},
        where level=2{text width=15em,font=\normalsize,}{},
        [
        {Memory Optimizations for LLM Training}, ver 
                [
                {Activation Recomputation}, fill=blue!10
                    [
                    {Dynamic Evicting}, fill=yellow!10
                        [
                        DTR~\cite{marisa2021dtr}{, }
                        MegTaiChi~\cite{hu2022megtaichi}{, }
                        Coop~\cite{zhang2024coop},
                        leaf, text width=26em 
                        ]
                    ]
                    [
                    {Static Evicting}, fill=yellow!10
                        [
                        Checkmate~\cite{jain2020checkmate}{, }
                        LoongTrain~\cite{gu2024loongtrain}{, }
                        Yuan et al.~\cite{yuan2024accelerating}{, }\\
                        Selective Checkpointing~\cite{korthikanti2022ReducingActivation}{, } 
                        DistFlashAttn~\cite{li2023LightSeqSequence} \\
                        ,
                        leaf, text width=26em 
                        ]
                    ]
                ] 
                [
                {Redundancy Reduction}, fill=blue!10
                    [
                    {Fully Sharding}, fill=yellow!10
                        [
                        ZeRO~\cite{rajbhandari2020zero}{, }
                        FSDP~\cite{zhao2023PyTorchFSDP},
                        leaf, text width=26em
                        ]
                    ]
                    [
                    {Partially Sharding}, fill=yellow!10
                        [
                        ZeRO++~\cite{wang2023ZeROExtremely}{, }
                        MiCS ~\cite{zhang2022mics}{, }
                        PaRO~\cite{wu2023rethinking}{, }
                        RTP~\cite{luo2023rtp}{, } \\
                        AMSP~\cite{chen2023amsp},
                        leaf, text width=26em
                        ]
                    ]
                ] 
                [
                {Defragmentation}, fill=blue!10 
                    [
                    {Tensor-based Defragmentation}, fill=yellow!10
                        [
                        ROAM~\cite{shu2023roam}{, }
                        ZeRO-R~\cite{rajbhandari2020zero}{, }
                        Imanishi et al.~\cite{imanishi2024heuristic}{, } \\
                        MegTaiChi~\cite{hu2022megtaichi}{, } 
                        Coop~\cite{zhang2024coop}
                        ,
                        leaf, 
                        text width= 26em
                        ]
                    ]
                    [
                    {VMM-based Defragmentation}, fill=yellow!10
                        [
                        GMLake~\cite{guo2024gmlake}{, }
                        Expandable Segments~\cite{torch2023expandable}
                        ,
                        leaf, 
                        text width= 26em
                        ]
                    ]
                ] 
                [
                {Offloading}, fill=blue!10
                    [
                    {CPU Offloading}, fill=yellow!10
                        [
                        \textbf{Static Offloading:}~
                        L2L~\cite{pudipeddi2020training}{, }
                        ZeRO-Offload~\cite{ren2021zero}{, } \\
                        Elixir~\cite{huang2023Elixir}{, }
                        Yuan et al.~\cite{yuan2024accelerating}\\
                        \textbf{Dynamic Offloading:}~
                        TSPLIT~\cite{nie2022TSPLIT}{, } 
                        PatrickStar~\cite{fang2022parallel}{, } \\
                        Mobius~\cite{feng2023Mobius}{, }
                        Harmony~\cite{li2022harmony}{, } 
                        TMOF~\cite{lin2023tensor}{, } \\
                        STRONGHOLD~\cite{sun2022stronghold}
                        ,
                        leaf, 
                        text width= 26em
                        ]
                    ]
                    [
                    {SSD Offloading}, fill=yellow!10
                        [
                        ZeRO-Infinity~\cite{rajbhandari2021zero}{, }
                        Angel-PTM~\cite{nie2023AngelPTM}{, } \\
                        Smart-Infinity~\cite{jang2024smart}{, }
                        Fuyou~\cite{liao2024Adding}{, }
                        MoESys\cite{yu2024moesys,shen2022se},
                        leaf, 
                        text width= 26em
                        ]
                    ]
                ]   
        ]
        \end{forest}
    }
    
    \caption{Studies on memory optimizations for distributed LLM training.}
    \label{taxonomy:memory}
\end{figure*}

Memory consumption during the training of LLMs can be categorized into four key components: model states, activations, temporary buffers, and memory fragmentation.

\begin{itemize}
    \item \textbf{Model States:} Model states encompass the memory consumed by the optimizer states, gradients, and model parameters. In mixed-precision training~\cite{micikevicius2017mixed}, model parameters and activations are stored in 16-bit precision. When training a model with $\Phi$ parameters, $4\Phi$ bytes are needed to store parameters and gradients. The 32-bit copies of the parameters, momentum, and variance each require $4\Phi$ bytes, totaling $12\Phi$ bytes. Therefore, the overall memory requirement for storing model states is $16\Phi$ bytes. 
    
    \item \textbf{Activations:} Activations refer to the tensors generated during the forward pass. These tensors are essential for gradient computation during the backward phase.
    
    \item \textbf{Temporary Buffers:} Temporary buffers are used to store intermediate results. For example, operations such as gradient AllReduce often fuse gradients in a bucket into a single flattened buffer before applying the operation to enhance throughput.
    
    \item \textbf{Memory Fragmentation:} Memory fragmentation can lead to scenarios where memory requests fail despite having a large amount of available memory. This occurs because usable memory can become fragmented, and there is insufficient contiguous memory to satisfy the memory request \cite{rajbhandari2020zero}.
\end{itemize}

To address memory constraints of LLM training, various memory-efficient techniques have been proposed. These include activation recomputation strategies, which trade increased computation for reduced memory usage; redundancy reduction methods that minimize data duplication across training processes; defragmentation techniques that optimize memory allocation and deallocation to reduce fragmentation and improve memory utilization; and swap and offload approaches that leverage CPU memory and NVMe SSDs to supplement GPU memory. Figure \ref{taxonomy:memory} outlines the taxonomy of these optimizations for memory-efficient  LLM training.

\subsection{Activation Recomputation}
\label{subsec_activation_recomputation}

During the backward phase of model training, activations are essential for computing gradients. As model sizes increase, the memory required to store these activations during training can exceed GPU memory capacity, thereby limiting the scale of the models that can be trained. Activation recomputation~\cite{chen2016training} provides a solution by strategically discarding certain activations during the forward pass and recomputing them as needed during the backward pass. This approach has become a de facto method for reducing memory consumption in LLM training. The key to effective activation recomputation is balancing the memory savings against the additional computational overhead. 

We categorize these methods into two primary approaches: static evicting and dynamic evicting. Static evicting methods typically involve the formulation of evicting strategies tailored to specific model architectures or modules. In contrast, dynamic evicting methods make decisions in real-time without prior knowledge of the model. Although static approaches necessitate modifications for new models, the structure of the majority of LLMs share similar architectures, enabling the general application of these strategies during LLM training. Despite their inherent flexibility, dynamic evicting methods have not been widely adopted in the training of LLMs. Nevertheless, we still explore some related works in this section for further reference.

\subsubsection{Static Evicting}
\label{subsubsec_static_evicting}

Static evicting involves establishing a fixed plan for discarding activations during the forward pass and later recomputing them during the backward pass. Checkmate~\cite{jain2020checkmate} formulates this activation recomputation problem as a mixed integer linear program to determine the optimal rematerialization plan for static deep learning models. However, Checkmate struggles to scale to large models like LLMs due to the vast search space. 

Recently, several works have proposed customized activation recomputation policies tailored for LLM training. Selective-checkpointing~\cite{korthikanti2022ReducingActivation} selectively discards the activations of memory-intensive attention modules. FlashAttention~\cite{dao2022flashattention} fuses the attention module into a single kernel, and also employs selective-checkpointing to reduce memory consumption. DistFlashAttn~\cite{li2023LightSeqSequence} addresses the high computation overhead in long sequences caused by the recomputation of attention modules, employing a rematerialization-aware gradient checkpointing strategy. Specifically, DistFlashAttn places checkpoints at the output of the FlashAttention kernel instead of at the Transformer layer boundary, thereby removing recomputation in the attention module during the backward pass and only requiring storage of its output.
LoongTrain~\cite{gu2024loongtrain} introduces selective-checkpoint++, which further optimizes the checkpointing process, particularly for training with long sequences, by adding attention modules to a \textit{whitelist}. This method saves the attention output and softmax statistics (\texttt{softmax\_lse}). During the forward pass, it saves the outputs of the modules in the whitelist, and during the backward pass, it retrieves these stored outputs instead of recomputing them, continuing the computation graph and thus reducing the need for recomputing attention.

Unlike recent works that predominantly focus on handcrafted checkpointing policies on attention modules for LLM training, Yuan et al.~\cite{yuan2024accelerating} carefully measure the minimum computation cost required to reconstruct each activation tensor during model training. They derive a Pareto frontier of memory and computation costs by enumerating all possible checkpointing methods. From this Pareto frontier, they select a solution that optimally balances computation and memory costs.

\subsubsection{Dynamic Evicting}
\label{subsubsec_dynamic_evicting}

Dynamic evicting makes real-time decisions on which activations to discard and recompute based on the current state of the training process. DTR~\cite{marisa2021dtr} proposes a greedy online algorithm to heuristically evict and rematerialize tensors at runtime for both static and dynamic models. MegTaiChi~\cite{hu2022megtaichi} introduces a dynamic tensor evicting that leverages the access patterns of tensors tracked at runtime. Coop~\cite{zhang2024coop} proposes to mitigate the memory fragmentation issue caused by activation recomputing methods due to evicting tensors without considering their contiguity. Coop employs an efficient sliding window algorithm to ensure that only contiguous memory blocks are evicted, thereby minimizing memory fragmentations.

\subsection{Redundancy Reduction}
\label{subsec_redundancy_reduction}

Traditional data parallel approaches replicate the entire model state across all GPUs, which leads to substantial redundant memory usage. Redundancy reduction techniques are proposed to optimize memory usage by eliminating or reducing memory redundancies on each device. These techniques often seek to balance memory efficiency with the induced communication overhead, thereby facilitating training of larger scale or batch size with acceptable costs.

\subsubsection{Fully Sharding}
\label{subsubsec_full_sharding}

The Zero Redundancy Optimizer (ZeRO)~\cite{rajbhandari2020zero} optimizes memory redundancies by fully sharding model states across all GPUs through three stages: ZeRO-1, ZeRO-2, and ZeRO-3. 
ZeRO-1 globally distributes optimizer states across all GPUs. During the training, each GPU conducts independent forward and backward propagation to compute gradients, which are subsequently synchronized across all GPUs within the data parallel group using an {ReduceScatter} operation. Each GPU is responsible for updating specific shard of the model parameters. Following this, the updated model parameter shards are collected from other GPUs using an {AllGather} operation, ensuring that all GPUs have the latest model parameters. ZeRO-1 reduces the memory consumption of optimizer states from $12\Phi$ to ${12\Phi}/{N}$, where $N$ is the size of data parallelism. 
Building upon ZeRO-1, ZeRO-2 further shards the gradients across all GPUs and each GPU only updates its parameter shard, reducing the memory required for holding gradients from $2\Phi$ to ${2\Phi}/{N}$.
ZeRO-3 partitions parameters in addition to optimizer states and gradients. Each GPU only holds a part of the parameters. When the parameters from remote GPUs are needed for the upcoming computation, they are collected by an {AllGather} operation and discarded afterward. In ZeRO-3, each GPU holds only the weights, gradients, and optimizer states corresponding to its specific parameter partition, reducing the overall memory consumption from $16\Phi$ to ${16\Phi}/{N}$.  ZeRO is widely adopted by numerous frameworks, such as DeepSpeed~\cite{rajbhandari2022deepspeed}, PyTorch-FSDP~\cite{zhao2023PyTorchFSDP}, and ColossalAI~\cite{li2023colossal}.

\subsubsection{Partially Sharding}
\label{subsubsec_partially_sharding}

ZeRO faces communication challenges since the latency of collective communication operations increases with the communication scale. There exists a trade-off between memory utilization and communication cost in distributed LLM training. Optimizing communication overhead can be achieved by sharding the model states across smaller groups of GPUs, which are smaller sets of GPUs within a large GPU cluster. This approach reduces inter-node communications and communication scales, though it may lead to higher memory usage due to increased redundancy of model states. The key is to balance the communication scale with memory utilization~\cite{chen2023amsp}.

Several approaches building upon the ZeRO framework have been proposed to address the communication inefficiencies while improving memory utilization. 
ZeRO++~\cite{wang2023ZeROExtremely} partitions all model states globally across all devices following ZeRO-3 and further introduces a secondary shard of parameters within subgroups of GPUs. In the forward phase, it collects parameters leveraging the primary shard across all GPUs and maintains a secondary shard of parameters within subgroups, typically within the same node. During the backward phase, it collects parameters from this secondary shard, reducing communication scale and inter-node communications. Additionally, ZeRO++ uses quantization to compress parameters and gradients, effectively diminishing communication volume with a trade-off in accuracy. MiCS \cite{zhang2022mics} and FSDP~\cite{zhao2023PyTorchFSDP} shards all model state components within subgroups and replicates them across subgroups, thereby reducing communication scale and consequently communication latency, leading to enhanced training performance. AMSP \cite{chen2023amsp} and PaRO \cite{wu2023rethinking} incorporate three flexible sharding strategies, including Full-Replica, Full-Sharding, and Partial-Sharding, allowing each component within the model states to independently choose a sharding strategy. AMSP formulates an optimization problem to find the optimal sharding strategy that minimizes communication costs under memory constraints. In addition, AMSP proposes a customized communication and computation overlap strategy, incorporating these flexible sharding strategies to achieve optimized training efficiency. RTP (Rotated Tensor Parallelism)\cite{luo2023rtp} seeks to minimize memory duplication by strategically sharding activations and rotating weights/gradients.


\subsection{Defragmentation}
\label{subsec_defragmentation}
GPU memory fragmentation refers to the scattered, unusable chunks of GPU memory that arise between adjacent tensors. This problem is particularly pronounced during the training of LLMs due to the varying lifetimes of different tensors and the inefficient memory allocation and deallocation schemes of general deep learning frameworks, such as PyTorch~\cite{paszke2019pytorch} and TensorFlow~\cite{abadi2016tensorflow}. Furthermore, memory optimization techniques like recomputation and offloading exacerbate the issue by introducing more frequent and irregular memory allocation and deallocation requests~\cite{guo2024gmlake, shu2023roam, zhang2024coop}. The fragmentation problem could cause high peak memory and out-of-memory (OOM) errors limiting the batch size and overall training efficiency. Defragmentation efforts are proposed to mitigate these issues through memory management techniques.


\subsubsection{Tensor-based Defragmentation} 
\label{subsubsec_tensor_based_defrag}

Deep learning frameworks typically use a caching allocator with a memory pool to enable fast memory allocation and deallocation without requiring device synchronization. Several approaches have been proposed to reduce memory fragmentation based on the tensor allocation and deallocation scheme in the caching allocator.
ROAM~\cite{shu2023roam} co-optimizes the execution order of operators and tensor allocation by considering tensors' lifetimes and sizes. It introduces an efficient tree-based algorithm to search for an execution plan that maximizes tensor reuse and reduces data fragmentation. ROAM has been evaluated in single-GPU scenarios, specifically with the largest model being the 1.5B GPT-2 XL~\cite{radford2019language}, but it has not yet been tested in distributed training scenarios with larger models, where computation graphs can become significantly larger.
Imanishi et al.~\cite{imanishi2024heuristic} present an offline optimization approach by modeling tensor allocation as a 2D bin-packing problem. In this model, each tensor allocation is represented as a vertically movable rectangle, reflecting periodic allocation patterns during model training. They propose a heuristic algorithm using simulated annealing to optimize the topological ordering of allocations, aiming to minimize fragmentation. While effective, this method may struggle with scalability issues when applied to LLMs due to the high number of allocations and complex patterns involved.
MegTaiChi~\cite{hu2022megtaichi} and  Coop~\cite{zhang2024coop} consider memory fragmentation issues when evicting activation tensors for reducing memory consumption.  


\subsubsection{VMM-based Defragmentation}
\label{subsubsec_vmm_defrag}
GMLake~\cite{guo2024gmlake} and PyTorch expandable segments~\cite{torch2023expandable} propose to mitigate fragmentation by utilizing the virtual memory management (VMM) functions of the low-level CUDA driver application programming interface. This low-level API provides developers with direct control over the GPU's virtual memory operations, such as reserving, mapping, and managing virtual memory addresses. Building on this, GMLake~\cite{guo2024gmlake} introduces a virtual memory stitching mechanism that consolidates non-contiguous memory blocks into larger ones through virtual memory address mapping, minimizing data movement and copying. Similarly, PyTorch's expandable segments~\cite{torch2023expandable} enable allocated memory segments to be expanded to larger sizes for reuse. Both approaches are transparent to different models and memory-efficient training techniques and can be seamlessly integrated into existing deep learning frameworks. Furthermore, GMLake demonstrates excellent scalability on multi-GPUs with minimal overhead and does not require modification to user code. PyTorch-v2.1 has also integrated expandable segments.

\subsection{Offloading}
\label{subsec_offloading}

To enable efficient training of LLMs on fewer GPUs, various works leveraging swap and offload methods have been proposed. These techniques transfer parts of the computation and data from GPU memory to external resources, which are inexpensive and slower but enjoy vast capacity.

\subsubsection{CPU Offloading}
\label{subsubsec_cpu_offloading}

Numerous studies have proposed methods to efficiently utilize CPU memory to enhance distributed LLM training. These techniques can be broadly categorized into two main approaches: {Static Offloading} and {Dynamic Offloading}.

\phb{Static Offloading.}~Static offloading methods involve a predetermined allocation of model components between GPU and CPU memory. L2L~\cite{pudipeddi2020training} manages and moves tensors layer by layer. L2L synchronously fetches tensors required for the upcoming computational layers into GPU memory while keeping the tensors for the remaining layers stored in CPU memory. L2L allows scaling the models to arbitrary depth but fails to scale across multi-GPUs. In contrast, ZeRO-Offload~\cite{ren2021zero} concentrates on multi-GPU training. It holds model parameters on GPU, and stores optimizer states and gradients on CPU memory. In addition, it offloads optimizer update computation to the CPU. This method enables the training of up to 70B models with 16 V100s. However, ZeRO-Offload can leave some GPU memory unused and suffers from slow CPU optimizer updates~\cite{huang2023Elixir}. To address this issue, Elixir~\cite{huang2023Elixir} employs a search engine to find the optimal combination of memory partitioning and offloading by leveraging pre-runtime model profiling. Unlike ZeRO-Offload, Elixir effectively utilizes all available GPU memory by partitioning both the model states and optimizer chunks between GPU and CPU. Mobius~\cite{feng2023Mobius} tackles multi-GPU training on commodity servers with limited inter-GPU bandwidth and high communication contention by introducing a pipeline parallelism scheme. This scheme assigns each GPU multiple stages and dynamically swaps them between GPU and CPU memory. Additionally, Mobius optimizes communication through prefetching and cross-mapping to reduce overhead and contention.
Yuan et al.~\cite{yuan2024accelerating} propose to mitigate the activation bottleneck by offloading and reloading activations at the granularity of pipeline stages while maximizing the overlap between activation transmission with computation, thereby avoiding slowing the training process. Compared to other offloading efforts, this work focuses more on improving the balance between computation and memory utilization rather than training with extremely tight memory budgets.

\subsubsection{Dynamic Offloading}
\label{subsubsec_dynamic_offloading}

Dynamic offloading methods adaptively allocate partitions of model or tensors between GPU and CPU memory based on real-time optimization of memory utilization and data transmission. STRONGHOLD~\cite{sun2022stronghold} proposes to dynamically offload model states between GPU and CPU memory and maintain a suitable working window size to minimize GPU stalls during offloading. Harmony~\cite{li2022harmony} employs a heuristic-based scheduler to map computation and model states to physical devices. Harmony reduces the overhead for offloading with reduced swaps and fast peer-to-peer swaps. TMOF~\cite{lin2023tensor} introduces disjoint swapping and bidirectional overlapping coordination mechanisms to prevent PCIe channel contention in swapping and offloading. For MoE models, MPipeMoE\cite{zhang2024mpmoe} designs an adaptive and memory-efficient pipeline parallelism algorithm. Specifically, MPipeMoE employs efficient memory reusing strategies by eliminating memory redundancies and an adaptive selection component to decide whether to offload or recompute the required tensors to reduce memory requirements. 

To facilitate better memory management, some studies have proposed systems that break tensors into finer-grained units.  TSPLIT~\cite{nie2022TSPLIT} and PatrickStar~\cite{fang2022parallel} are two dynamic memory management systems that optimize peak GPU memory usage. TSPLIT splits tensors into micro-tensors and performs operations at the micro-tensor level, enabling precise and dynamic memory operations. %
PatrickStar organizes model data into memory chunks that are dynamically distributed between CPU and GPU memory and optimizes CPU-GPU data transmission as well as bandwidth utilization. 
Additionally, TSPLIT uses a model-guided planning algorithm to find optimal memory configurations for each tensor, while PatrickStar employs runtime memory tracing, chunk eviction strategies, and device-aware operator placement to further minimize data movement between CPU and GPU.

\subsubsection{SSD Offloading}
\label{subsubsec_ssd_offloading}

To facilitate the training of trillion-scale LLMs, where methods solely relying on CPU offloading are insufficient, several works have been proposed for offloading data to both CPU memory and NVMe SSDs during training. 
ZeRO-Infinity~\cite{rajbhandari2021zero} offloads all the partitioned model states to CPU or NVMe memory and offloads activation only to CPU memory. This method supports training models with up to 32T parameters on 32 nodes (a total of 512 V100s). However, the CPU offloading for activations still requires extensive CPU memory. For instance, approximately 0.76 TB of CPU memory is needed to store activation checkpoints for training a 10T model, and around 4 TB for 100T models. Fuyou~\cite{liao2024Adding} focuses on training LLMs on commodity servers with limited CPU memory capacity and a single GPU. Compared to ZeRO-Infinity, Fuyou further offloads the activations to SSDs and incorporates SSD-CPU communication as an additional optimization dimension. It also proposes a synchronous out-of-core CPU optimizer that overlaps with the backward propagation stage and introduces an automatic activation swapping mechanism, thereby maximizing GPU utilization.
Smart-Infinity~\cite{jang2024smart} proposes to reduce the secondary storage bandwidth requirements by using near-storage processing devices for parameter update. 
MoESys\cite{yu2024moesys,shen2022se} combines various storage devices (GPU, CPU memory, and SSDs) to save the sparse parameter states and dense parameter states and propose a 2D prefetch scheduling strategy to MoE training so that the computation of parameters can be overlapped with the scheduling.

\section{Communication Optimizations}
\label{sec:comm}

\begin{figure}[t]
    \centering
    \includegraphics[width=0.95\linewidth]{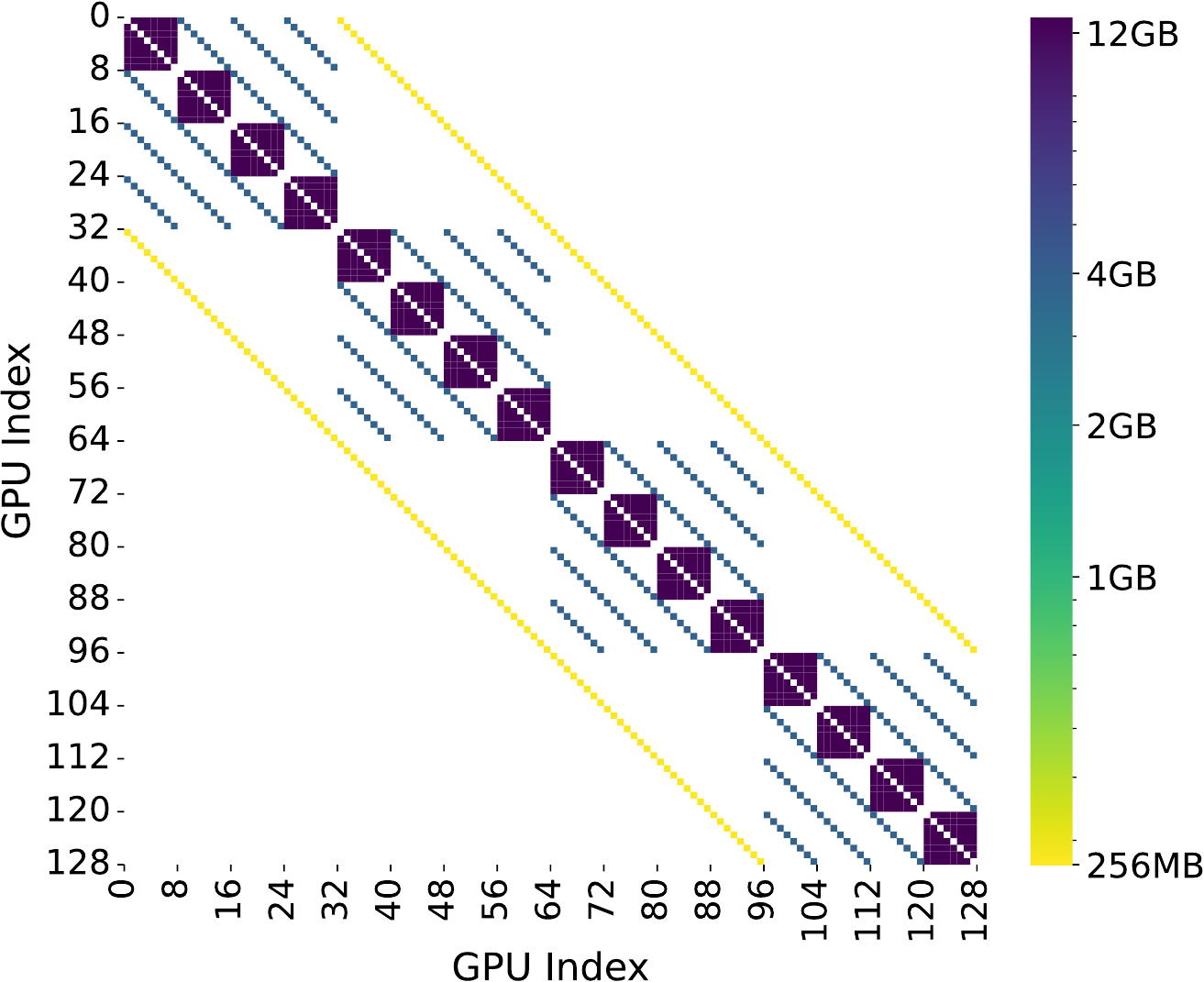}
  \caption{Communication traffic heatmap for InternLM-2 102B pre-training using 128 GPUs during a single iteration, with tensor parallelism (TP) size 8, pipeline parallelism (PP) size 4, data parallelism (DP) size 4 and ZeRO stage 1 (ZeRO-1) size 4. The prioritization of topology arrangement is TP \textgreater DP/ZeRO-1 \textgreater PP. There are four different data traffic loads: \circled{1} the AllReduce of TP; \circled{2}\circled{3}  ReduceScatter/AllGather of DP/ZeRO-1; \circled{4} Send/Recv of PP. The communication for TP utilizes the fully-connected topology of NVSwitch, resulting in sixteen dense square traffic patterns along the diagonals in the diagram, with each pattern representing a node. The cross-node communication traffic for DP and ZeRO-1 are shown in the diagram as six symmetric diagonal lines within the four 32x32 rectangular topologies. It is important to note that DP/ZeRO-1 also involves intra-node communication traffic, which accumulates into the same heatmap grid as TP. Due to its relatively small communication volume, PP forms two yellow lines on the heatmap at coordinates ((32, 0), (128, 96)) and ((0, 32), (96, 128)). (In this diagram, all communications use the ring-based collective  algorithm)} 
  \label{fig:communication_traffic}
\end{figure}

\tikzstyle{my-box}=[
    rectangle,
    draw=black,
    rounded corners,
    text opacity=1,
    minimum height=1.5em,
    minimum width=5em,
    inner sep=2pt,
    align=center,
    fill opacity=.5,
    line width=0.5pt,
]
\tikzstyle{leaf}=[my-box, minimum height=1.5em,
    fill=hidden-red!10, text=black, align=left,font=\normalsize,
    inner xsep=2pt,
    inner ysep=4pt,
    line width=0.8pt,
]

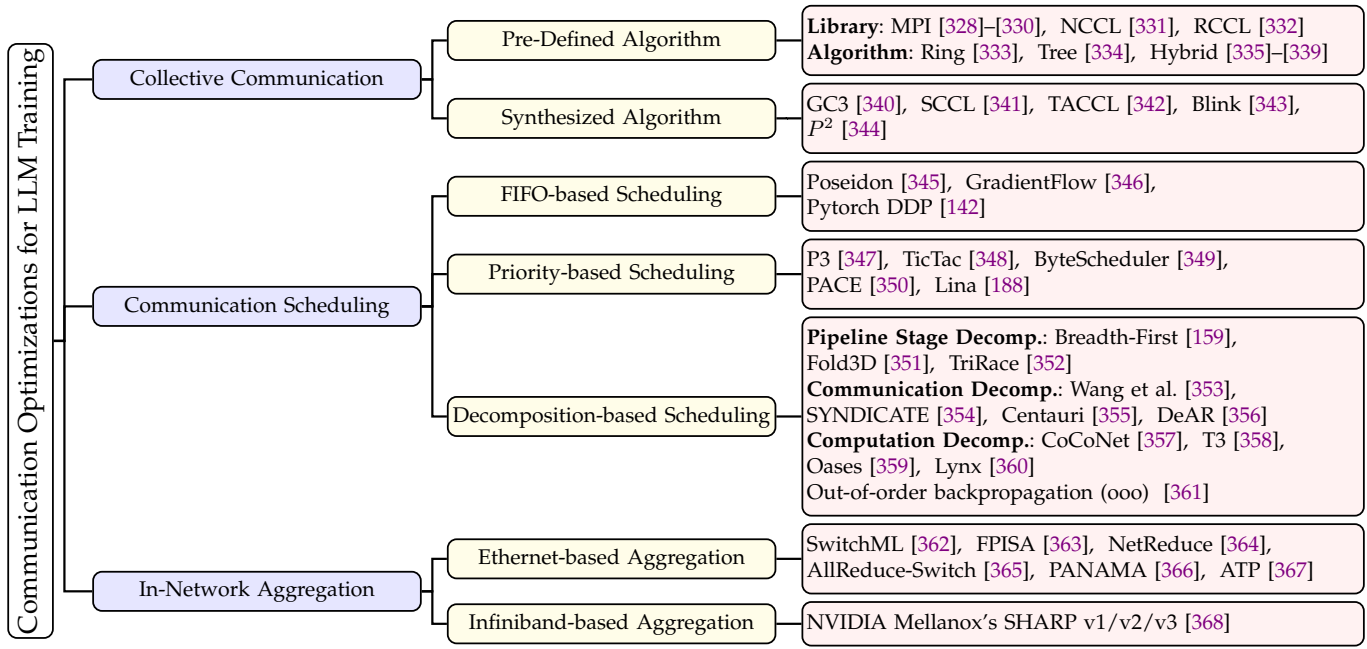
\begin{figure*}[htpb]
    \centering
    \resizebox{\textwidth}{!}{
        \begin{forest}
        forked edges,
        for tree={
            grow=east,
            reversed=true,
            anchor=base west,
            parent anchor=east,
            child anchor=west,
            base=center,
            font=\large,
            rectangle,
            draw=black,
            rounded corners,
            align=left,
            text centered,
            minimum width=4em,
            edge+={black, line width=1pt},
            s sep=3pt,
            inner xsep=2pt,
            inner ysep=3pt,
            line width=0.8pt,
            ver/.style={rotate=90, child anchor=north, parent anchor=south, anchor=center},
        },
        where level=1{text width=15em,font=\normalsize,}{},
        where level=2{text width=15em,font=\normalsize,}{},
        [
        {Communication Optimizations for LLM Training}, ver
                [
                {Collective Communication }, fill=blue!10
                    [
                    {Pre-Defined Algorithm}, fill=yellow!10
                        [
                        \textbf{Library}: MPI \cite{gabriel2004open_mpi,mpich,panda2013mvapich}{, }
                        NCCL \cite{nccl2016}{, }
                        RCCL \cite{rccl2018} \\
                        \textbf{Algorithm}: Ring \cite{patarasuk2009bandwidth}{, }
                        Tree \cite{Doublebinarytrees}{, }
                        Hybrid \cite{jia2018highly,2DTorusAllreduce,dong2021accl,cho2019blueconnect,luo2020plink}, 
                        leaf, 
                        text width= 26em
                        ]
                    ]
                    [
                    {Synthesized  Algorithm}, fill=yellow!10
                        [
                        GC3 \cite{cowan2022gc3}{, }
                        SCCL \cite{cai2021synthesizing}{, } 
                        TACCL \cite{shah2023taccl}{, }
                        Blink~\cite{wang2019blink}{, } \\
                        $P^2$~\cite{xie2022synthesizing},
                        leaf, 
                        text width= 26em
                        ]
                    ]
                ]
                [
                {Communication Scheduling}, fill=blue!10
                    [
                    {FIFO-based Scheduling}, fill=yellow!10
                        [
                        Poseidon \cite{zhang2017poseidon}{, }
                        GradientFlow \cite{sun2019gradientflow}{, } \\
                        Pytorch DDP \cite{li2020PyTorchDistributed},
                        leaf, 
                        text width= 26em
                        ]
                    ]
                    [
                    {Priority-based Scheduling}, fill=yellow!10
                        [
                        P3 \cite{jayarajan2019p3}{, }
                        TicTac~\cite{hashemi2018tictac}{, }
                        ByteScheduler~\cite{peng2019bytescheduler}{, } \\
                        PACE~\cite{bao2020preemptive}{, }
                        Lina \cite{li2023accelerating},
                        leaf, 
                        text width= 26em
                        ]
                    ]
                    [
                    {Decomposition-based Scheduling}, fill=yellow!10
                        [
                        \textbf{Pipeline Stage Decomp.}: 
                        Breadth-First \cite{lamy2023breadth}{, } \\
                        Fold3D \cite{li2023fold3d}{, }
                        TriRace \cite{li2024trirace} \\
                        \textbf{Communication Decomp.}: 
                        Wang et al.~\cite{wang2022OverlapCommunication}{, } \\
                        SYNDICATE~\cite{mahajan2023syndicate}{, } 
                        Centauri~\cite{chen2024centauri}{, } 
                        DeAR~\cite{zhang2023dear} \\
                        \textbf{Computation Decomp.}: 
                        CoCoNet~\cite{jangda2022coconet}{, }
                        T3~\cite{pati2024t3}{, } \\
                        Oases~\cite{li2023oases}{, } 
                        Lynx~\cite{chen2024lynx} \\
                        Out-of-order backpropagation (ooo)  ~\cite{oh2022out-of-order},
                        leaf, 
                        text width= 26em
                        ]
                    ]
                ]
                [
                {In-Network Aggregation}, fill=blue!10
                    [
                    {Ethernet-based Aggregation}, fill=yellow!10
                        [
                        SwitchML \cite{sapio2021switchml}{, }
                        FPISA \cite{yuan2022unlocking}{, }
                        NetReduce \cite{liu2020netreduce}{, } \\
                        AllReduce-Switch \cite{liu2021allreduceswitch}{, } 
                        PANAMA \cite{gebara2021panama}{, }
                        ATP \cite{lao2021atp},
                        leaf, 
                        text width= 26em
                        ]
                    ]
                    [
                    {Infiniband-based Aggregation}, fill=yellow!10
                        [
                        NVIDIA Mellanox's SHARP v1/v2/v3 \cite{graham2016sharp},
                        leaf, 
                        text width= 26em
                        ]
                    ]
                ]
        ]
        \end{forest}
    }
    \caption{Studies on communication optimizations for distributed LLM training.}
    \label{taxonomy:communication}
\end{figure*}

Different parallelism mechanisms introduce varying patterns of network communication traffic. For instance, tensor parallelism requires AllReduce operations across the tensor parallelism ranks. Data parallelism, on the other hand, necessitates AllReduce operations for gradient synchronization across data parallelism ranks at the end of each iteration. Pipeline parallelism involves passing activation values to the next stage at the end of each stage. Typically, training frameworks place tensor or sequence parallel communication groups, which demand high bandwidth, within high-bandwidth domains (e.g., the same node), while placing data parallel or pipeline parallel communication groups, which have lower bandwidth requirements, between high-bandwidth domains. Fig.~\ref{fig:communication_traffic} shows the communication heatmap of LLM training in practice and well reflects the data traffic brought by different parallel strategies. From this heatmap, it can be observed that the LLM training communication traffic exhibits a clear pattern and hierarchy, with most communication occurring within smaller scopes, and only a little fraction of the traffic crossing the entire cluster. This insight has inspired approaches like rail-optimized topology~\cite{rail-optimized-topology}, which reduces unnecessary core switches to cut costs.

This section introduces systems and techniques for optimizing the collective communication performance of distributed LLM training. As shown in Fig.~\ref{taxonomy:communication}, we first discuss collective communication libraries, which utilize both pre-defined and synthesized algorithms. Next, we explore communication scheduling techniques designed to reorganize communication operations to overlap with computation, thereby reducing delays and accelerating the training process. Finally, we delve into in-network aggregation (INA), which leverages the computational capabilities of network devices to perform aggregation operations, such as summing gradients of deep learning models.

Compressing model parameters and gradients effectively reduces communication overhead during distributed LLM training. Various studies explore sparse communication and quantization approaches. For example, ZeRO++ \cite{wang2023ZeROExtremely} adopt quantization on weights to shrink down each model parameter from FP16 to INT8 data type before communicating. However, these works typically involve lossy sparsification or quantization techniques. We do not survey lossy data compression techniques in this section, as they are beyond the scope of this work.

\subsection{Collective Communication}
\label{collective_communication}

The Message Passing Interface (MPI) is a widely adopted programming model for large-scale scientific applications on parallel computing architectures. MPI has several implementations, including OpenMPI \cite{gabriel2004open_mpi}, MPICH2 \cite{mpich}, and MVAPICH \cite{panda2013mvapich}. These libraries provide a variety of CUDA-aware primitives such as AllReduce, AllGather, and ReduceScatter, which are essential for distributed LLM training. In practice, current training frameworks prefer collective communications tailored to specific AI accelerators with pre-defined or synthesized algorithms.

\subsubsection{Pre-Defined Collective Communication Algorithm}
\label{pre-defined}

NVIDIA's NCCL \cite{nccl2016} and AMD's RCCL \cite{rccl2018} are highly optimized libraries that typically outperform MPI-based collective communication libraries on their respective AI accelerators. These libraries usually select pre-defined algorithms to perform collectives based on conditions such as network topology and input tensor size.

\phb{Ring Algorithm.} The Ring algorithm is used for collective communications like AllReduce to move data across all GPUs. With this algorithm, the input tensor is split into multiple chunks and transferred one by one during the operation. This pipeline reduces the idle time that each device spends waiting for data. Baidu used the bandwidth-optimal ring AllReduce algorithm \cite{patarasuk2009bandwidth} for distributed deep learning model training. Horovod \cite{alex2018horovod} replaced the Baidu ring-AllReduce implementation with NCCL and designed a user-friendly interface for distributed training.

\phb{Tree Algorithm.} The latency of the Ring algorithm  increases with the number of GPU devices \cite{sun2019gradientflow}. The Double Binary Tree algorithm \cite{Doublebinarytrees} was proposed to solve this problem. Double binary trees rely on the fact that half or fewer ranks in a binary tree are nodes and half or more ranks are leaves. Therefore, a second tree can be built using leaves as nodes and vice-versa for each binary tree. This algorithm is implemented in MPI-based libraies, NCCL and RCCL.

\phb{Hybrid Algorithm.}
Several approaches propose using hybrid algorithms to  handle collective communication tasks on training clusters with heterogeneous intra-node and inter-node communication bandwidth.
Two-level AllReduce \cite{jia2018highly} divides a single AllReduce operation into three  steps: intra-node  Reduce utilizing PCIe/NVLINK, inter-node  AllReduce utilizing network, and intra-node Broadcast.  2D-Torus AllReduce \cite{2DTorusAllreduce} and ACCL \cite{dong2021accl}  decompose a single AllReduce operation into three phases: intra-node ring-based ReduceScatter, inter-node tree-based AllReduce, and intra-node ring-based AllGather. 
BlueConnect \cite{cho2019blueconnect}  breaks down a single AllReduce operation into numerous parallelizable ReduceScatter and AllGather operations. Each operation can be mapped to different network fabrics, leveraging the best-performing pre-defined implementation for each specific fabric. Plink \cite{luo2020plink} could probes network topology and efficiently generates two-level hybrid communication plans, exploiting locality in datacenter networks.

\subsubsection{Synthesized Collective Communication Algorithm}
\label{synthesized}

Several approaches have emerged that synthesize collective communication algorithms and kernels specifically tailored to the hardware topology, aiming to outperform generic pre-defined algorithms in many cases.
GC3 \cite{cowan2022gc3} introduces a data-oriented domain-specific language (DSL) for designing custom collective communication algorithms. It includes an optimizing compiler that translates these algorithms into executable forms optimized for specific hardware configurations.
SCCL \cite{cai2021synthesizing}  encodes the  collective communication  synthesis problem as an SMT (Satisfiability Modulo Theories) formula. This approach aims to derive exact schedules for Pareto-optimal algorithms, optimizing both latency and bandwidth utilization.
TACCL \cite{shah2023taccl} formulates the problem of finding optimal communication algorithms as a mixed integer linear program (MILP). It leverages a communication sketch abstraction to efficiently gather essential information and reduce the search space, with the goal of minimizing overall execution time.
Blink~\cite{wang2019blink} dynamically constructs a topology with suitable link capacities by probing available link sets for each job at runtime. Using this topology, it optimizes communication rates through the creation of packet generation trees, and generating CUDA code.
$P^2$~\cite{xie2022synthesizing}  utlizes parallel matrices to partition the parallel axis at the system level, thereby generating topology-aware parallel placement and reduction strategies. By simulating and predicting communication costs, this method reduces the number of actual evaluations required.

\subsection{Communication Scheduling}
\label{communication-scheduling}

Communication scheduling in distributed training reorganizes communication operations to overlap with computation, thereby reducing delays and accelerating the training process. The key concept of communication scheduling involves reordering communication operations based on the data dependencies of parallel training. Hybrid parallel LLM training necessitates multidimensional communication scheduling schemes to manage communications generated by data, pipeline, tensor, and sequence parallelism, as well as their combinations.

\subsubsection{FIFO-based Scheduling}
\label{fifo-based}

During the backward phase, rather than waiting for all gradient calculations to complete before initiating communication, communication can begin as soon as each gradient is ready. This wait-free backpropagation approach leverages a dependency-directed acyclic graph to manage tasks efficiently. Poseidon \cite{zhang2017poseidon} employs a First-In-First-Out (FIFO) queue to schedule  AllReduce operators, ensuring that each layer starts its communication once its gradients are generated. 
Motivated by the efficiency of collective communications on large tensors, GradientFlow \cite{sun2019gradientflow} and Pytorch DDP \cite{li2020PyTorchDistributed} fuse multiple sequential  AllReduce communication operations into a single operation. This method avoids transmitting a large number of small tensors over the network by waiting for a short period of time and then combining multiple gradients into one  AllReduce operation during the backward phase.

\subsubsection{Priority-based Scheduling}
\label{priority-base}

The FIFO scheme is often sub-optimal because the generated communication sequence of in the backward phase differs from the computation sequence in the forward phase. This mismatch can lead to communication blocking computation, even when overlap is enabled. Consequently, many approaches employ priority queues to schedule communication operators efficiently.
P3 \cite{jayarajan2019p3} schedules  AllReduce operations at a finer granularity, overlapping gradient communication of the current layer with the forward computation of the next layer.  Unlike FIFO queue-based scheduling, this method divides layers into fixed-sized slices and prioritizes synchronizing slices based on the order in which they are processed in forward propagation. Therefore, the first layer gets the highest priority, with priority decrementing towards the end. 
When utilizing the parameter-server architecture for distributed model training, TicTac~\cite{hashemi2018tictac} prioritizes transfers that accelerate the critical path within the underlying computational graph.

ByteScheduler~\cite{peng2019bytescheduler} and PACE~\cite{bao2020preemptive}  are proposed to generalize priority-based communication scheduling across  training frameworks. Specifically,  ByteScheduler~\cite{peng2019bytescheduler} introduces a unified abstraction to facilitate communication scheduling without disrupting the original dependencies within framework engines. ByteScheduler achieves good performance by using Bayesian optimization to automatically tunes two critical parameters: partition size and credit size.  
PACE~\cite{bao2020preemptive}  implements preemptive communications by segmenting primitive  AllReduce operations into smaller pieces.  The preempted  AllReduce operators can be resumed at a later time. This preemption strategy  prevents head-of-line blocking of large communication tensors.  Additionally, PACE uses a dynamic programming approach to  fuse small communication tensors to reduce the overhead caused by handling a large number of small tensors, thereby achieving more efficient bandwidth utilization.

To improve bandwidth efficiency in MoE systems, Lina~\cite{li2023accelerating}  prioritizes All-to-All operations over AllReduce. Typically, expert-parallel (All-to-All) and data-parallel (AllReduce) processes use separate CUDA streams, causing potential overlap and bandwidth sharing without coordination. Lina breaks tensors into smaller chunks,  ensures All-to-All operations get full bandwidth while allowing AllReduce micro-ops to run during idle time. In addition, micro-ops enable overlap  All-to-All operations with expert computations.

\subsubsection{Decomposition-based Scheduling}
\label{decomposition-based}

Several advancements have focused on decomposing communication and computation operations into fine-grained tasks, reordering these operations with greater flexibility  to maximize overlap and optimize execution efficiency.

\phb{Pipeline Stage Decomposition.} When using conventional pipeline parallelism, each GPU stores a contiguous segment of layers. Breadth-First \cite{lamy2023breadth} further splits these contiguous stages into finer-grained stages distributed across different GPUs, forming a loop by connecting the first and last GPUs, so each GPU is assigned multiple stages. This allows the given micro-batch to reach the end of the pipeline earlier, reducing pipeline bubbles. Breadth-First uses a breadth-first scheduling strategy to achieve greater computation-communication overlap. 
Fold3D \cite{li2023fold3d} employs an all-in-all-out scheduling strategy to overlap the pipeline's gradient synchronization process with computation. This involves further folding model fragments within the pipeline, where each device contains two model fragments, allowing one fragment's gradient synchronization to overlap with another fragment's forward or backward computation. 

Asynchronous pipeline parallelism relaxes data dependencies between gradients and parameter updates. Leveraging this characteristic, TriRace \cite{li2024trirace} postpones parameter updates to maximize computation overlap with gradient communication. Additionally, TriRace decomposes bidirectional P2P communication between pipeline stages into two separate unidirectional operations and prioritizes them based on critical path analysis.

\phb{Communication Decomposition.}  Communication primitives could be decomposed into fine-grained operations with high scheduling flexibility.  Wang et al.~\cite{wang2022OverlapCommunication} decomposed communication operations (e.g., AllGather and ReduceScatter), into a series of fine-grained peer-to-peer collections. In addition, computational operations (e.g., Einstein Summation) were divided into fine-grained tasks, each performing a part of the computation. This decomposition creates more opportunities for overlapping  communication with computation.
SYNDICATE~\cite{mahajan2023syndicate} segments communication operations into smaller sub-operations, termed Motifs, and employs a Central Optimizer using Markov Chain Monte Carlo search to achieve optimal overlap execution plans. 
Centauri~\cite{chen2024centauri} adopts a different approach by using Primitive Partition, Group Partition, and Workload Partition to decompose communication operations into fine-grained atomic operations. These operations are then scheduled using Workload-aware Scheduling, Backward Scheduling, and Elastic Scheduling to maximize overlap efficiency. DeAR~\cite{zhang2023dear} also decomposes communication primitives, specifically breaking down  AllReduce  into AllGather and ReduceScatter. This decomposition allows subsequent operations to overlap with the forward propagation process of the model, thus eliminating the need to wait for the completion of both communication steps.

\phb{Computation Decomposition.} When using tensor parallelism, an AllReduce communication is required to synchronize the matrix multiplication outputs in the forward phase. CoCoNet~\cite{jangda2022coconet} facilitates the overlap of matrix multiplication and AllReduce by partitioning the output into smaller blocks and immediately initiating the AllReduce kernel after computing each result block within the matrix multiplication kernel. To minimize waiting time for the AllReduce kernel, the data blocks are fed into the matrix multiplication kernel in a carefully scheduled order. T3~\cite{pati2024t3} applies a hardware-software co-design approach, which transparently overlaps matrix multiplication with communication while minimizing resource contention. At the hardware level, T3 introduces a track-and-trigger mechanism to orchestrate the producer's compute and communication activities. Additionally, it employs compute-enhanced memories to handle the attendant compute operations required by the communication processes.

The backward pass generates two types of gradients: the output gradient, which is used to calculate the gradients of the preceding layer, and the weight gradient, which is used to update the layer’s weight parameters. These weight gradients need to be synchronized with other ranks using AllReduce. Conventional frameworks simultaneously perform gradient computation for both weights and outputs. Out-of-order backpropagation (ooo-backprop) ~\cite{oh2022out-of-order} decouples the gradient computations for weights and outputs, scheduling the weight gradient computations flexibly out of their original order. This allows more critical computations to be prioritized and scheduled accordingly. Consequently, ooo-backprop optimizes overall performance by scheduling communications based on this out-of-order computation strategy. This scheme is also used by Zero Bubble \cite{qi2023zero} to reduce the bubble rate of pipeline parallelism.

With activation checkpointing enabled, training frameworks need to recompute activations during the backward phase. This recomputation also involves AllReduce communication when using tensor parallelism. Oases~\cite{li2023oases} reduces redundant communication in recomputation by always placing  AllReduce communication as the last forward communication operation of a recomputation unit, and further splits the batch into smaller sub-batches, allowing the communication and computation of two batches to overlap. Lynx~\cite{chen2024lynx} also exploits the potential of recomputation and communication overlap, using two recomputation scheduling algorithms, OPT and HEU, to search for the optimal or near-optimal recomputation scheduling strategy, achieving the best overlap and training performance.

\subsection{In-Network Aggregation}
\label{in-network}

In-network aggregation (INA) uses the computational capabilities of network devices to perform aggregation operations like summing gradients of deep learning models. This technique has been previously proposed to accelerate big data processing. Notably, frameworks like NetAgg \cite{mai2014netagg}, SwitchAgg \cite{yang2019switchagg}, and CamDoop \cite{costa2012camdoop} have demonstrated significant performance advantages by executing data aggregation at switch-attached high-performance middleboxes or servers within a direct-connect topology.
Many approaches have been proposed to apply in-network aggregation to  deep learning model training, aiming to reduce the data exchanged between nodes during  AllReduce operations on gradients in the backward phase \cite{zhu2024when}. 

\subsubsection{Ethernet-based Aggregation} 
\label{ethernet-based}

Many Ethernet-based in-network aggregation  systems depend on programmable switches, and can be leveraged for distributed LLM training.  SwitchML \cite{sapio2021switchml} supports offloading collective communication operations to programmable network switches during the backward phase of distributed training. Since complete model updates can exceed the storage capacity of a switch, SwitchML streams the aggregation through the switch, processing the aggregation function on a limited number of vector elements at a time. 
There are two limitations of SwitchML. First, when dealing with floating-point operations, SwitchML cannot directly perform collective communications (such as  AllReduce ) for floating-point tensors. Instead, it converts floating-point values into 32-bit integers using a block floating-point-like approach. Second, SwitchML is primarily implemented on DPDK, and while there is an RDMA-capable implementation, it is difficult to integrate with training frameworks.

To better facilitate distributed model training, FPISA \cite{yuan2022unlocking} implements floating-point computation as a P4 \cite{bosshart2014p4} program running directly on a programmable switch. 
Therefore, training frameworks could offload collective communication operations on FP16 tensors to switches without converting them to 32-bit integers.
NetReduce \cite{liu2020netreduce} supports in-network aggregation compatible with RoCE, fully utilizing the congestion control and reliability design of RoCE without the need for costly network protocol processing stacks in switches. NetReduce is prototyped with an FPGA board attached to an Ethernet switch. 
AllReduce-Switch \cite{liu2021allreduceswitch} is closely related to NetReduce and compatible with its network protocol. It introduces a novel switch architecture tailored for in-network aggregation tasks and has implemented a prototype using FPGA hardware. PANAMA \cite{gebara2021panama} and ATP \cite{lao2021atp} have also contributed to the field with their in-network aggregation frameworks designed for shared environments. PANAMA focuses on optimizing network load by managing bandwidth allocation among multiple active training jobs concurrently. It addresses the challenge that traditional congestion controls may not adequately support simultaneous training operations. ATP, on the other hand, enables multiple concurrent tenants to run several jobs simultaneously, emphasizing the support for diverse workloads in shared environments.

Certain works are tailored for specific training workloads, making them unsuitable for distributed LLM training. For example, Libra \cite{pan2022libra} is designed for sparse model training using a parameter server architecture. It offloads the aggregation of frequently updated parameters to programmable switches, leaving infrequently updated parameters to be handled by servers. This approach effectively reduces server load. On the other hand, iSwitch \cite{li2019iswitch} is designed for parameter aggregation in reinforcement learning training tasks. Although its FPGA-based implementation supports native floating point operations, it operates at a significantly lower bandwidth. Furthermore, iSwitch stores an entire gradient vector during aggregation, which is feasible for reinforcement learning workloads but does not scale well for large-scale models, especially LLMs.

\subsubsection{Infiniband-based Aggregation} 
\label{infiniband-based}

NVIDIA Mellanox's Scalable Hierarchical Aggregation Protocol (SHARP) \cite{graham2016sharp} is a proprietary in-network aggregation scheme available in certain InfiniBand switches and NVIDIA GPUs. Built on InfiniBand, SHARP leverages link-layer flow control and lossless guarantees and employs dedicated on-chip FPUs for collective offloading. SHARPv1 was introduced on InfiniBand EDR switches, and SHARPv2 was enhanced on InfiniBand HDR switches with features such as support for collective communications (e.g., Barrier, Reduce,  AllReduce and Broadcast), integer and floating-point operations (16/32/64 bits), and GPUDirect RDMA. SHARPv2 also uses streaming aggregation for large vector reductions at line rate, integrates with NCCL, and is easily usable by existing training frameworks. Enabled on the latest InfiniBand NDR switches, SHARP is production-ready for distributed LLM training and has been deployed in many training clusters.
In addition to Infiniband, NVIDIA's NVSwitch-v3 \cite{nvswitch} also integrates SHARP  to speed up collective operations in GPU-based clusters.



\section{Fault Tolerance}
\label{sec:resi}

LLM training typically involves extended training periods ranging from weeks to months, utilizing clusters of tens of thousands of GPUs. The vast array of components involved, spanning from the underlying infrastructure to training system optimizations, necessitates robust fault tolerance mechanisms to ensure the reliability of training processes. This is because a single point of failure in any part of the system can result in a suspension of the training process due to the synchronous nature of the training. In this section, we first present a failure analysis in LLM training, then investigate the approaches for fast failure detection and recovery.

\tikzstyle{my-box}=[
    rectangle,
    draw=black,
    rounded corners,
    text opacity=1,
    minimum height=1.5em,
    minimum width=5em,
    inner sep=2pt,
    align=center,
    fill opacity=.5,
    line width=0.5pt,
]
\tikzstyle{leaf}=[my-box, minimum height=1.5em,
    fill=hidden-red!10, text=black, align=left,font=\normalsize,
    inner xsep=2pt,
    inner ysep=4pt,
    line width=0.8pt,
]

\begin{figure*}[htpb]
    \centering
    \resizebox{\textwidth}{!}{
        \begin{forest}
        forked edges,
        for tree={
            grow=east,
            reversed=true,
            anchor=base west,
            parent anchor=east,
            child anchor=west,
            base=center,
            font=\large,
            rectangle,
            draw=black,
            rounded corners,
            align=left,
            text centered,
            minimum width=4em,
            edge+={black, line width=1pt},
            s sep=3pt,
            inner xsep=2pt,
            inner ysep=3pt,
            line width=0.8pt,
            ver/.style={rotate=90, child anchor=north, parent anchor=south, anchor=center},
        },
        where level=1{text width=15em,font=\normalsize,}{},
        where level=2{text width=15em,font=\normalsize,}{},
        [
        {Fault Tolerance for LLM Training}, ver
                [
                {Anomaly Detection}, fill=blue!10 
                    [
                    {Statistical Monitoring}, fill=yellow!10
                        [
                        Healthd in TPUv4~\cite{zu2024ResiliencyScaleManaging}{, }
                        MegaScale~\cite{jiang2024MegaScaleScaling}{, }
                        C4~\cite{dong2024boosting}{, } \\
                        Vela~\cite{gershon2024infrastructure}{, } 
                        Unicron~\cite{he2023UnicronEconomizing}{, }
                        Transom~\cite{wu2023TRANSOMEfficient}{, }
                        NCCLX~\cite{llama3}{, } \\
                        NCCL flight recorder~\cite{ansel2024pytorch},
                        leaf, 
                        text width= 26em
                        ]
                    ]
                    [
                    {Proactive Validation}, fill=yellow!10
                        [
                        Lightweight Test in MegaScale~\cite{jiang2024MegaScaleScaling}{, }
                        SuperBench~\cite{xiong2024superbench}{, } \\
                        Vela~\cite{gershon2024infrastructure}{, } 
                        Preflight Check in TPUv4~\cite{zu2024ResiliencyScaleManaging},
                        leaf, 
                        text width= 26em
                        ]
                    ]
                ]
                [
                {Checkpointing-Based Recovery}, fill=blue!10
                    [
                    {Persistent Checkpoiting}, fill=yellow!10
                        [
                        \textbf{Synchronous:}{ }
                        DeepSpeed~\cite{rasley2020deepspeed}{, }
                        Varuna~\cite{athlur2022VarunaScalable}{, } \\
                        JIT-Checkpointing~\cite{gupta2024just}{, } 
                        Flash-Checkpoint~\cite{dlrover2024antgroup}{, } \\
                        Universal Checkpointing~\cite{lian2024universal} \\
                        \textbf{Snapshot-Stall:}{ }
                        Check-N-Run~\cite{eisenman2022CheckNRuncheckpointing}{, } \\
                        TorchSnapshot~\cite{TorchSnapshot}\\
                        \textbf{Asynchronous:}{ }
                        DeepFreeze~\cite{nicolae2020deepfreeze}{, }
                        CheckFreq\cite{mohan2021checkfreq}{, } \\
                        LightCheck~\cite{chen2023cost}{, } 
                        DataStates-LLM~\cite{maurya2024datastates}{, } \\
                        FastPersist~\cite{wang2024fastpersist},
                        leaf, 
                        text width= 26em
                        ]
                    ]
                    [
                    {In-Memory Checkpoiting}, fill=yellow!10
                        [
                        Gemini~\cite{wang2023GEMINIFast}{, }
                        REFT~\cite{wang2023ReliableEfficientInMemory},
                        leaf, 
                        text width= 26em
                        ]
                    ]
                ] 
                [
                {Checkpointing-Free Recovery}, fill=blue!10
                    [
                    {Live Migration}, fill=yellow!10
                        [
                        Parcae~\cite{duan2024ParcaeProactive}{, }
                        Oobleck~\cite{jang2023OobleckResilient},
                        leaf, 
                        text width= 26em
                        ]
                    ]
                    [
                    {Module Redundancy}, fill=yellow!10
                        [
                        Bamboo~\cite{thorpe2023BambooMakinga}{, }
                        SlipStream~\cite{gandhi2024slipstream}{, }
                        SWARM~\cite{ryabinin2023SWARMParallelism},
                        leaf, 
                        text width= 26em
                        ]
                    ]
                ]
            ]
        \end{forest}
    }
    
    \caption{Studies on fault tolerance techniques for distributed LLM training.}
    \label{taxonomy:resilience}
\end{figure*}
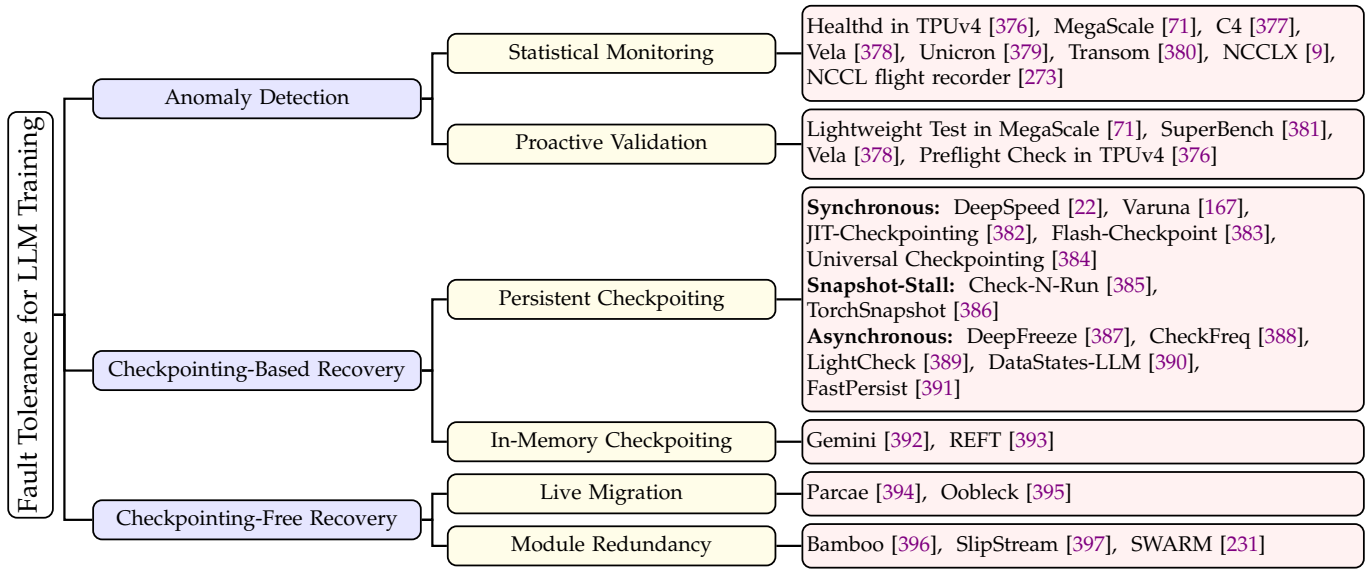

\subsection{LLM Failure Analysis}

Empirical evidence from various sources underscores the frequency of failures in LLM training. For example, the training of Bloom experiences 1-2 GPU failures per week on average on a cluster with 384 GPUs~\cite{le2023bloom}. Meta's comprehensive training records~\cite{meta2022opt175B} of 175B OPT model on 992 A100 GPUs document over 40 interruptions within a two-week period, attributed to hardware, infrastructure, and other external factors. More recent studies further highlight this issue. Acme~\cite{hu2024CharacterizationLarge} reported failure occurrences every 1-2 days on average in their training process using over 1,000 A100 GPUs. ByteDance's MegaScale project~\cite{jiang2024MegaScaleScaling}, utilizing 12,288 Ampere GPUs, experiences over 100 times failures in several weeks. Meta's LLaMA3 experiences 466 job interruptions during a 54-day period of pre-training on a cluster of 16,384 H100 GPUs~\cite{llama3}. The frequent failure is primarily attributed to the immense complexity and scale of these systems and extended training periods. The whole training system encompasses a vast array of components as we investigated in previous sections. Moreover, the synchronized training further exacerbates this issue, as errors in any single node can cause the entire job to fail, making the system particularly vulnerable to even isolated hardware or software faults. Even a seemingly low $1.5\%$ daily failure rate for a single node, as observed in Alibaba's cluster~\cite{dlrover2024antgroup}, translates to a staggering $84.8\%$ daily failure rate when scaled to a system with 1,000 GPUs. However, the trend of scaling up the training system continues to grow, emphasizing the concomitant challenges for fault tolerance mechanisms to maintain system reliability.


The reasons behind these failures are multifaceted and stem from various components of the LLM training system. According to Acme~\cite{hu2024CharacterizationLarge}, the most severe impact comes from hardware failures, such as issues with GPU (e.g., CUDA-Error, ECC-Error), NVLink, and network system (e.g., NCCL-Timeout-Error, Connection-Error). Similar observations are also delivered in Alibaba C4~\cite{dong2024boosting}. C4 further observes that the majority of errors (about $82.5\%$) are confined to specific nodes or even individual devices, although most errors observed by users are NCCL error. LLaMA3 pre-training~\cite{llama3} also reports that $78\%$ of the failures are hardware issues. Moreover, the latest generation GPUs (A100 and H100) tend to exhibit high error rates, likely due to rapid development, rushed delivery, and increased power consumption~\cite{dong2024boosting, gao2023empirical}. Beyond hardware, software-related issues in distributed training frameworks, data preprocessing pipelines, or library dependencies can lead to crashes or unexpected behavior~\cite{hu2024CharacterizationLarge, gao2023empirical, gershon2024infrastructure}. The complex nature of the models themselves can introduce instabilities such as loss spikes, numerical overflow or underflow, gradient explosions, or optimization difficulties~\cite{meta2022opt175B, chowdhery2022PaLMscaling}. External factors like power outages or cooling system failures in data centers further contribute to system instabilities. For example, the high temperature in the cluster server room also tends to result in GPU overheating, which can cause NVLink-Error or ECC-Error~\cite{hu2024CharacterizationLarge} or unstable training speed~\cite{llama3}.

These high frequent and multifaceted LLM failures lead to significant waste of GPUs. This inefficiency manifests in two primary ways: failure recovery and performance degradation. 
First, LLM training jobs periodically save checkpoints during runtime to maintain progress. Upon failure, system maintainers must first locate and diagnose the issue before restarting the training by rolling back to previous checkpoints. Some hardware failures, however, can be challenging to detect proactively and often require considerable time to diagnose and recover from, resulting in prolonged stalls in LLM training. 
Second, stragglers in the cluster, caused by network link failures~\cite{dong2024boosting} or abnormal computational slowdowns~\cite{jiang2024MegaScaleScaling}, can significantly decrease the MFU, further compounding the overall training inefficiency.
The training of Meta's 175B OPT model exemplifies these inefficiencies~\cite{meta2022opt175B}.
While the ideal training time was estimated at about 25 days based on the MFU, the actual training lasted approximately 57 days. This means that a staggering $56\%$ of the total time was wasted handling various failures, underscoring the severe impact of system instabilities on resource utilization and training efficiency in LLM training.

\subsection{Anomaly Detection}
\label{subsec:AnomalyDetection}

Rapid detection and diagnosis of LLM failures are crucial for maintaining training stability and efficiency. This process, known as anomaly detection, primarily employs two approaches: statistical monitoring and proactive validation.

\subsubsection{Statistical Monitoring}
\label{subsubsec:StatisticalMonitoring}

Statistical monitoring is a systematic approach to observing and analyzing various metrics and indicators throughout the LLM training process. This method involves collecting, processing, and interpreting data to identify anomalies or deviations from expected behavior. In a typical setup, each GPU is assigned a dedicated monitoring process responsible for collecting basic information and runtime statistics~\cite{jiang2024MegaScaleScaling, he2023UnicronEconomizing, gershon2024infrastructure}. These statistics are then transmitted to a central monitor node as heartbeat messages for further analysis. Nodes that fail to send heartbeat messages are considered to have failed. The primary objective of this monitoring system is to detect anomalies promptly, allowing for quick recovery that minimize training interruptions and maintain overall efficiency.

Most runtime statistics monitored in LLM training are hardware-related, encompassing both GPU and network metrics. Recent works~\cite{jiang2024MegaScaleScaling, he2023UnicronEconomizing, gershon2024infrastructure} collect GPU-related statistics with NVIDIA DCGM~\cite{nvidia-dcgm}, including SM block utilization, SM occupancy, SM pipe utilization, PCIe traffic rate, NVLink traffic rate and etc. One frequently occurring issue is GPU memory row-remapping, which seamlessly replaces known degraded memory cells with sparse ones in hardware. Vela~\cite{gershon2024infrastructure} detects this by leveraging the \texttt{DCGM\_FI\_DEV\_ROW\_REMAP\_PENDING} statistics from DCGM. Megascale~\cite{jiang2024MegaScaleScaling} and Transom~\cite{wu2023TRANSOMEfficient} also detect errors by analyzing errors occurred in training logs.

In addition to GPU metrics, network statistics are crucial for monitoring distributed training performance. MegaScale~\cite{jiang2024MegaScaleScaling} tracks RDMA traffic metrics to detect potential anomalies. It also develops visualization system to identify inefficiency GPUs manually. Unicron~\cite{he2023UnicronEconomizing} detects errors like NCCL timeout, TCP timeout, and task hangs with delayed notification during training. C4~\cite{dong2024boosting} gathers connection specifics such as RDMA IP and QP numbers, along with message statistics including counts, sizes, and durations of transfers at the transport layer to detect training slowdowns and hangs. Collective communication activities can also be monitored with PyTorch’s built-in NCCL flight recorder~\cite{ansel2024pytorch}, which captures collective metadata and stack traces into a ring buffer for later diagnosis. Meta further co-designs NCCLX~\cite{llama3} with PyTorch, allowing PyTorch to access its internal state for fast and accurate failure detection. NCCLX traces the kernel and network activities of each NCCLX communication, which can help diagnose communication issues. Vela~\cite{gershon2024infrastructure} implements an enhanced Multi-NIC health checker that collects node network bandwidth data for all 2-node pairs on every port. This information can be utilized to detect nodes with degraded RoCE/GDR performance.  Leveraging the key characteristics of LLMs training as prior knowledge, Transom~\cite{wu2023TRANSOMEfficient} develops machine learning algorithms to do anomaly detection. 

Statistical monitoring also enables the resiliency of Google's TPUv4 supercomputer~\cite{zu2024ResiliencyScaleManaging}. Each TPUv4 machine is equipped with a \texttt{healthd} daemon that performs real-time monitoring of ICI (Inter-Chip Interconnect) links, PCIe links and TPU ASIC. Detected severe symptoms will notify cluster scheduler for appropriate action, such as evicting affected jobs or rescheduling them.


\subsubsection{Proactive Validation}
\label{subsubsec:ProactiveValidation}

Proactive validation offers an alternative to reactive troubleshooting based on online statistical monitoring, aiming to validate the training system before failures occur. However, a trade-off exists between validation test time and accuracy, as comprehensive validation can significantly impact effective training time. MegaScale~\cite{jiang2024MegaScaleScaling} introduces a suite of lightweight tests, including intra-network host and NCCL tests, to diagnose a wide spectrum of potential failures. Vela~\cite{gershon2024infrastructure} employs a two-tiered strategy with lightweight tests running periodically on every node and more intrusive tests executed only when nodes are idle. Google's TPUv4 supercomputer implements a preflight check~\cite{zu2024ResiliencyScaleManaging} before user jobs, consisting of an end-to-end check and an intent-driven checker for hardware health. SuperBench~\cite{xiong2024superbench} resents a comprehensive benchmark suite for evaluating individual hardware components, incorporating a selector to balance validation time against potential issue-related penalties.

\subsection{Checkpoint-Based Recovery}
\label{subsection:Checkpoint-basedRecovery}

Periodically saving the model states, i.e., \emph{checkpointing}, and resuming computation from the latest checkpoint after failures happen is the common practice for fault tolerant LLM training. However, this presents a dilemma: frequent checkpointing incurs high I/O overhead, while infrequent checkpointing results in substantial progress loss when failures occur. To address this dilemma, fast persistent and in-memory checkpointing approaches have been designed.


\subsubsection{Persistent Checkpointing}
\label{subsection:PersistentCheckpoint}

Persistent checkpointing involves saving model states to non-volatile storage, e.g. SSD and remote cloud storage, ensuring data persistence across system failures. The process typically consists of two phases: first, the \emph{snapshot} phase copies model states from GPU to CPU memory, and second, the \emph{persist} phase writes the snapshots to persistent storage devices. Despite considerable I/O overhead due to the low bandwidth of storage devices, persistent checkpointing remains a widely used approach for fault tolerance due to its ease-of-use and reliability. Advanced persistent checkpointing approaches have been proposed to reduce training stall, thereby enabling more frequent checkpointing without significant performance penalties.

\phb{Synchronous Checkpointing.}
To keep consistency of model parameters, DeepSpeed's default synchronous checkpointing~\cite{rasley2020deepspeed} and Varuna~\cite{athlur2022VarunaScalable} periodically stalls the training process to perform checkpointing to persistent storage synchronously on data parallel rank 0. This approach results in GPU idle time during both the snapshot and persist phases, leading to resource underutilization. Recognizing that most failures are attributable to a single GPU or network device, JIT-Checkpointing~\cite{gupta2024just} proposes an alternative strategy. It takes just-in-time checkpoints immediately after failures occur, allowing training to resume from these JIT checkpoints. This approach significantly reduces the cost of wasted GPU time, limiting it to at most one mini-batch iteration of work. DLRover Flash-Checkpoint~\cite{dlrover2024antgroup} accelerates the migration efficiency utilizing a distributed caching service. Universal Checkpointing~\cite{lian2024universal} introduces a universal checkpoint representation to decouple the distributed checkpoints storage from parallelism techniques. Universal Checkpointing can seamlessly transform checkpoints from one parallelization strategy to another upon demands. 

\phm{Snapshot-Stall Checkpointing.}
To reduce LLM training stalls while maintaining checkpoint consistency, Check-N-Run~\cite{eisenman2022CheckNRuncheckpointing} decouples the snapshot and persist phases. It achieves atomic checkpointing by stalling training only during the snapshot phase and asynchronously persisting snapshots using dedicated background CPU processes. TorchSnapshot~\cite{TorchSnapshot} further optimizes this process through tensor chunking and multi-threaded disk writing. By creating chunked snapshots, TorchSnapshot allows the persist phase to begin earlier through parallel writing, thereby reducing overall training stall time.
MegaScale~\cite{jiang2024MegaScaleScaling} and InternEvo~\cite{cai2024internlm2} also adopt a snapshot-stall approach for fast checkpointing and recovery. The snapshot phase stalls training for several seconds to capture the model states, while the persist phase asynchronously transfers checkpoints from CPU memory to a distributed file system. MegaScale optimizes the recovery process by designating a single worker within the data parallel group to read from the distributed file system, thus mitigating the low bandwidth bottleneck. This worker then broadcasts the checkpoint data to other GPUs, enabling faster and more efficient recovery across the entire system. To save storage space, InternEvo also asynchronously moves checkpoints from expensive hot storage to cheaper cold storage.

\phm{Asynchronous Checkpointing.}
Asynchronous checkpointing aims to minimize training stall by executing the snapshot and persist phases concurrently with training. DeepFreeze~\cite{nicolae2020deepfreeze} applies both lightweight (snapshot) and heavy (persist) persistence strategies in the background, sharding checkpoints across data-parallel GPUs to distribute I/O workload. CheckFreq~\cite{mohan2021checkfreq} carefully pipelines the snapshot and persist phases with subsequent iteration's forward and backward passes, ensuring snapshot completion before the next parameter update. It also dynamically tunes checkpointing frequency to balance recovery costs and runtime overhead. LightCheck~\cite{chen2023cost} exploits inter-iteration data dependencies, introducing layer-wise checkpointing pipeline to reduce stall. DataStates-LLM~\cite{maurya2024datastates} addresses slow host memory allocation by pre-allocating pinned host memory for snapshots and utilizes efficient computation, snapshot, and persist layer-wise pipelining. FastPersist~\cite{wang2024fastpersist} identifies risks in fully asynchronous persist phases and synchronizes them with the next iteration's parameter update. It improves SSD bandwidth utilization through double-buffering pinned memory and reduces hardware contention by using a subset of data-parallel ranks for checkpoint writing.

\subsubsection{In-Memory Checkpointing}
\label{subsubsection:In-MemoryCheckpointing}

The low bandwidth of remote persistent storage severely restricts the frequency of checkpointing, in-memory checkpointing addresses the limitations by storing checkpoints in the memory of other compute nodes or dedicated in-memory storage systems, significantly reducing I/O overhead and enabling higher checkpointing frequencies. Gemini~\cite{wang2023GEMINIFast} proposes checkpointing to CPU memory for faster failure recovery, along with a checkpoint placement strategy to minimize checkpoint loss and a traffic scheduling algorithm to reduce interference with training. REFT~\cite{wang2023ReliableEfficientInMemory} asynchronously caches model states to host memory and in-memory storage like Redis, bypassing checkpoint I/O and enabling high checkpointing frequency. It also leverages erasure coding to implement RAIM5 (inspired by RAID5 with ``Disk'' replaced by ``Memory'') that protects data against node failures. While these approaches significantly advance fault tolerance for LLM training by enabling more frequent checkpointing without performance penalties, they may not provide the same long-term data persistence as traditional storage-based methods. Consequently, hybrid approaches combining both in-memory and persistent checkpointing is necessary for comprehensive fault tolerance strategies.

\subsection{Checkpoint-Free Recovery}
\label{subsection:Checkpoint-FreeRecovery}

Checkpoint-free recovery methods aim to minimize training stalls by eliminating the need to restart and roll back to previous checkpoints when failures occur. These techniques depend on automatic failure detection mechanisms to identify issues promptly. When a failure is detected, checkpoint-free approaches automatically address the problem and allow the training process to continue without interruption. By avoiding the time-consuming process of loading from a checkpoint and repeating computations, these methods can significantly reduce downtime and improve overall training efficiency. Checkpoint-free recovery strategies can be broadly categorized into two main approaches: live migration and module redundancy.

\subsubsection{Live Migration}
\label{subsubsection:Live-Migration}

Live migration leverages the inherent redundancy present in distributed LLM training setups, particularly the model replicas across different data parallel pipelines, to restore  model states in case of failure. When a failure is detected, live migration approaches dynamically reconfigure the parallelization strategy using the remaining healthy instances or by incorporating new instances into the training cluster. The current model states are then transferred to these reconfigured nodes, allowing the training process to continue with minimal interruption.
Parcae~\cite{duan2024ParcaeProactive} proposes three distinct migration mechanisms, each with different communication overheads, to efficiently transfer model states between varying parallelization strategies. Oobleck~\cite{jang2023OobleckResilient} takes a pipeline template-based approach to live migration. It maintains a set of predefined pipeline templates and, upon detecting a failure, swiftly instantiates new heterogeneous pipelines based on these templates.

\subsubsection{Module Redundancy}
\label{subsubsection:Module-Redundancy}

Module redundancy, like live migration, also leverages the redundancy of model states. However, instead of restoring the latest model states across different GPUs, this approach continues training by routing computation to redundant modules. Bamboo~\cite{thorpe2023BambooMakinga} places a redundant pipeline stage in the GPU holding an adjacent pipeline stage within the same pipeline. This redundant stage performs redundant computations during training, utilizing pipeline bubbles, and is activated as a normal stage upon failure. SlipStream~\cite{gandhi2024slipstream} leverages the redundancy across model replica pipelines, routing the computation of failed nodes to nodes in different data parallel pipelines. SWARM~\cite{ryabinin2023SWARMParallelism} proposes a similar solution but focuses more on poorly connected, heterogeneous, and unreliable devices. In addition to redundant computation, SWARM also incorporates instance migration to rebalance the pipeline, combining aspects of both redundant computation and live migration approaches.

\label{sec:workloads}

\section{Conclusion and Outlooks}
\label{sec:conclusion}

The rise of LLMs has transformed AI, enabling applications like personal assistants, code generation, and scientific research. Models such as GPT, LLaMA, and Gemini have set new standards, but training these massive models, exemplified by LLaMA-3's 54-day process on 16,384 GPUs, presents challenges in scalability, efficiency, and reliability. Managing vast GPU clusters requires innovative hardware and networking solutions. Efficient training demands optimizing computation, communication, and memory use. Reliability involves robust mechanisms to detect and recover from failures over long training periods. This survey reviews recent advancements in LLM training systems and infrastructure, highlighting approaches to enhance scalability, efficiency, and reliability.

Traditional digital circuit-based computing systems, guided by Moore’s Law and Dennard Scaling, are facing significant physical and economic constraints in meeting the computational demands for training and deploying LLMs. Consequently, the AI industry necessitates innovative solutions. One promising approach is large-scale optoelectronic integration technology, which replaces traditional digital circuits with integrated silicon photonics to enhance computational and communication capabilities \cite{NewDataCenterComputingParadigm}. This optoelectronic hybrid data center technology combines optical computing with optical networks, increasing single-node computing power and the efficiency of large-scale distributed computing. Several works have proposed leveraging optical networks for LLM training. For instance, TopoOpt \cite{weiyang2023topoopt} optimizes both the optical network topology and parallelization strategies in distributed training, enhancing computational and communication efficiency. TPUv4 \cite{jouppi2023TPUv4}  uses Optical Circuit Switches to dynamically reconfigure its 3D-Torus interconnect topology, improving data flow for the intensive communication patterns in LLM training. Additionally, Taichi \cite{xu2024large}  explores a distributed diffractive-interference hybrid photonic computing architecture to effectively scale the optical neural network to the million-neuron level with 160 tera-operations per second per watt (TOPS/W) energy efficiency.   
The future may necessitate a paradigm shift in LLM training and inference towards silicon photonics. However, this transition will require extensive innovation across system design and implementation.



\bibliographystyle{IEEEtran}
\bibliography{IEEEabrv, references}

\end{document}